\pdfoutput=1
\documentclass[preprint,12pt]{elsarticle}

\usepackage{mathtools}
\usepackage{amssymb}
\usepackage{amsmath}
\usepackage[T1]{fontenc}
\usepackage{float}
\usepackage{xcolor}

\usepackage{siunitx}
\usepackage{tikz}
\usepackage{upgreek}

\usepackage{lineno}

\usepackage{subcaption}
\usepackage{mwe}
\usepackage{pdfpages}

\usepackage{hyperref}
\hypersetup{colorlinks,allcolors=black}
\usepackage{comment}
\usepackage{physics}
\usepackage{accents}

\newcommand*{\dt}[1]{%
  \accentset{\mbox{\large\bfseries .}}{#1}}

\graphicspath{{figures/}{figures/dEdx/}}

\journal{Nuclear Instruments and Methods}

\begin{document}

\begin{frontmatter}





\title{Characterization of Charge Spreading and Gain of Encapsulated Resistive Micromegas Detectors for the Upgrade of the T2K Near Detector Time Projection Chambers}

\author[saclay]{D.~Atti\'e}
\author[ifae]{O.~Ballester}

\author[ifj]{M.~Batkiewicz-Kwasniak}
\author[lpnhe]{P.~Billoir}
\author[lpnhe]{A.~Blondel}
\author[saclay]{S.~Bolognesi}
\author[saclay]{R.~Boullon}
\author[saclay]{D.~Calvet}
\author[ifae,pilar]{M.~P.~Casado}

\author[bari]{M.G.~Catanesi}
\author[legnaro]{M.~Cicerchia}
\author[padova]{G.~Cogo}
\author[saclay]{P.~Colas}
\author[padova]{G.~Collazuol}

\author[padova]{D.~D'Ago}
\author[lpnhe]{C.~Dalmazzone}
\author[saclay]{T.~Daret}
\author[saclay]{A.~Delbart}
\author[ifae,annalisa]{A.~De Lorenzis}
\author[cern]{R.~de Oliveira}
\author[cern]{S.~Dolan}
\author[wut]{K.~Dygnarowicz}
\author[lpnhe]{J.~Dumarchez}

\author[saclay]{S.~Emery-Schrenk}
\author[saclay]{A.~Ershova}
\author[saclay]{G.~Eurin}
\author[padova]{M.~Feltre}
\author[padova]{C.~Forza}

\author[padova]{L.~Giannessi}
\author[lpnhe]{C.~Giganti}
\author[legnaro]{F.~Gramegna}
\author[padova]{M.~Grassi}
\author[lpnhe]{M.~Guigue}

\author[aachen]{P.~Hamacher-Baumann}
\author[saclay]{S.~Hassani\fnref{fnref1}}
\author[saclay]{D.~Henaff}

\author[padova]{F.~Iacob}
\author[ifae]{C.~Jes\'{u}s-Valls}
\author[saclay]{S.~Joshi\fnref{fnref2}}
\author[wut]{R.~Kurjata}

\author[padova]{M.~Lamoureux}
\author[napoli]{A.~Langella}
\author[saclay]{J.~F.~Laporte}
\author[warwick]{K.~Lachner}

\author[napoli]{L.~Lavitola}
\author[saclay]{M.~Lehuraux}
\author[padova,trieste]{S.~Levorato}
\author[padova]{A.~Longhin}
\author[ifae]{T.~Lux}

\author[bari]{L.~Magaletti}
\author[legnaro]{T.~Marchi}
\author[padova]{M.~Mattiazzi}
\author[cern]{M.~Mehl}
\author[lpnhe]{L.~Mellet}
\author[padova]{M.~Mezzetto}
\author[cern]{L.~Munteanu}

\author[wut]{W.~Obrębski}
\author[lpnhe]{Y.~Orain}
\author[padova]{M.~Pari}
\author[lpnhe]{J.-M.~Parraud}
\author[bari]{C.~Pastore}
\author[padova]{A.~Pepato}
\author[lpnhe]{E.~Pierre}
\author[ifae]{C.~Pio Garcia}
\author[cern]{O.~Pizzirusso}
\author[lpnhe]{B.~Popov}
\author[saclay]{J.~Porthault}
\author[ifj]{H.~Przybiliski}
\author[padova]{F.~Pupilli}

\author[aachen]{T.~Radermacher}
\author[bari]{E.~Radicioni}
\author[saclay]{C.~Riccio}
\author[trieste]{L.~Rinaldi}
\author[saclay]{F.~Rossi}
\author[aachen]{S.~Roth}
\author[lpnhe]{S.~Russo}
\author[wut]{A.~Rychter}

\author[saclay]{Ph.~Schune}
\author[padova]{L.~Scomparin}
\author[aachen]{D.~Smyczek}
\author[aachen]{J.~Steinmann}
\author[ifj]{J.~Swierblewski}

\author[cern]{A.~Teixeira}
\author[lpnhe]{D.~Terront}
\author[aachen]{N.~Thamm}
\author[lpnhe]{F.~Toussenel}
\author[bari]{V. Valentino}
\author[ifae]{M.~Varghese}

\author[saclay]{G.~Vasseur}
\author[cern]{E.~Villa}
\author[lpnhe]{U.~Virginet}
\author[saclay]{C.~Vuillemin}

\author[lpnhe]{U.~Yevarouskaya}
\author[wut]{M.~Ziembicki}
\author[lpnhe]{M.~Zito}
%

\address[saclay]{IRFU, CEA, Universit\'e Paris-Saclay, Gif-sur-Yvette, France}
\address[ifj]{H. Niewodniczanski Institute of Nuclear Physics PAN, Cracow, Poland}
\address[lpnhe]{LPNHE, Sorbonne Universit\'e, CNRS/IN2P3, Paris, France}
\address[bari]{INFN sezione di Bari, Universit\`a di Bari  e Politecnico di Bari, Italy}
\address[legnaro]{INFN: Laboratori Nazionali di Legnaro (LNL), Padova , Italy}
\address[padova]{INFN Sezione di Padova and Universit\`a di Padova, Dipartimento di Fisica e Astronomia, Padova, Italy}
\address[aachen]{RWTH Aachen University, III.~Physikalisches Institut, Aachen, Germany}
\address[ifae]{Institut de F\'isica d’Altes Energies (IFAE) - The Barcelona Institute of Science and Technology (BIST), Campus UAB, 08193 Bellaterra (Barcelona), Spain}
\address[wut]{Warsaw University of technology, Warsaw, Poland}
\address[napoli]{INFN Sezione di Napoli and Universit\`a di Napoli Federico II, Dipartimento di Fisica, Napoli, Italy}
\address[annalisa]{Qilimanjaro Quantum Tech, Barcelona 08007, Spain}
\address[pilar]{Departament de Física, Universitat Autònoma de Barcelona}
\address[cern]{CERN, European Organization for Nuclear Research, Geneva, Switzerland}
\address[warwick]{University of Warwick, Department of Physics, Coventry, United Kingdom}
\address[trieste]{INFN Sezione di Trieste, via Valerio 2 - 34127 Trieste, Italy}


\cortext[cor1]{Corresponding authors}
\fntext[fnref1]{samira.hassani@cea.fr}
\fntext[fnref2]{shivam.joshi@cea.fr}


\begin{abstract}
An upgrade of the near detector of the T2K long baseline neutrino oscillation experiment is currently being conducted. This upgrade will include two new Time Projection Chambers, each equipped with 16 charge readout resistive Micromegas modules.\\
 A procedure to validate the performance of the detectors at different stages of production has been developed and implemented to ensure a proper and reliable operation of the detectors once installed. A dedicated X-ray test bench is used to characterize the detectors by scanning each pad individually and to precisely measure the uniformity of the gain and the deposited energy resolution over the pad plane. An energy resolution of about 10\% is obtained.\\
A detailed physical model has been developed to describe the charge dispersion phenomena in the resistive Micromegas anode. The detailed physical description includes initial ionization, electron drift, diffusion effects and the readout electronics effects. The model provides an excellent characterization of the charge spreading of the experimental measurements and allowed the simultaneous extraction of gain and $RC$ information of the modules.
\end{abstract}


\end{frontmatter}
\newpage
\tableofcontents
\section{Introduction}

The study of neutrino oscillations entered the precision era with long-baseline experiments based on accelerator beams, like T2K and NOvA. In these experiments, neutrino oscillations are measured by comparing the neutrino fluxes and spectra measured at near detectors, placed nearby the neutrino source, and at far detectors, placed hundreds of kilometres away from the source. 
The future of neutrino oscillation studies is promising. In particular, the analysis of neutrino and antineutrino oscillations at T2K provides first exciting hints of CP violation in the leptonic sector~\cite{Abe:2019vii}. Such results are still limited by the available statistics, therefore T2K will start a new data taking phase, with increased beam power, in 2023.

The increase in statistics is posing unprecedented challenges: the precision measurements require accurate modelling of neutrino interactions and of the detector response. In order to cope with these challenges, a new generation of subdetectors for the T2K near detector, ND280, have been developed.
The goal of these new subdetectors is to improve the near detector performance~\cite{T2K:2019bbb}, to measure the neutrino flux and to constrain the neutrino interaction cross-sections~\cite{Dolan:2021hbw} in order to reduce the uncertainty on the number of predicted events at far detector (Super-Kamiokande) below 4\%. This goal is achieved by modifying the upstream part of the detector, adding a new highly granular scintillator detector (Super-FGD)~\cite{Blondel:2020hml}, two new horizontal high angle Time Projection Chambers (HA-TPC) and six Time Of Flight planes~\cite{Korzenev:2021mny}. 

Each endplate of the HA-TPC will be instrumented with eight Encapsulated Resistive Anode Micromegas (ERAM). In the case of resistive anode Micromegas, a resistive layer is deposited onto the segmented anode in order to spread the charge onto several adjacent pads. This way, the spatial resolution for a given segmentation is improved. Additionaly, this method improves  the Micromegas stability and protects the electronics against sparking events. 
Detector prototypes for the new TPCs have been successfully tested at CERN and DESY test-beams~\cite{Attie:2021yeh,Attie:2019hua,Attie:2022smn} validating the detector technologies and their performance.  The test beam data allowed for spatial and $\mathrm{d}E/\mathrm{d}x$ resolutions to be determined as a function of the angle of the track with respect to the ERAM plane for all the drift distances of interest. A spatial resolution better than 800 $\mu$m and a $\mathrm{d}E/\mathrm{d}x$ resolution better than 10\% are measured for all the incident angles and all the drift distances.  \\

The paper is organized as follows, section \ref{sec:ERAM Production} explains in detail the design and production process of the ERAM. Section \ref{sec:Characterisation} describes the architecture of the readout electronics and its corresponding response function. A calibration of the electronics is used to verify the electronics model and its linearity of response. A procedure referred to as mesh pulsing, is used to validate the functioning of the ERAM modules, is detailed in section \ref{sec:meshPulsing}. A description of the X-ray test bench setup, used for detailed characterization of ERAM response, is given in section \ref{sec:setup}. A model for gain extraction in ERAM along with its application on actual X-ray data is demonstrated in section \ref{sec:gain_Fe55}. A signal model combining the charge spreading phenomena and electronics response is explained in section \ref{sec:simul_RCGain}. The model is then applied on X-ray data through simultaneous fit of waveforms. Section \ref{sec:RCmap} is dedicated to the study of the $RC$ map of ERAMs and their features, and comparing the values with the expected ones. Systematic uncertainties on $RC$ are also studied. Section \ref{sec:Gainmap-simulfit} shows the comparison between the gain obtained using two different methods. Also, gain non-uniformity within some ERAMs and within a pad is explained. In section \ref{sec:Gain_correction}, the effect of environmental conditions on the gain is studied. Finally, conclusions from this paper are drawn in section \ref{sec:conclusion}.

\label{sec:introduction}
\section{ERAM Production}
\label{sec:ERAM Production}
\begin{figure}[hbt!]
     \centering
     \begin{subfigure}[b]{0.36\textwidth}
         \centering
         \includegraphics[width=1.5\textwidth]{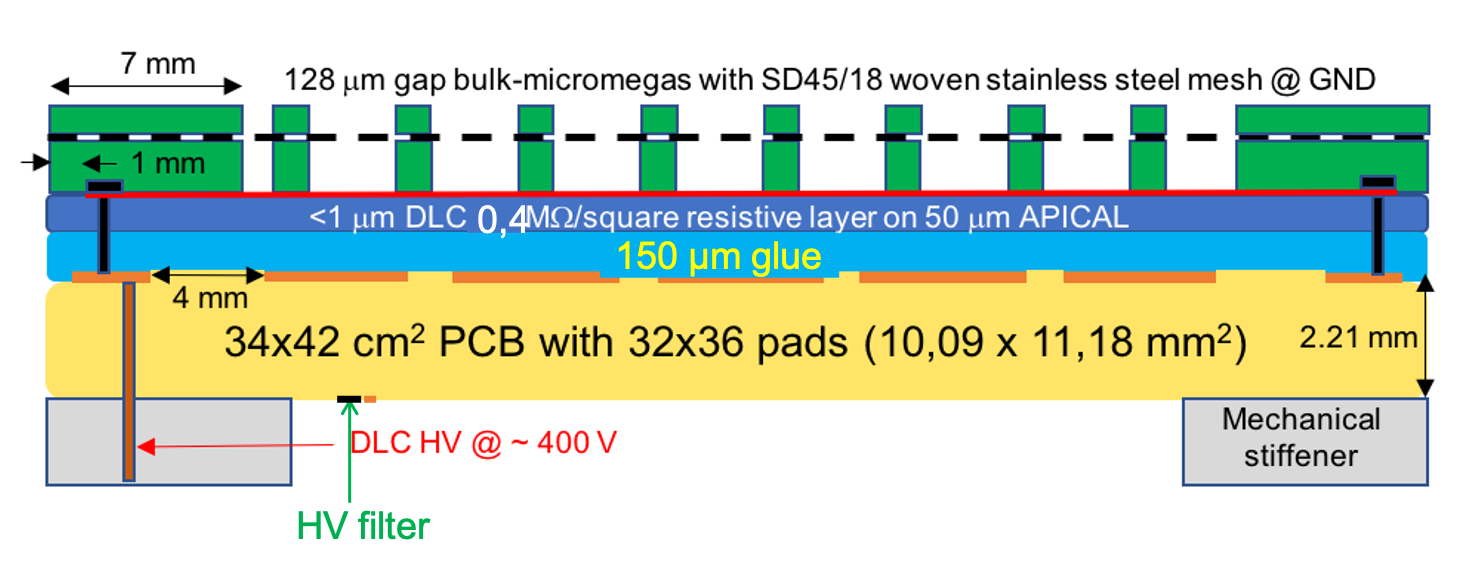}
     \end{subfigure}
     \hfill
     \begin{subfigure}[b]{0.45\textwidth}
         \centering
         \includegraphics[width=\textwidth]{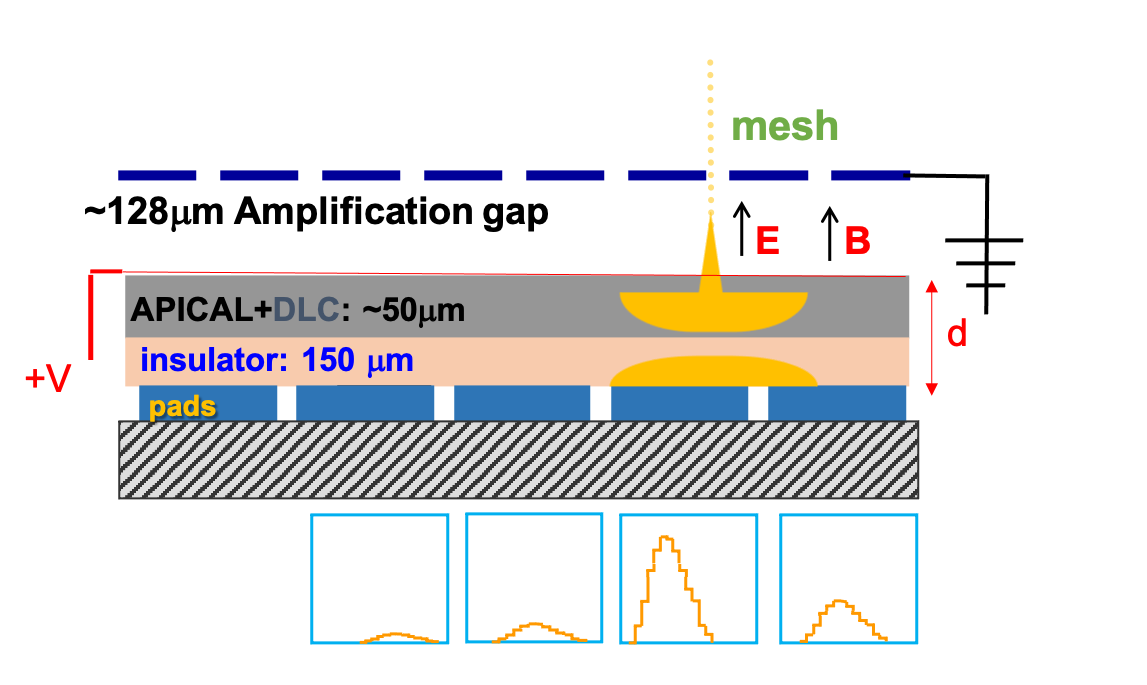}
     \end{subfigure}
        \caption{A sketch of the resistive Micromegas concept and characteristics.}
        \label{fig:sketch_ERAM}
\end{figure}

The ERAM module consists of a resistive Micromegas detector glued on an aluminium frame on which the readout electronics is fixed directly on its backside. The $42\times34$ cm$^2$ detector has 1152 pads of 11.18$\times$10.09~mm$^2$ disposed in a matrix of 36 pads along $x$ direction and 32 pads along $y$ direction. The pad plane is covered by a resistive layer made of an insulated $50~\mu$m Apical polyimide foil (pressed with $150~\mu$m glue), on which diamond-like carbon (DLC) is deposited by electron beam sputtering. This resistive layer technology enables to spread the charge over several pads in order to improve the spatial resolution. It can also improve the Micromegas stability and protect the electronics against sparking events. To guarantee a charge dispersion over at least two pads, a DLC surface resistivity $R$ of about 400~k$\Omega/\square$ was chosen. A schematic cross section of the ERAM detector and its specifications are presented in Figure~\ref{fig:sketch_ERAM}.
The production of the ERAMs, at the time of this paper writing is well underway, with 21 detectors produced and fully characterized out of the 32 (40 including spares) necessary for the equipment of the two endplates of the two HA-TPCs. A versatile infrastructure has been setup for the quality control: notably two test-benches, one for cosmic rays data-taking and one for test and calibration with X-ray photons from a $^{55}$Fe source. Both are fully instrumented with cooling, readout electronics and DAQ.

In Fall 2018, the global design of the upgraded ND280 detector was fixed. 
Two prototypes named MM1 (MM1-DLC1 and MM1-DLC2) with a 75~$\upmu$m thick glue layer and a final DLC resitivity close to 200~k$\Omega$/$\Box$ were produced.
One of these prototypes was mounted on a small TPC with a drift distance of 15 cm and extensively tested with cosmic rays and at DESY with an electron beam inside the 0.2 T PCMAG (Persistent Current Superconducting Magnet) in June 2019~\cite{Attie:2019hua}.
The different runs aimed at studying the impact of the main parameters
(selection of optimal peaking time for the readout electronics and anode voltage). 
The test allowed to validate the ERAM global design, including the PCB and some design choices for the final front-end electronics (removal of the external spark protection diodes at AFTER chip inputs and compatibility of some components for operation in a magnetic field of the expected strength).
 In general, the ERAM prototype demonstrated excellent robustness.
 It was operated at up to 380~V, collecting a large amount of signals, and without any damage either on the detector, or on the electronics.
 The DESY test beam allowed to further characterize the charge spreading, the resistive foil uniformity and to ensure that the performance satisfies the ND280 upgrade requirements with the final pad size.

 Following all the validations obtained with the MM0 and MM1 prototypes, the optimisation of the $RC$ constant of the detector needed to be tackled, where $R$ is the surface resistivity of the layer and $C$ is the capacitance per unit surface determined by the spacing between the anode and readout planes.
 Since a lower DLC resistivity means a less efficient spark protection, the DLC's resistivity specification was set to 400~$\pm$~60~k$\Omega$/$\Box$ (final DLC resistivity at the very end of the manufacturing process). 
It should be noted that mastering the $R$ value precisely is not an easy task at production level. Moreover the process of detector assembling (inducing pressing and warming of the DLC foil) causes a reduction of the resistivity value.
In order to better control the resistivity, a process of annealing ("baking" of the DLC foil in an oven) has been developed at CERN: with two hours of annealing at a temperature higher than the one reached during the manufacturing process (e.g. $ 200^\circ C $), a drop in resistivity of a factor of approximately 2 has been found and proven to be reproducible for different foils. 
This procedure allows to accommodate a larger value of resistivity at production/sputtering process and it avoids an uncontrolled drop of resistivity at the stage of detector production. Indeed, a second cycle of annealing at the same or lower temperature of the same foil does not change the resistivity further.

The last and critical specification is the glue layer thickness which determines the capacitance $C$ of the continuous $RC$ network. The glue is glass fiber tissue impregnated with epoxy resine. A trade-off must be found between charge spreading, signal amplitude and detector stability.
Two new pre-series ERAM prototypes were produced with the same batch of DLC foils as the MM1 detectors but with a thicker glue layer of 200~$\upmu$m. 

In terms of reproducibility, the metrology measurements showed less than 3~$\upmu$m difference in  glue's thickness between the two MM1 and the two pre-series ERAM prototypes. 
Their final resistivity was also well controlled. 
The two pre-ERAM prototypes have been extensively tested with cosmic rays using the mini-TPC. 

Some stability issues have been observed with the pre-series prototypes.
These instabilities were due to a defect in the mechanical assembly of the module. 
Corrections were done in the mechanical design and the assembly procedure. 
A Production Readiness Review triggered the launch of the production of the ERAM modules in November 2020. 
The $RC$ of the detector was chosen with an intermediate value of the glue thickness of 150 $\mu$m as a trade-off between the track reconstruction performances and the safe operation of the detector at a DLC high voltage working point of around 350~V. 
The first ERAM detector with final design (labelled ERAM-01) was produced and the series production of the ERAM modules was resumed as soon as this first module was validated with cosmic rays and a $^{55}$Fe X-ray source. 

In 2021, the TPC prototype was equipped with the first ERAM module and the HA-TPC readout electronics chain and it has been exposed to the DESY Test Beam~\cite{Attie:2022smn} in order to measure spatial and $\mathrm{d}E/\mathrm{d}x$ resolutions. Spatial resolution better than 800 $\mu$m and $\mathrm{d}E/\mathrm{d}x$ resolution better than 10\% were obtained for all the incident angles and for all the drift distances of interest. All the main features of the data are correctly reproduced by the simulation and the performance fulfills the requirements for the HA-TPCs of T2K.
\section{Readout Electronics}
\label{sec:Characterisation}
\begin{figure}[hbt!]
  \centering
  \includegraphics[width=0.85\textwidth]{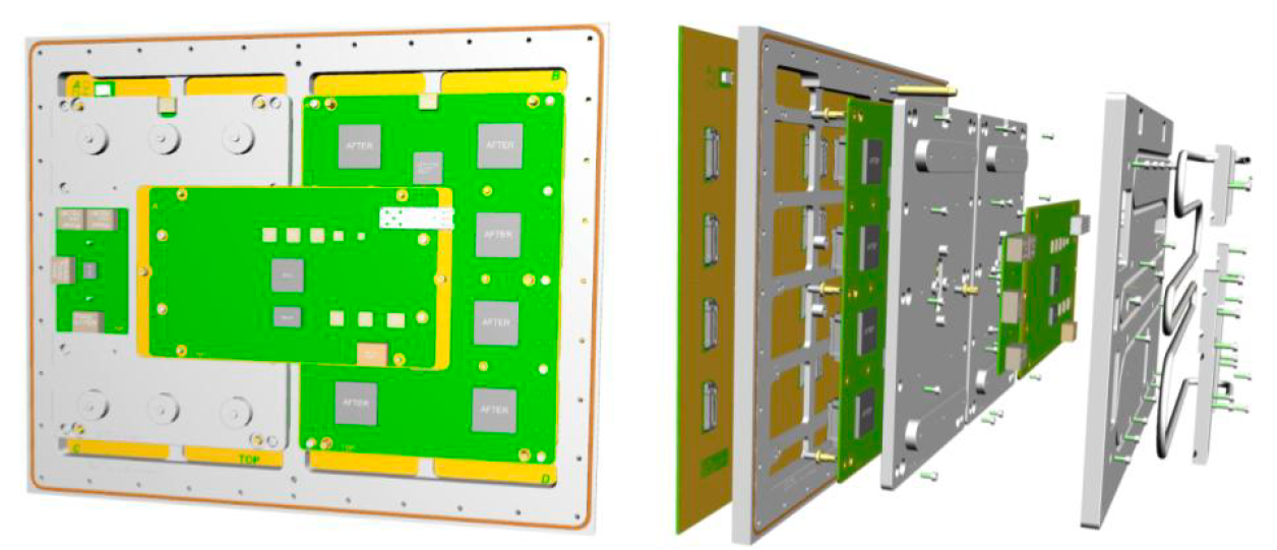}
  \caption{A CAD model of the AFTER chip-based electronics mounted on the detector (left) and in exploded view (right) to also visualise the shielding and cooling system pipe.}
  \label{fig:fecPhoto}
\end{figure}

The new readout electronics is based on the AFTER chip~\cite{AFTERREF} operated at a sampling frequency of 25~MHz, a peaking time of 200 or 412~ns and a gain such that 4096~ADC counts are obtained for a pulse current carrying 120~fC charge. Each ASIC reads 72 electronic channels connected to an array of $9 \times 8$ pads. A Front End Card (FEC) hosts eight AFTER ASICs and performs the digitization of the pad signals. Two FECs are needed to read a single ERAM module and are directly plugged on the detector PCB as illustrated in Figure~\ref{fig:fecPhoto}. Finally, a Front End Mezzanine (FEM) comes on top of the two FECs and synchronizes signal digitization with a master clock. Shielding covers are designed to allow the electronics to be placed and operated, along with the TPCs, inside the magnetic field. The temperature of the electronics is kept under control thanks to a cooling system circulating cold water throughout copper pipes in a close proximity to the ASICs. 
 Finally, the data coming from all 16~ERAM modules of a HA-TPC are sent via a set of optical fibers to a custom made board called the Trigger and Data Concentrator Module (TDCM)~\cite{Calvet:2018lac}.
\subsection{Description of the AFTER chip}
\label{subsec:DescriptionElectro}
\begin{figure}[hbt!]
  \centering
  \includegraphics[width=0.9\textwidth]{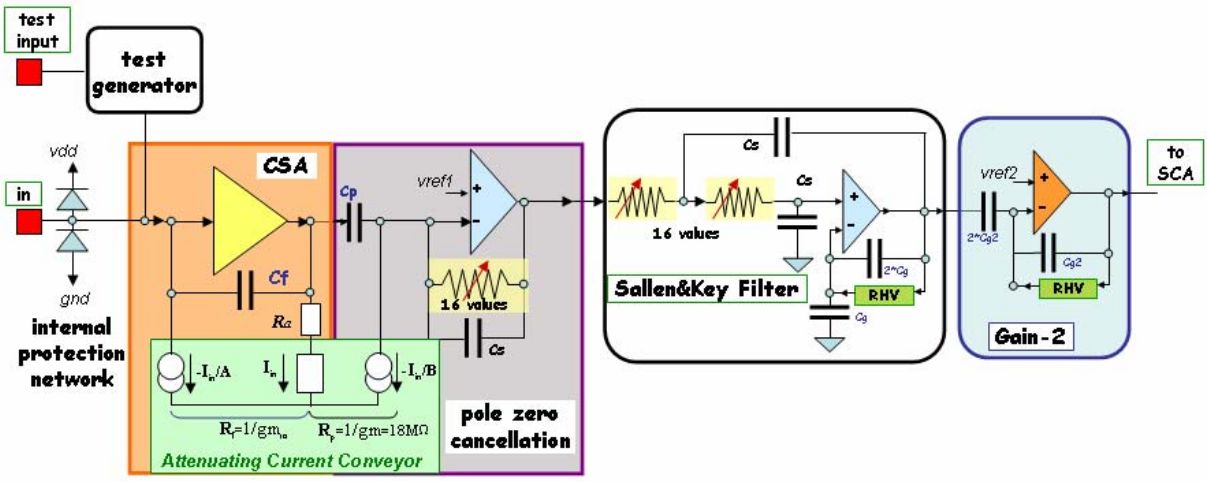}
  \caption{Architecture of the input signal condition stage of the AFTER chip. All 72 channels are identical.}
  \label{fig:AFTERFIG}
\end{figure}

The architecture of the AFTER chip~\cite{AFTERREF} is shown in Figure~\ref{fig:AFTERFIG}.
It is  composed  of a charge integration stage (CSA), a pole zero compensation stage, a Sallen \& Key filter, an amplifier, and an analog memory composed of a 511-cell switched capacitor array (SCA). Upon trigger, the SCA is frozen and all cells are sequentially digitized by an external Analog-Digital converter.

Assuming ideal operational amplifiers and exact pole zero compensation, 
the
response to a Dirac current pulse, prior to the late stage of discretization, is
found to be 
proportional to the function:
\begingroup
\small
\begin{equation}
f  ( t; w_s, Q)
=
 e^{ -w_{s}t }
+
   e^{ -   \frac{w_{s}t}{2 Q}}    
  \left[
  \sqrt{ \frac{ 2Q- 1 }{ 2Q +1 } }
  \sin \left( \frac{w_{s}t }{2} \sqrt{4-\frac{1}{Q^2}} \right)
-
  \cos \left( \frac{w_{s}t }{2} \sqrt{4-\frac{1}{Q^2}} \right)
  \right]
\label{equ:elecResponse}
\end{equation}
\endgroup
where $w_s$ and $Q$ depend on the values of the circuit resistances and capacitances~\cite{Luca:LUCAREF}.
Both parameters will be adjusted to fit electronics calibration data.

Most usefully, a proportionality factor can be written as:
\begin{equation}
ADC^{D}(t; w_s, Q) = \frac{ADC_{o}}{Q_{o}} \frac{f  ( t; w_s, Q)}{f_{max}(w_s, Q)}
  \label{equ:ElResponse2Dirac}
\end{equation}
where $f_{max}(w_s, Q)$ is the maximal value of the function $f(t;w_s, Q)$
\footnote{
It can be shown that $f_{max}$ does actually depend on $Q$ only.
}.
Parameterised in this way, the response to a Dirac pulse current carrying the charge $Q_{o}$, $i(t)= Q_{o} \delta(t)$, is $ ADC^D(t; w_s, Q ) = ADC_{o} \frac{f  ( t; w_s, Q )}{f_{max}(w_s, Q )} $, the maximal value of which is $ADC_{o}$.
So this parameterization implements the electronics gain, i.e.
the proportionality between charge input and ADC output,
which has been set to:
\begin{equation}
Q_{o} = 120 \text{ fC}
\text{ and }
ADC_{o}= 4096 \text{ counts}.
  \label{equ:Charge2ADCCorrespondance}
\end{equation}

This Charge-ADC correspondence holds for Dirac current pulses only.
For a slower current, the maximal value of the electronics response cannot be converted into a proper charge according to eq.~\ref{equ:Charge2ADCCorrespondance}. 

\subsection{Validation of the electronic model using calibration data}
\label{subsec:ValidationElectronicModel}
To test and validate the electronics model described in section~\ref{subsec:DescriptionElectro}, calibration data sets are collected using an on-board pulser with peaking times of 200 and 412~ns, various amplitudes, and with or without the detector connected. Each ERAM is paired with two FECs and then calibrated for use in the experiment. 

The pulses are fitted using the function $f( t; w_s, Q)$ (eq. ~\ref{equ:elecResponse}) first to ensure that the model correctly describes the electronic shape, but also to extract the $w_s$ and $Q$ parameter values and study the uniformity of the electronic responses between ASICs. Ideally, the parameters $w_s$ and $Q$ can be fixed independently from the amplitude of the input signal so that, while fitting the waveforms later in the analysis, the amplitude factor can be taken into account from the charge deposition function and therefore the electronic shape is completely determined. 

\begin{figure}[hbt!]
     \centering
     \begin{subfigure}[b]{0.49\textwidth}
         \centering
         \includegraphics[width=\textwidth]{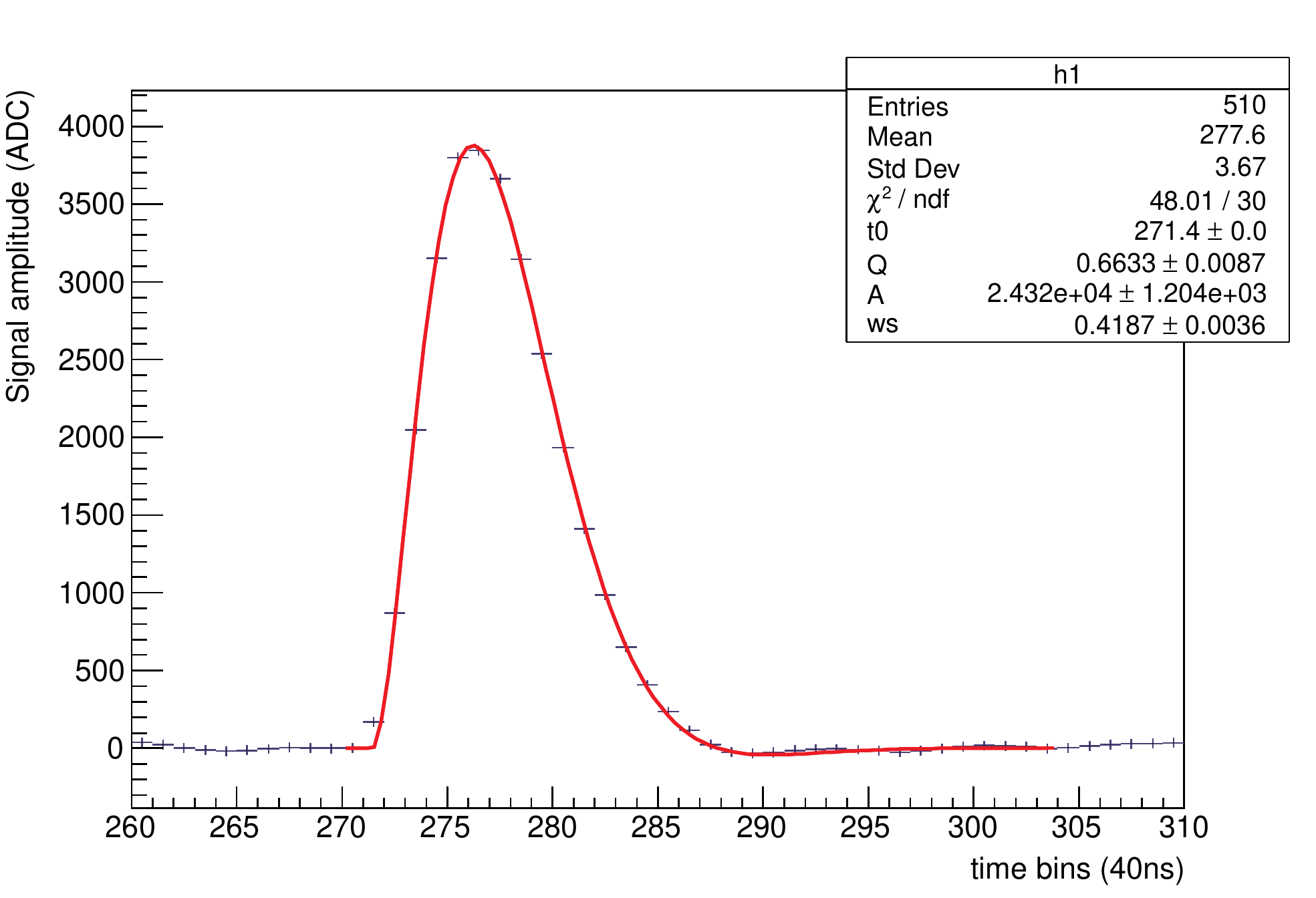}
          \caption{200ns peaking time signal}
         \label{fig:200fit}
     \end{subfigure}
     \hfill
     \begin{subfigure}[b]{0.49\textwidth}
         \centering
         \includegraphics[width=\textwidth]{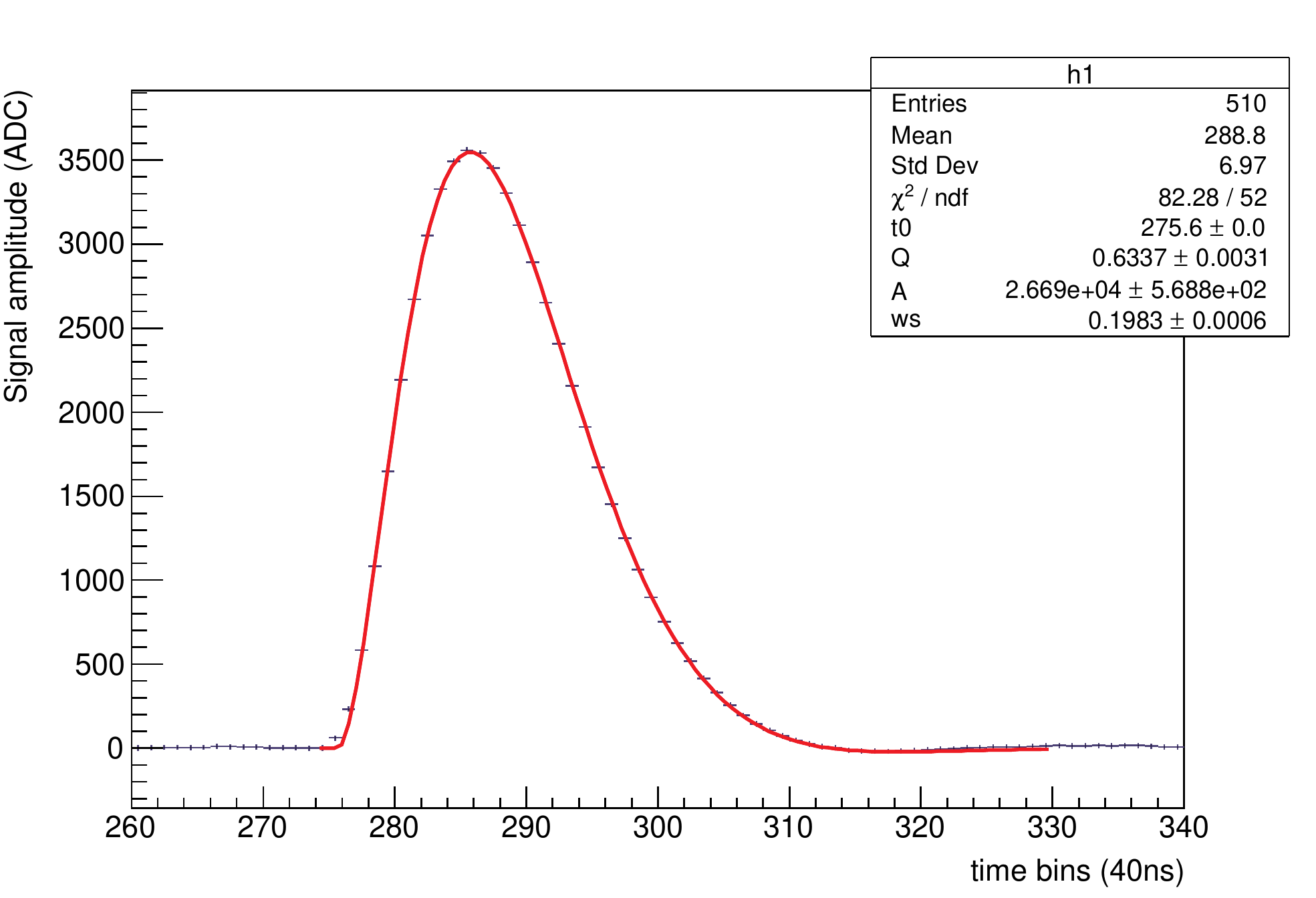}
         \caption{412ns peaking time signal}
         \label{fig:412fit}
     \end{subfigure}
        \caption{Output signals from FEC calibration, fitted with the electronics response function to extract $Q$ and $w_{s}$. The fitted amplitude $A$ and the $t_{0}$ are also shown. }
        \label{fig:CalibrationFitExamples}
\end{figure}

\begin{figure}[hbt!]
     \centering
     \begin{subfigure}[b]{0.45\textwidth}
         \centering
         \includegraphics[width=\textwidth]{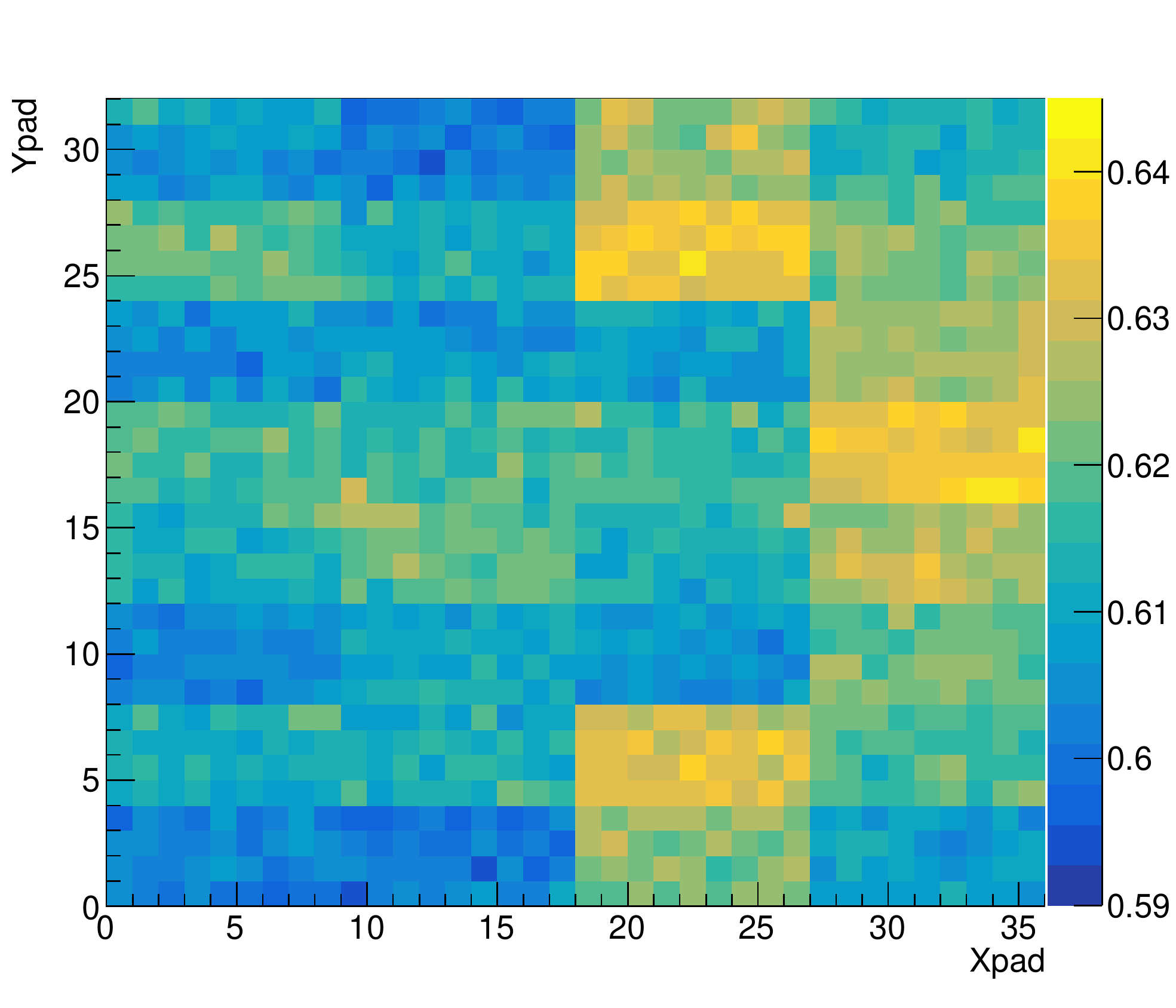}
     \end{subfigure}
     \hfill
     \begin{subfigure}[b]{0.47\textwidth}
         \centering
         \includegraphics[width=1.15\textwidth]{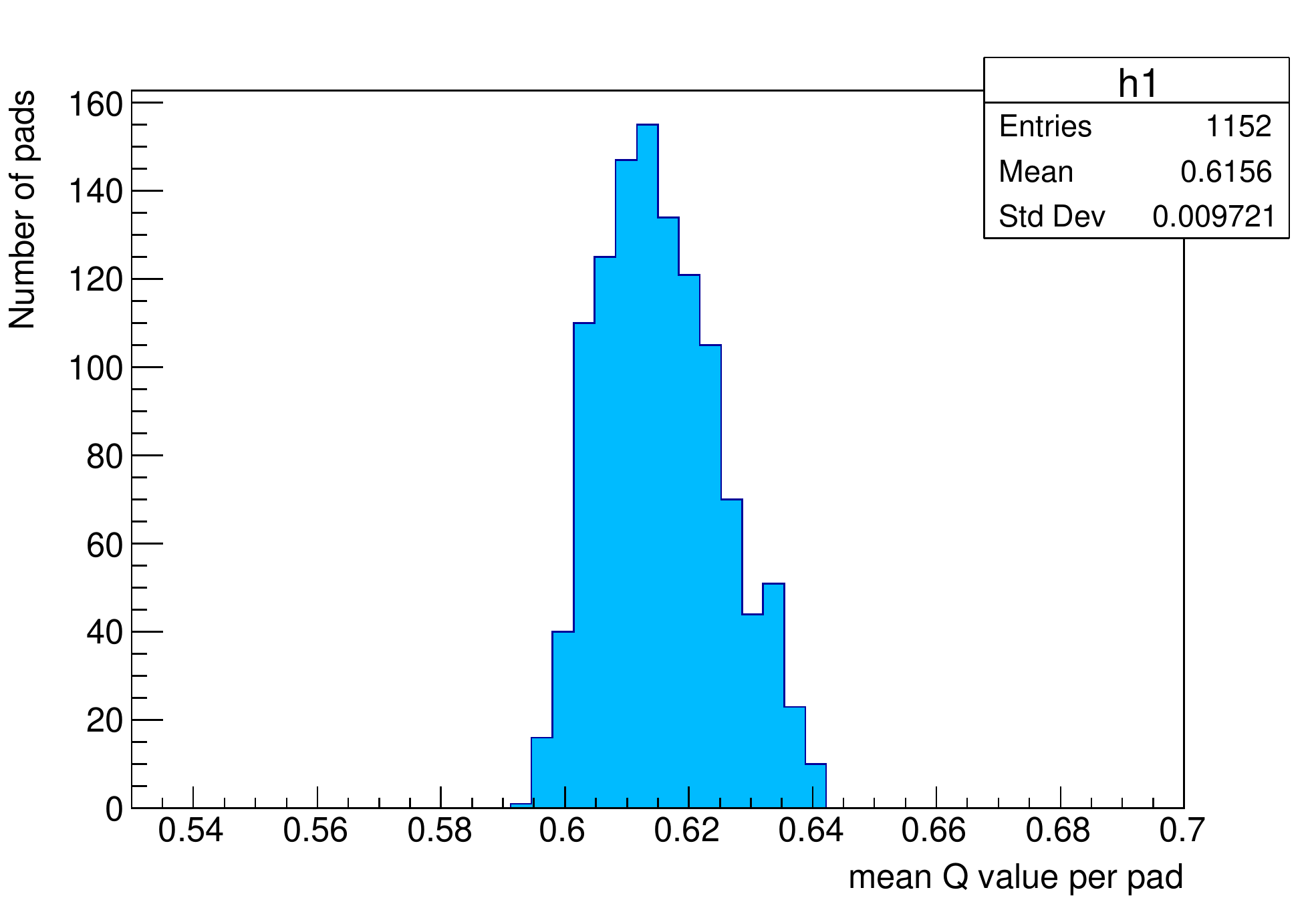}
     \end{subfigure}
        \caption{Mean values of electronics parameter $Q$ for each pad, mapped over entire ERAM (left) and represented in a 1D  distribution (right).  }
        \label{fig:ElectronicsResponse_Q}
\end{figure}

\begin{figure}[hbt!]
     \centering
     \begin{subfigure}[b]{0.45\textwidth}
         \centering
         \includegraphics[width=\textwidth]{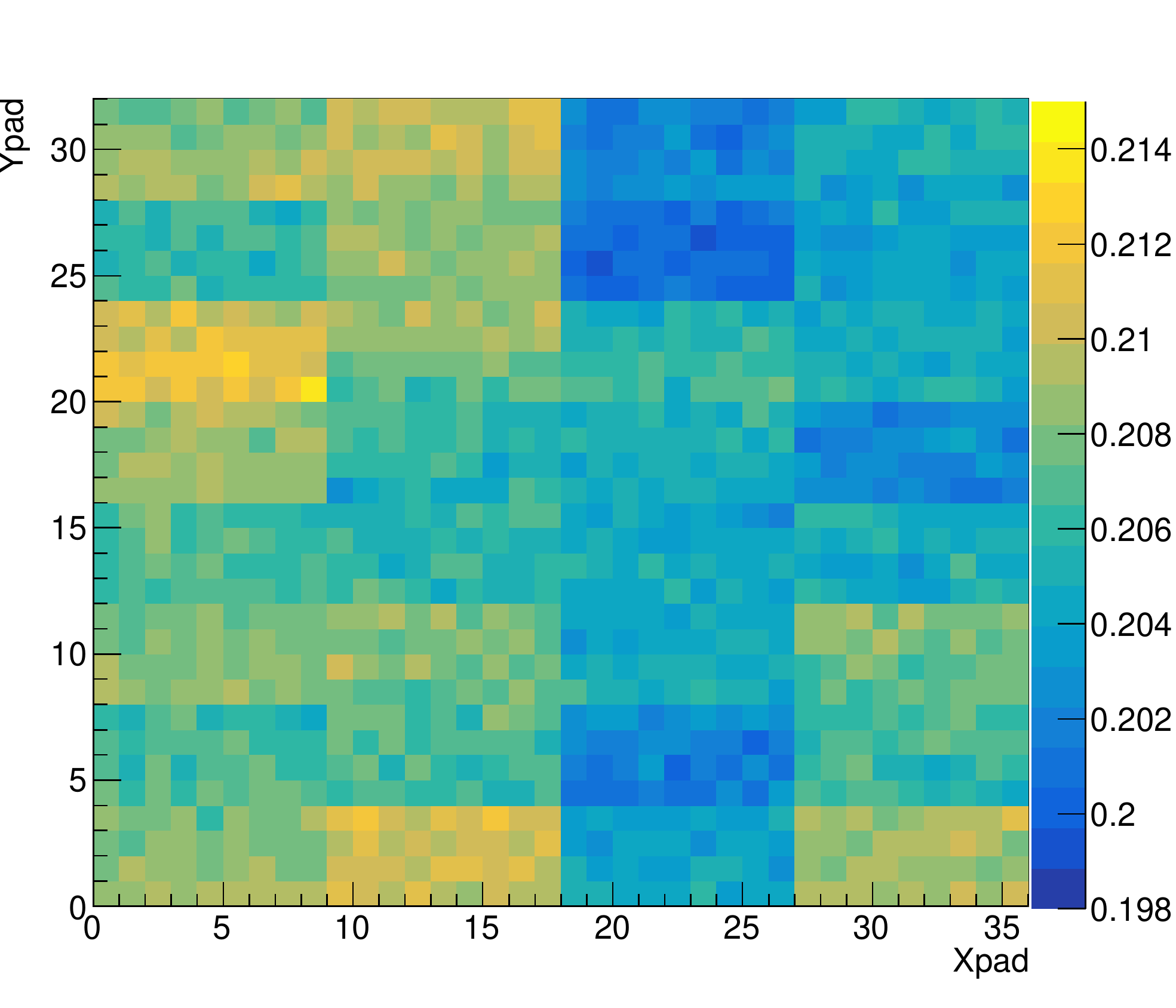}
     \end{subfigure}
     \hfill
     \begin{subfigure}[b]{0.47\textwidth}
         \centering
         \includegraphics[width=1.15\textwidth]{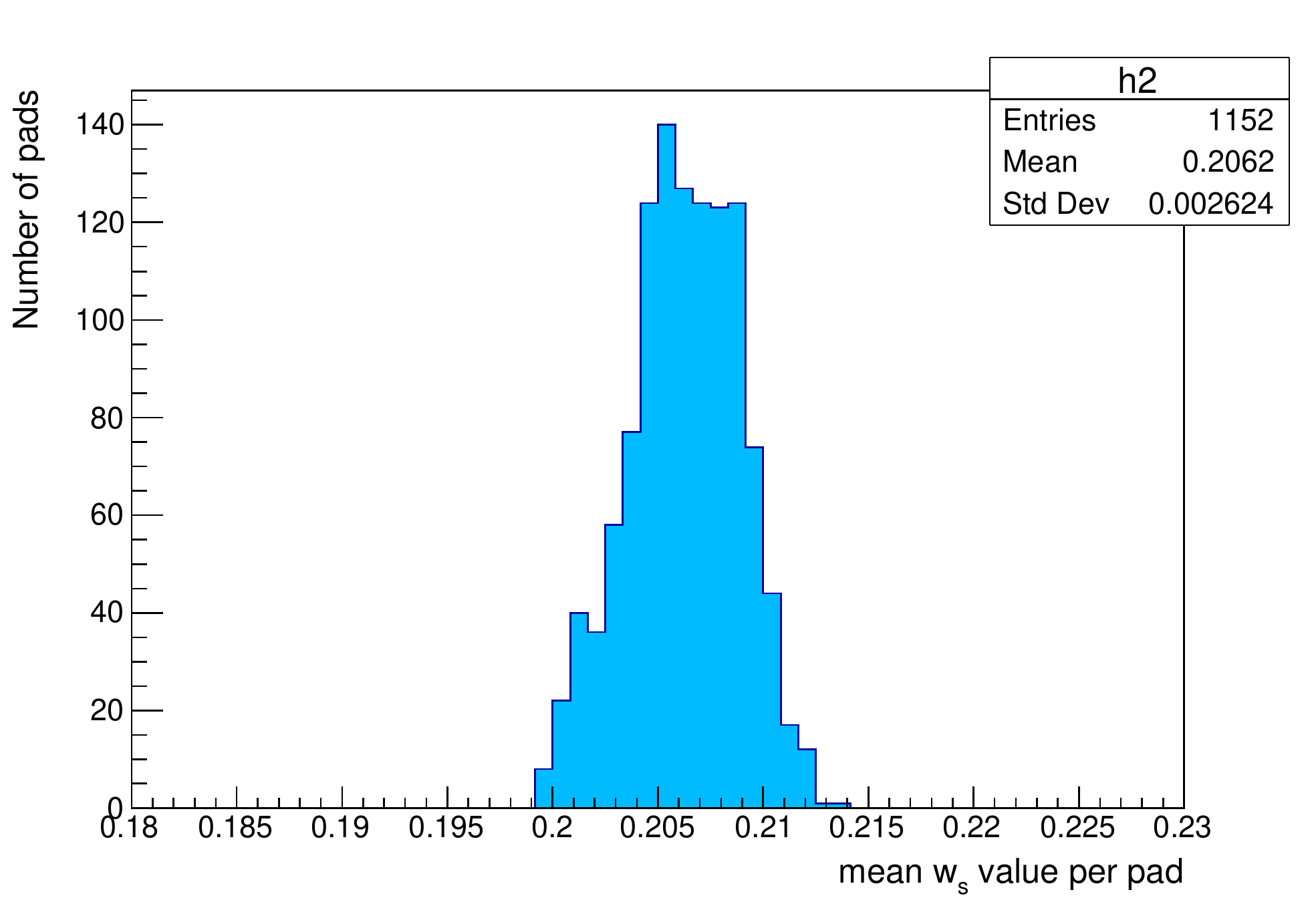}
     \end{subfigure}
        \caption{Mean values of the electronics parameter $w_{s}$ for each pad, mapped over entire ERAM (left) and represented in a 1D  distribution (right). }
        \label{fig:ElectronicsResponse_ws}
\end{figure}

Figure~\ref{fig:CalibrationFitExamples} shows examples of the calibration pulses fitted by the analytical model based on the simulation of the AFTER chip for 200 and 412~ns peaking times.
The analytical shape describes well the electronic response. The parameters $w_s$ and $Q$ are extracted for each pad separately as shown in Figures~\ref{fig:ElectronicsResponse_Q} and \ref{fig:ElectronicsResponse_ws} . The dispersion between any two ASICs is estimated to be lower than 10\%. 
The largest difference between $Q$ and $w_{s}$ values of adjacent pads belonging to two different ASICs can be up to 4.5\%, while that of two different pads within the same ASIC can be up to 3.5\%. 
An anti-correlation  of 85\% between $Q$ and $w_{s}$ is observed since both parameters act in the same direction on the width of the signal. This study allows to validate the consistency of the response of the  readout electronics with the analytical model.

\subsection{Electronics Response Linearity}
\label{subsec:ElectronicLinearity}
\begin{figure*}[hbt!]
        \centering
        \begin{subfigure}[b]{0.475\textwidth}
            \centering
            \includegraphics[width=1.15\textwidth]{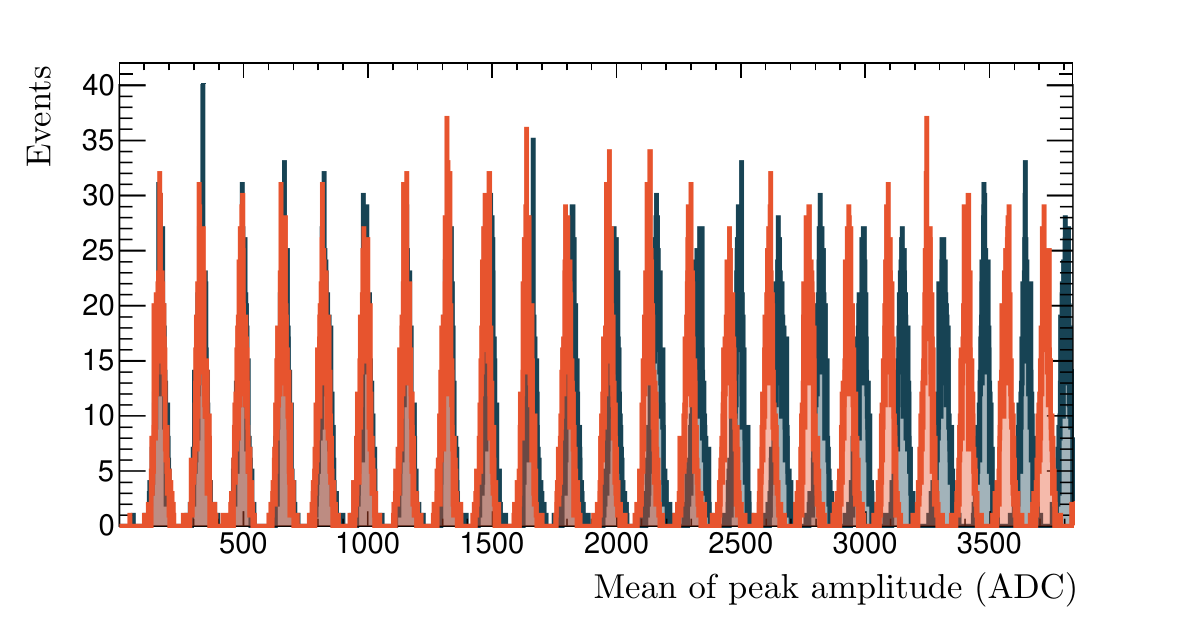}
            \caption[Network2]%
            {{\small Example of the maximum of amplitude of injected signals in two different channels.}}    
            \label{fig:Peaks1}
        \end{subfigure}
        \hfill
        \begin{subfigure}[b]{0.475\textwidth}  
            \centering 
            \includegraphics[width=\textwidth]{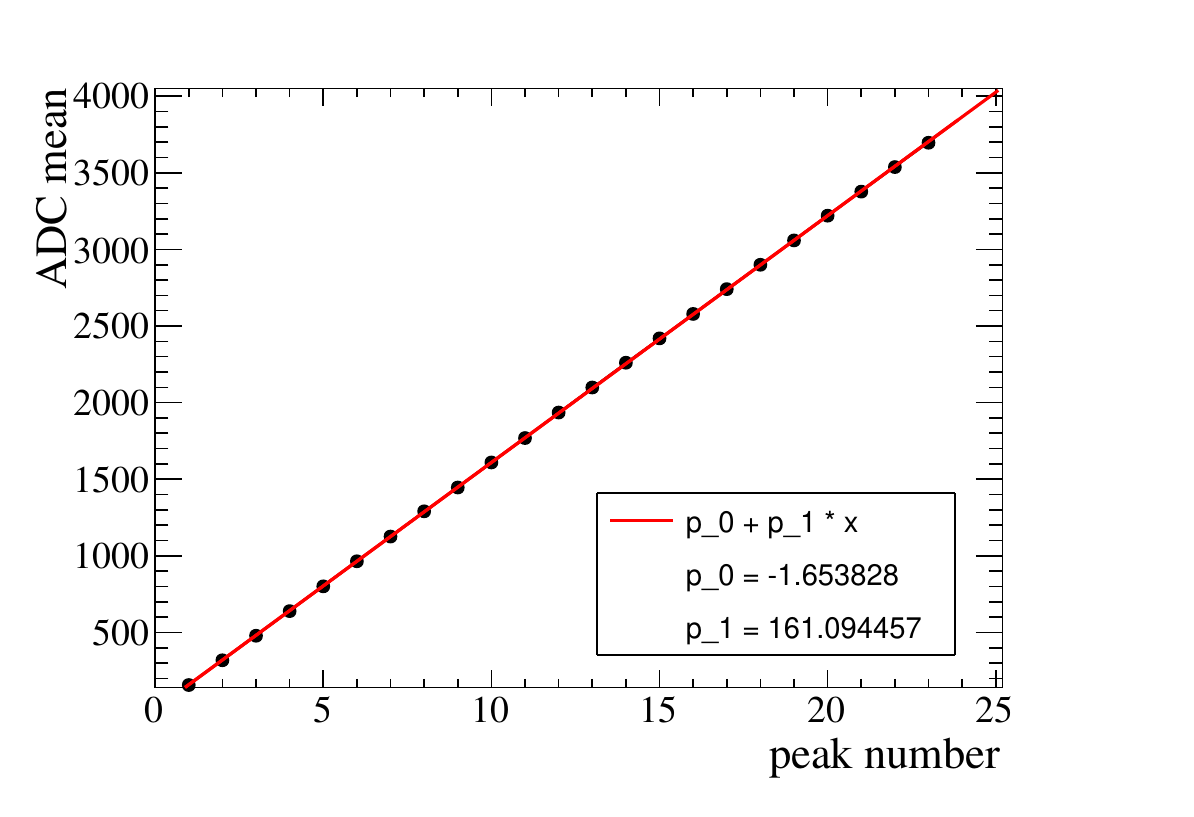}
            \caption[]%
            {{\small Example of calibration test in a given channel.}}    
            \label{fig:cal_ex}
        \end{subfigure}
        \vskip\baselineskip
        \begin{subfigure}[b]{0.475\textwidth}   
            \centering 
            \includegraphics[width=\textwidth]{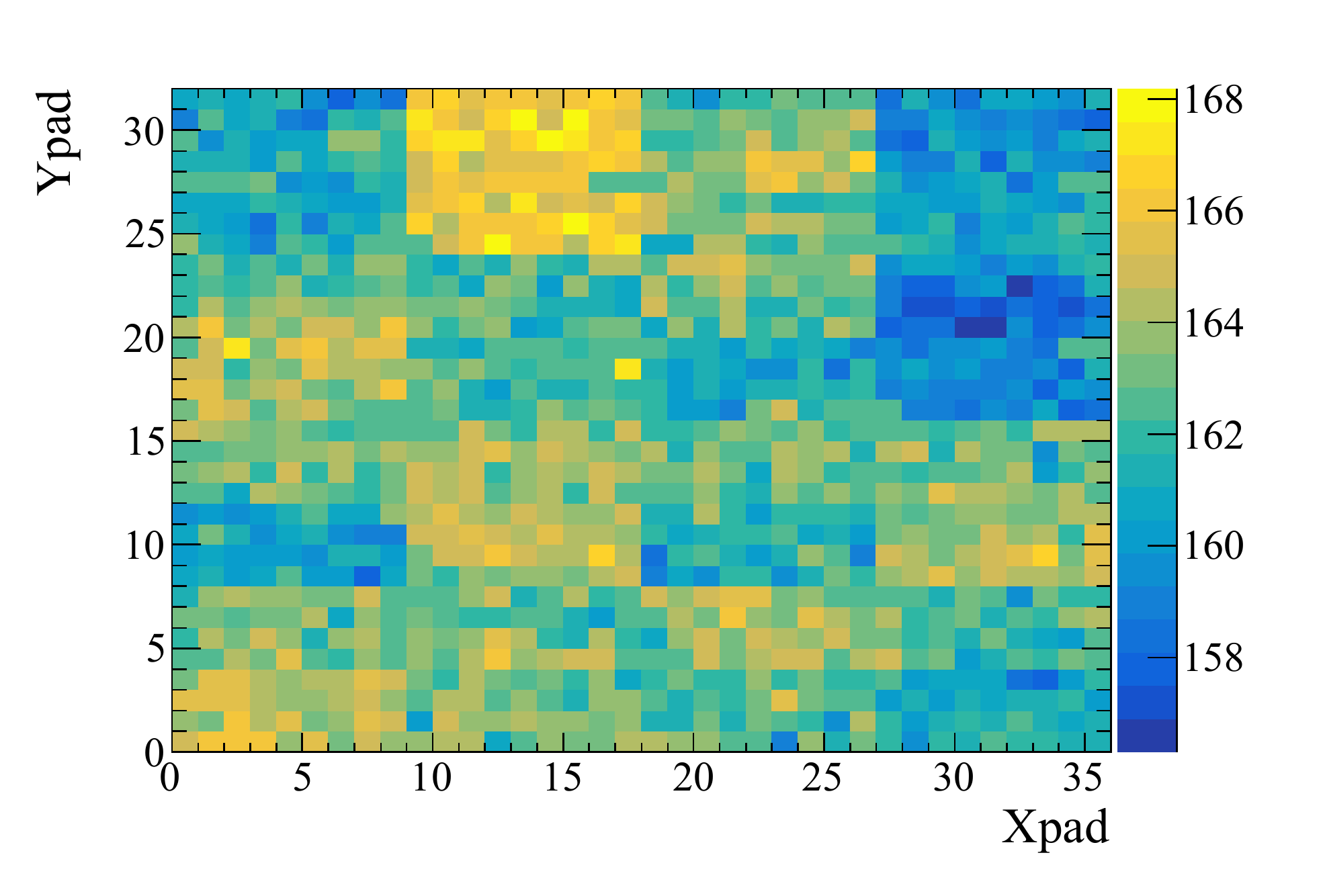}
            \caption[]%
            {{\small Map of the electronic linearity coefficients of two FECs.}}    
            \label{fig:p1_display_34_MEAN}
        \end{subfigure}
        \hfill
        \begin{subfigure}[b]{0.475\textwidth}   
            \centering 
            \includegraphics[width=\textwidth]{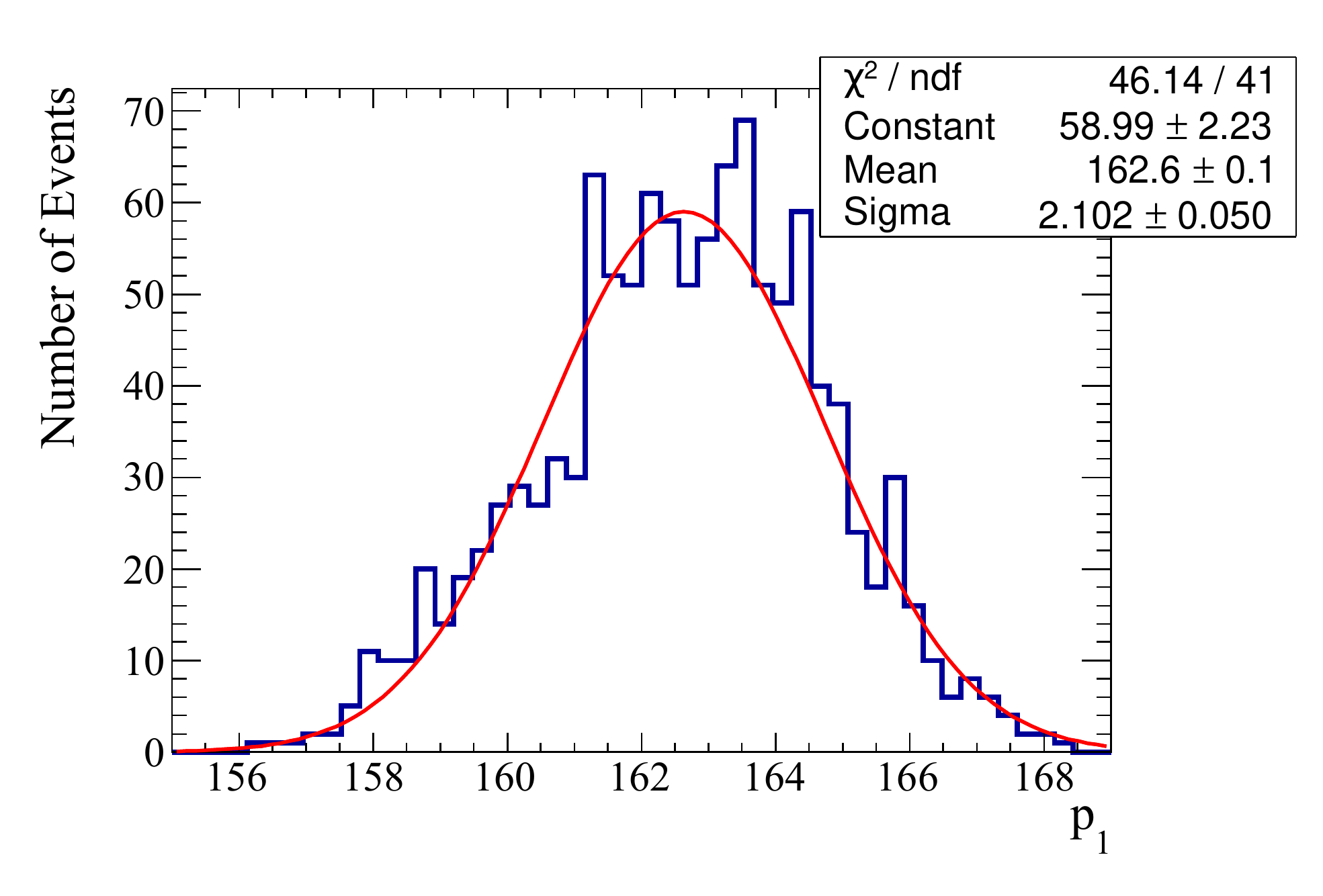}
            \caption[]%
            {{\small 1D distribution of the electronic linearity coefficients of two FECs.}}    
            \label{fig:p1_34}
        \end{subfigure}
        \caption[]
        {\small Linearity study of electronics response.} 
        \label{fig:fecLinearity}
    \end{figure*}

The calibration data are also used to study the electronics response linearity.
Figure~\ref{fig:Peaks1} shows an example of the maximum of amplitude of  injected signals in two different channels. We can see that the higher the injected amplitude, the larger the shift in the response between different channels. However, the shift is small (2\%-2.5\%) as expected.
Figure~\ref{fig:cal_ex} also shows an example of linearity test in a given channel. In this plot, each peak number is drawn as function of the mean value of the the maximum of amplitude obtained from a Gaussian fit. The linearity coefficients obtained from all the channels of two FECs (Figure~\ref{fig:p1_display_34_MEAN}) shows a uniform response with typical differences in linearity among neighboring ASICs smaller than 3\%. 
The 1D distribution of the linearity coefficients of two FECs is illustrated in Figure~\ref{fig:p1_34}.
\section{Mesh Pulsing}
\label{sec:meshPulsing}
The on-board DLC High Voltage (HV) filter located on the left corner and top layer of the ERAM PCB comes with an optional and independent mesh connection allowing to inject a signal on the Micromegas mesh (sustained 128~$\mu$m above the anode), thus inducing a signal on all pads simultaneously. This procedure (referred to as mesh pulsing) is used to validate the functioning of the ERAM modules. 

A signal is injected on the mesh to be read simultaneously on all pads. As the excitation is uniform, the expected response is also uniform over the pad plane, hence this procedure allows to quickly detect any localized defect. Eventual dead pads can also be identified during this test: in total four dead pads on a module are allowed as long as they are not adjacent. The mesh pulsing is repeated before and after gluing the detector onto the mechanical support.  A non-uniformity of the signal of 10-15\% is tolerated.

\begin{figure}[hbt!]
     \centering
     \begin{subfigure}[b]{0.45\textwidth}
         \centering
         \includegraphics[width=\textwidth]{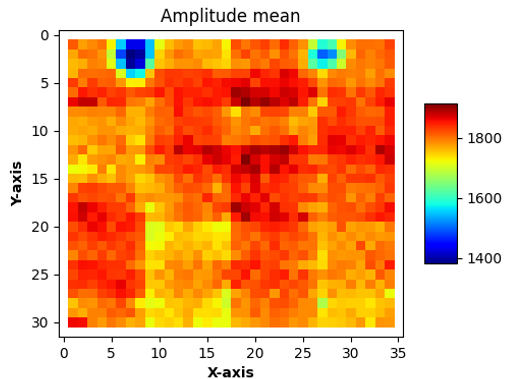}
     \end{subfigure}
     \hfill
     \begin{subfigure}[b]{0.45\textwidth}
         \centering
         \includegraphics[width=\textwidth]{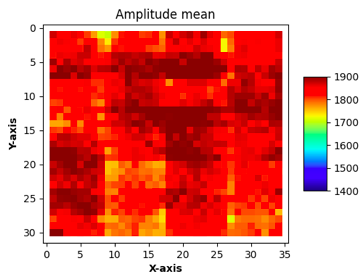}
     \end{subfigure}
        \caption{An example of mean amplitude seen by each pad of an ERAM before (left) and after (right) repair.}
        \label{fig:mesh_pulsing}
\end{figure}

A 300~mV square signal at 1~kHz is sent to the ERAM mesh through a 50~$\Omega$ adapted cable. The readout electronics DAQ is triggered with a NIM signal synchronized with the mesh pulsing. A Faraday cage is used and careful grounding is performed to avoid EMC noise. 
Figure~\ref{fig:mesh_pulsing} (left) shows the mean amplitude seen by each pad of a defective ERAM.  Two lower amplitude regions are observed. 
After investigation, it was found that the mesh was locally peeled off from the pillars due to dust during lamination. The detector was repaired and validated by the same procedure as shown in Figure~\ref{fig:mesh_pulsing} (right) where no localised problems are seen.
\section{X-ray Experimental Setup}
\label{sec:setup}
%
 An X-ray test bench is used for the ERAM characterization. 
It consists of an aluminium chamber with 3~cm drift distance and a robotic $x-y-z$ arm system on an optical breadboard of 120$\times$60~cm$^2$ holding a 280~MBq $^{55}$Fe radioactive source. 

A 20~$\upmu$m aluminised mylar window is taped on the chamber side opposite of the ERAM in order to let the X-rays penetrate the gas volume. A mesh cathode allows the electric field to be applied throughout the 3~cm drift volume towards the ERAM grounded mesh. The DLC voltage is set to 350~V. The settings chosen for the AFTER chip are a sampling period of 40~ns and a peaking time of 412~ns. 

Each step of the robot movement corresponds to a displacement of 0.1~mm, allowing for a precise positioning of the source in the center of each pad of an ERAM. After each installation of a new detector, an alignment procedure is performed in order to ensure the position of the source with respect to the center of each pad. A 1.5~mm diameter collimation hole in front of the source assures that the majority of photo-electron arrives on the targeted pad. About three minutes of scan per pad with a 100~Hz internal trigger is necessary to produce a $^{55}$Fe spectrum with enough statistics, so approx. 64~hours are needed to perform automatically a full scan of the 1152 pads. A random trigger is used. 

The test bench is fed with the so-called T2K gas mixture (Ar-CF$_4$-iC$_4$H$_{10}$ $\left[ 95\%-3\%-2\% \right]$) with a gas flow of 14~liters/hour.
A key aspect of the X-ray chamber is the gas tightness and more generally the environmental conditions inside the chamber such as pressure, humidity level, flow, temperature and gas purity.
Given the impact of the environmental conditions on the gain of the ERAMs, it is crucial to monitor the gas conditions.
A set of sensors has therefore been added in the gas loop at the exit of the chamber.
A Gas Monitoring Chamber~\cite{T2KND280TPC:2010nnd} identical to the ones deployed
at T2K’s ND280 detector has been added in the latest ERAM scans. It allows to control  the gas conditions in a similar way to what will be performed in ND280.

Due to potential risks linked to humidity during the operation of the ERAMs, a criterion on a maximum level of acceptable humidity in the chamber had to be defined, based on the measurements of the various humidity sensors available. The humidity should be kept less than 0.4\%.
\section{ Gain Extraction using $^{55}$Fe Source}
\label{sec:gain_Fe55}
\begin{figure}[hbt!]
  \centering
  \includegraphics[width=0.49\textwidth]{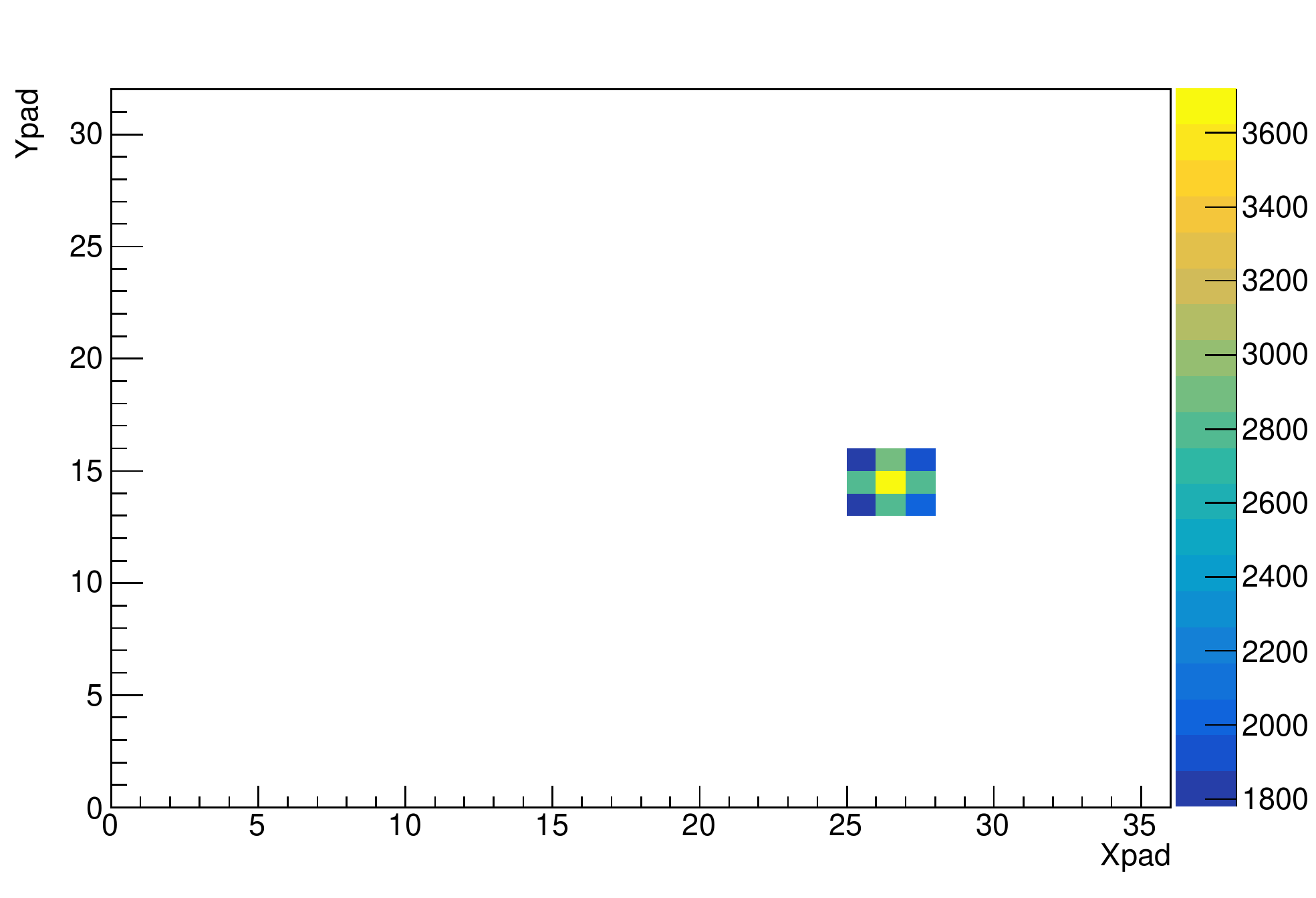}
  \caption{Display of X-ray events generated during the scan of a pad (number (26,14)) within three minutes. The scanned pad has the highest count rate. The counts recorded in the neighbouring pads are due to spreading of the initial charge. $X_\mathrm{pad}$ and $Y_\mathrm{pad}$ are the pad indices. }
  \label{fig:EventDisplayXrays}
\end{figure}

As mentioned in section \ref{sec:setup}, each pad of an ERAM is scanned using an X-ray beam produced by an $^{55}$Fe source placed inside a collimator. Each X-ray photo-electron causes an electron avalanche in the Micromegas amplification gap above the targeted pad (leading pad), which deposits charge into it. This initial charge eventually spreads into the neighbouring pads. 
Figure~\ref{fig:EventDisplayXrays} shows a display of all X-ray events during the scan of a given pad. It can be clearly seen that the scanned pad has highest count rate and that the charge is spread to its neighbouring pads. The arrival time and magnitude of charge induced in the adjacent pads are determined by the $RC$ value of the leading pad, the magnitude of the initial charge, and the position of the adjacent pad with respect to the position of initial charge deposition.
\subsection{Gain Calculation Model}
\label{subsec:GainCalculationModel}
The gain is defined as the ratio between the charge received on the anode $Q_{Anode}$ and the primary charge $Q_{Primary}$ deposited by an ionizing particle :
\begin{equation}
 G =\frac{Q_{Anode}}{Q_{Primary}}
 \label{equ:Gain-ChargeRatio}
\end{equation}
Capacitive coupling between the resistive layer and the conductive pads means that charges at some place on the resistive layer are compensated by opposite charges on the pads. These mirror charges are injected via  currents through the electronics circuits leading to readable ADC signals.

In X-ray events, the pads of interest are the pad with the highest amplitude signal (leading pad), and the eight pads surrounding it as illustrated in Figure~\ref{fig:EventDisplayXrays}. A method ~\cite{Attie:2022smn} to obtain the gain is to sum up the waveforms of the signals of these pads, pick up the maximum of the resulting waveform ($ADC^{\Sigma}_{max}$), and convert it to a charge ($Q^{\Sigma}_{max}$) according  to the conversion relation defined by  eq.~\ref{equ:Charge2ADCCorrespondance}. Gain is thus obtained as the ratio of this charge to the estimated primary charge deposited by the X-ray photon ($Q_{Primary}$) :
\begin{equation}
G 
= 
\frac{Q^{\Sigma}_{max}}{Q_{Primary}} 
= 
\frac{ADC^{\Sigma}_{max}}{ADC_{o}}\frac{Q_{o}}{Q_{Primary}} 
  \label{equ:GainMethod}
\end{equation}
This method, which implicitly identifies ADC counts with charges, works
due to a particular circumstance for the range of $RC$ values of the actual ERAM devices: 
the deposited charge spreading on the surface of the resistive layer
above the nine pads, is 
constant, to a good approximation, within the time scale of the electronics response. 
As far as the electronics is concerned, the total charge injected on the pads to neutralize the  charge  on the resistive layer, is constant and has been injected instantaneously i.e. the corresponding current is a Dirac pulse.
Then the sum of the signal of pads is the response of the equivalent electronics to a Dirac current pulse, the maximal amplitude of which can indeed be converted in charge according to eq.~\ref{equ:Charge2ADCCorrespondance}.

These considerations can be translated into a mathematical formulation as follows.
Let $ \rho(x,y,t) $
be the charge density on the resistive layer
and 
$Q_{i}(t) =\iint_{i} \rho(x,y,t) \mathrm{d}x\mathrm{d}y$
the charge sitting on the resistive layer at time $t$ opposite to a pad indexed by $i$.
The electronics response to this charge is the convolution of its
derivative with the electronics response to a Dirac pulse, eq.~\ref{equ:ElResponse2Dirac},
or equivalently the convolution of the charge with the derivative of this electronics response:
\begin{equation}
ADC_{i}(t) = 
\left( \frac{dQ_i}{dt} * ADC^{D}\right)(t)  
= 
\left(Q _i* \frac{d}{dt} ADC^D\right)(t)
\end{equation}
Therefore the sum of the responses of the nine above pads is:
\begin{equation}
ADC^{\Sigma}(t) = \sum_{i=1}^{9} ADC_{i}(t) 
=  
\left( \left[ \sum_{i=1}^{9} Q _i \right]* \dt{ADC^{D}}\right)(t)
  \label{equ:EquivResponse}
\end{equation}
where to derive the last term,
it has been assumed that the
electronics of the nine pads have identical responses.
From eq.~\ref{equ:EquivResponse} one derives that if 
the total charge is effectively constant
and so equal to the charge deposited on the resistive layer by the avalanche, $ \sum_{i=1}^{9} Q _i(t) \sim Q_{Primary} \, G \, \theta(t)$ where $\theta(t)$ is the Heaviside step function, 
it can be factorized out leading to 
\begin{equation}
ADC^{\Sigma}(t) \sim Q_{Primary}\,  G \, ADC^{D}(t)
  \label{equ:EquivResponse2}
\end{equation}
Therefore the maximal value of the sum of the responses of the nine pads is
\begin{equation}
ADC^{\Sigma}_{max}
\sim Q^{Primary}  \, G\, \frac{ADC_{o}}{Q_{o}}
\end{equation}
which is equivalent to eq.~\ref{equ:GainMethod}.

This proof relies only on the identity of the electronics responses of the nine pads. It does not rely  on the uniformity of  $RC$ over the resistive surface above the nine pads, but only on the assumption that, within the time scale of the electronics response, the charge initially deposited on the leading pad, did not escape significantly this extended surface.

\begin{figure}[hbt!]
     \centering
     \begin{subfigure}[b]{0.45\textwidth}
         \centering
         \includegraphics[width=1.15\textwidth]{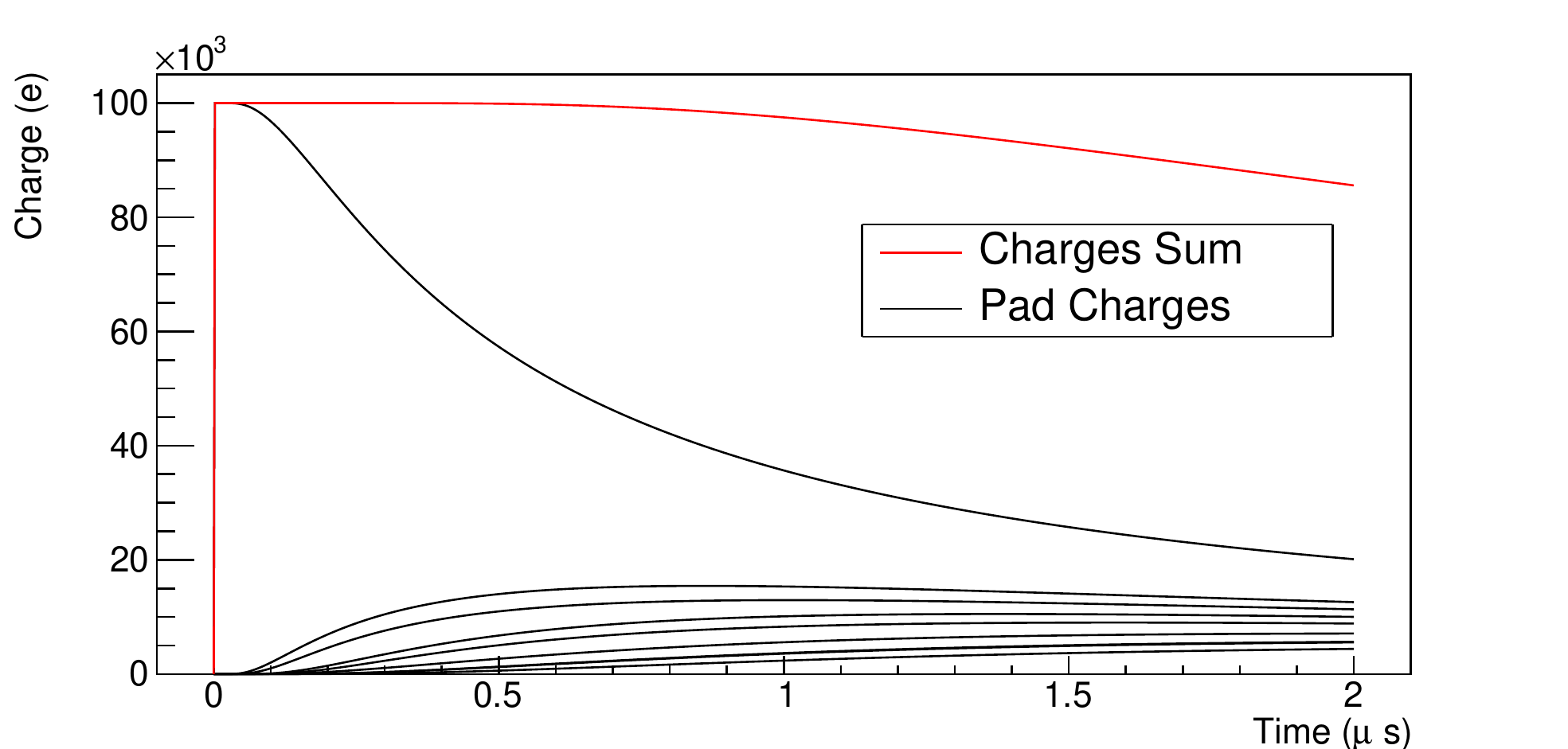}
         \caption{}
         \label{fig:Gain_02_Q_OffCenter}
     \end{subfigure}
     \hfill
     \begin{subfigure}[b]{0.45\textwidth}
         \centering
         \includegraphics[width=1.15\textwidth]{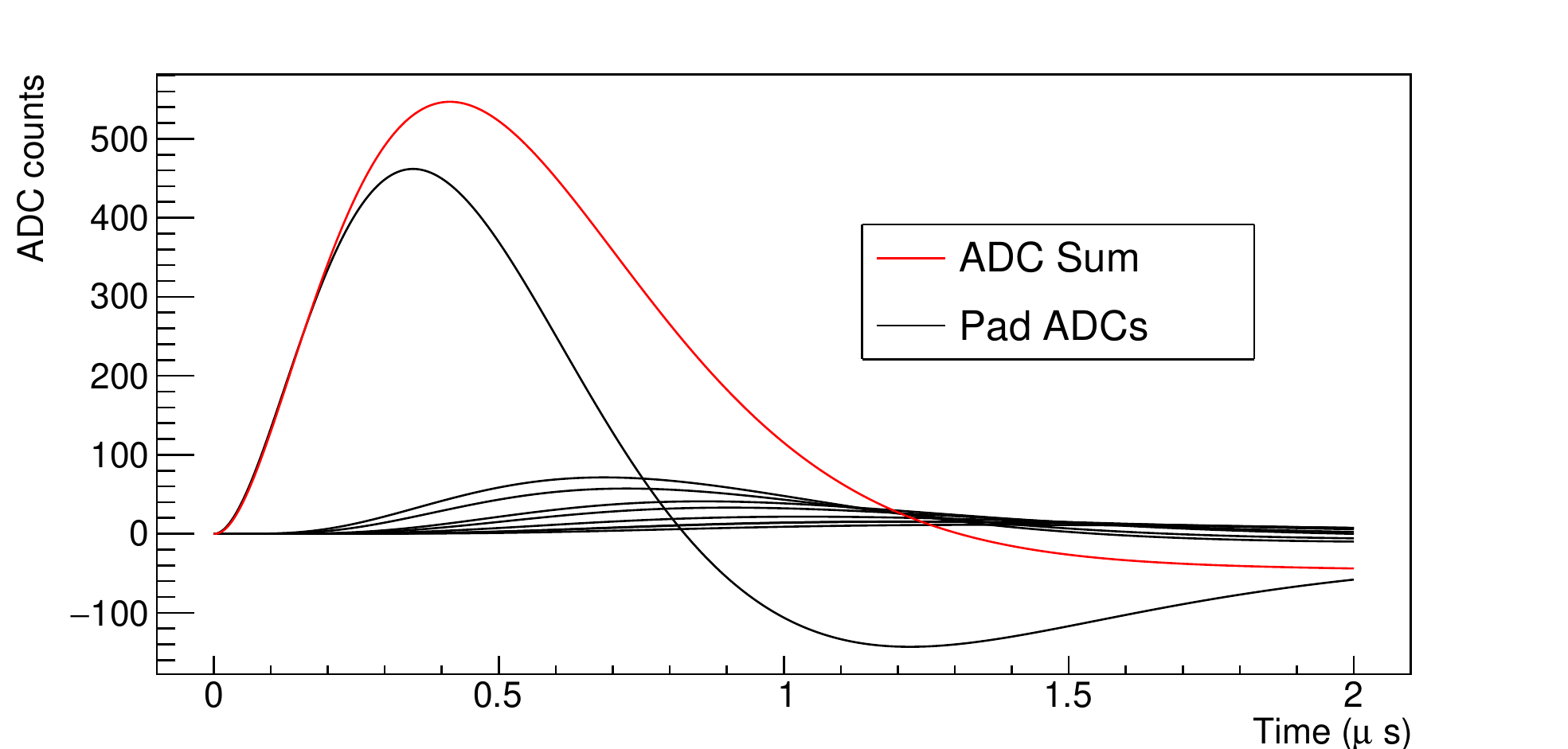}
         \caption{}
         \label{fig:Gain_02_A_OffCenter}
     \end{subfigure}
        \caption{
  Pad Charges in electron charge units (a) and ADCs (b) and their sums, for a deposit of $100$ primary electrons 
at 1 mm of the center of a pad both in X and Y directions,
 a gain of $10^3$ and a value of $RC$ of 50 ns/mm$^{2}$. 
}
        \label{fig:Gain_02_QA_OffCenter}
\end{figure}

Charges and ADC responses have been simulated for a deposit of $100$ primary electrons at 1 mm of the center of a pad in both the $x$ and $y$ directions, with a gain of $10^3$ and a value of $RC$ of 50 ns/mm$^{2}$. The results are shown 
in Figure~\ref{fig:Gain_02_Q_OffCenter} for the charges and their sum and in Figure~\ref{fig:Gain_02_A_OffCenter} for the ADC responses  and their sum. Although the individual charges on pads vary significantly, the sum over the nine pads varies far more  slowly over the time scale of the electronics responses displayed in Figure~\ref{fig:Gain_02_A_OffCenter}.

\begin{figure}[hbt!]
     \centering
     \begin{subfigure}[b]{0.45\textwidth}
         \centering
         \includegraphics[width=1.15\textwidth]{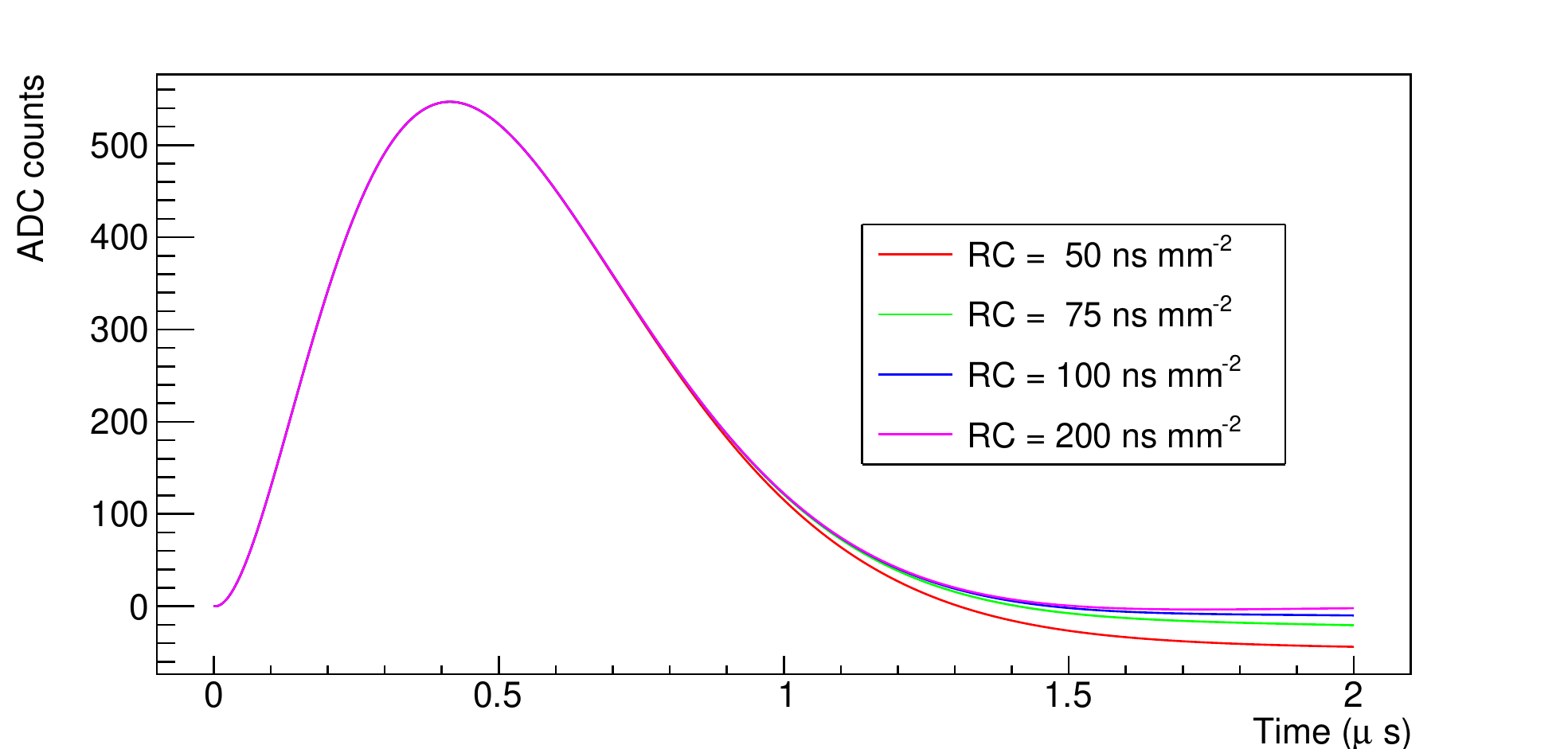}
         \caption{}
         \label{fig:Gain_02_A_VarRC}
     \end{subfigure}
     \hfill
     \begin{subfigure}[b]{0.45\textwidth}
         \centering
         \includegraphics[width=1.15\textwidth]{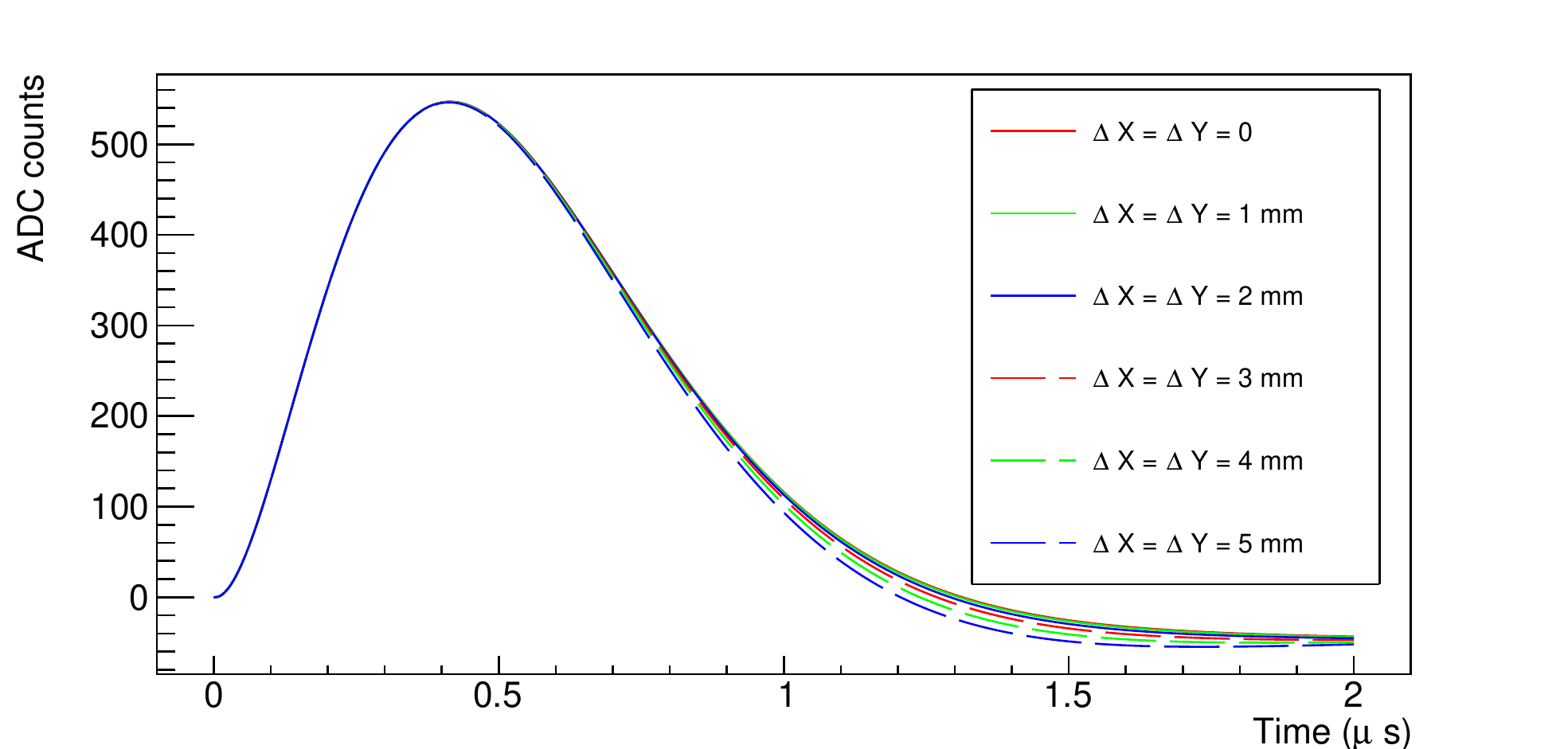}
         \caption{}
         \label{fig:Gain_02_A_VarTrue}
     \end{subfigure}
        \caption{
 Sum of the Pad ADCs  for a simulation
 for a gain of $10^3$,
 a deposit of $100$ primary electrons
 (a) at 1 mm of the center of a pad both in X and Y directions and various $RC$ values,
 (b) at various positions of coordinates $\Delta X$ and $\Delta Y$ w.r.t the center of a pad
 and fixed value of $RC$ of 50 ns/mm$^{2}$.
}
        \label{fig:Gain_02_A_VarRC_VarTrue}
\end{figure}

Equation~\ref{equ:EquivResponse2} is not exact. Indeed, Figure~\ref{fig:Gain_02_Q_OffCenter} shows that the total charge is not exactly constant.
However one only needs to check that the approximation is good enough for the maximal value of
the sum of the ADC to be effectively constant independent of the actual conditions.
Figure~\ref{fig:Gain_02_A_VarRC} shows that this is indeed the case for any value of $RC$ 
from 50~ns/mm$^{2}$ to  200~ns/mm$^{2}$, range which covers the actually measured values~\footnote{Incidentally, this shows that the non-uniformity of $RC$ over the nine pads is not an issue for this method, since even for the lowest $RC$ value, allowing the fastest charge escape, the total charge is effectively constant}.
Similarly, Figure~\ref{fig:Gain_02_A_VarTrue} shows that the approximation is effectively valid for any position of the initial charge deposit on the leading pad.

\begin{figure}[hbt!]
     \centering
     \begin{subfigure}[b]{0.45\textwidth}
         \centering
         \includegraphics[width=\textwidth]{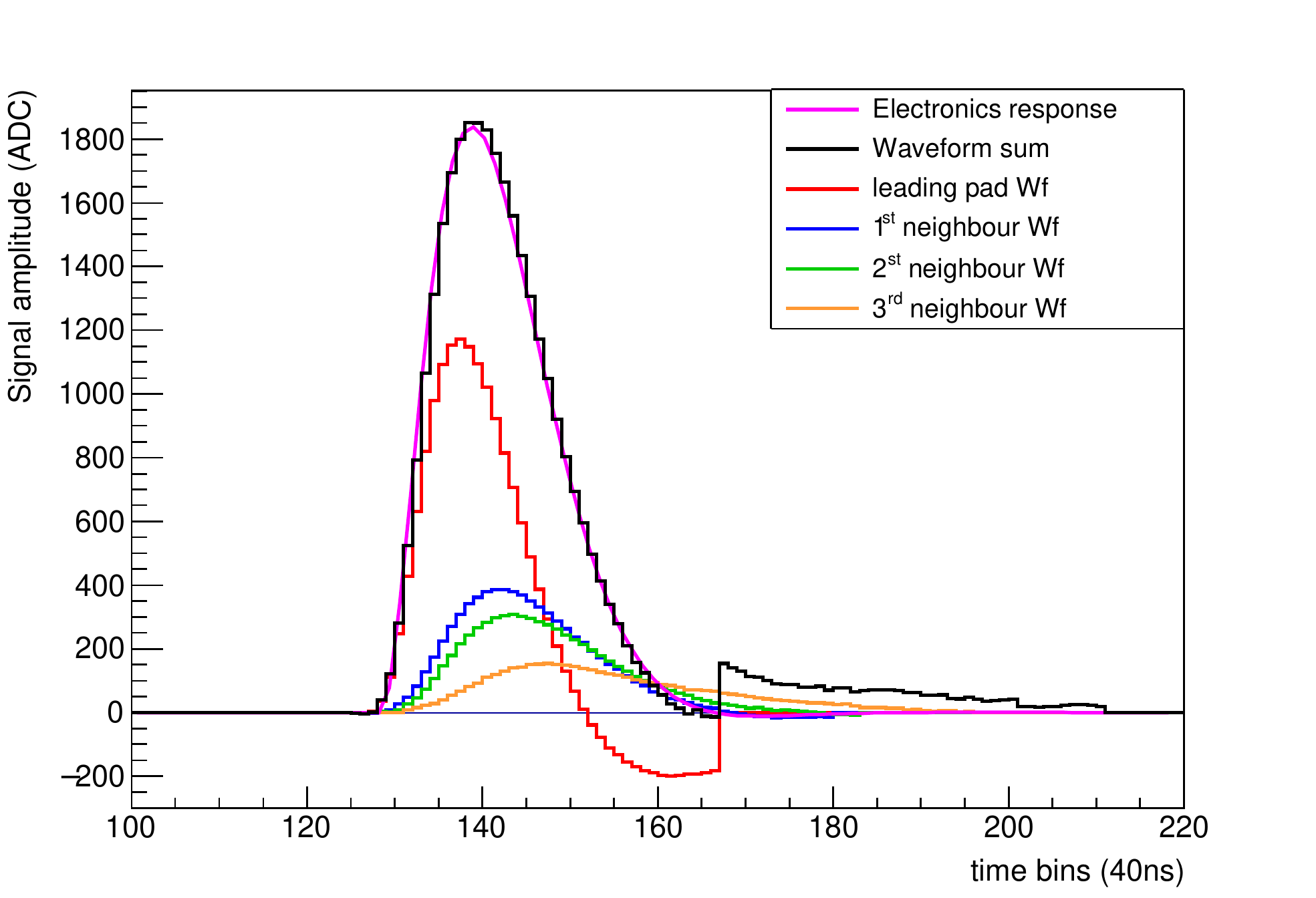}
     \end{subfigure}
     \hfill
     \begin{subfigure}[b]{0.45\textwidth}
         \centering
         \includegraphics[width=\textwidth]{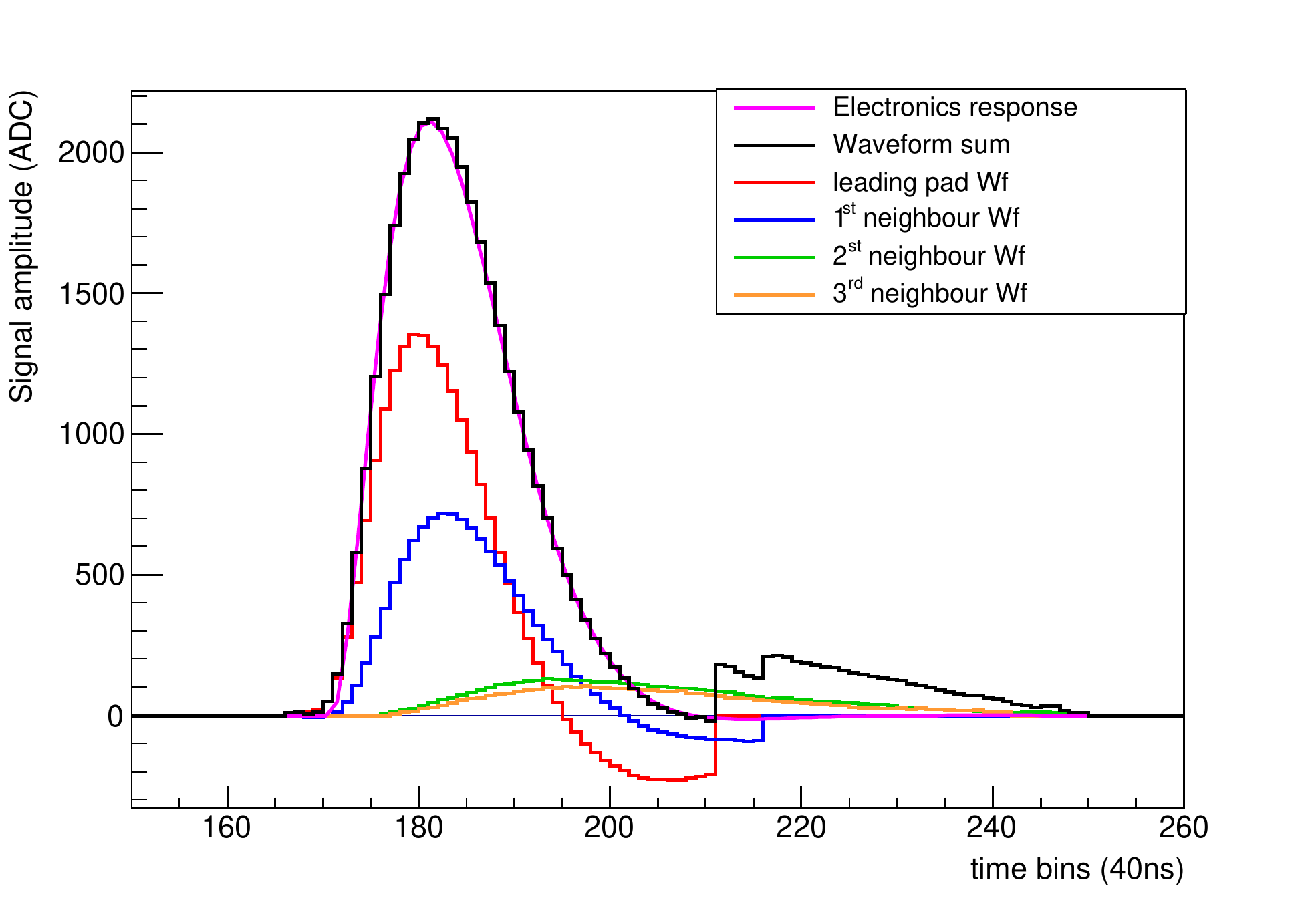}
     \end{subfigure}
        \caption{Two examples of comparison between sum of pad signals in an X-ray event and its equivalent electronics response normalized according to its simultaneous fit result.}
        \label{fig:ElecResp}
\end{figure}

Finally, the gain model presented here is validated using  X-ray data. Figure~\ref{fig:ElecResp} shows a comparison between the sum of pad signals generated in an X-ray event and its equivalent electronics response (normalized according to its simultaneous fit result detailed in section~\ref{subsec:simul_fit}). Excellent agreement is found between the data and the model.
\subsection{Gain results using X-ray data}
\label{subsec:GainCalculation_sum}
\begin{figure}[hbt!]
  \centering
  \includegraphics[width=0.5\textwidth]{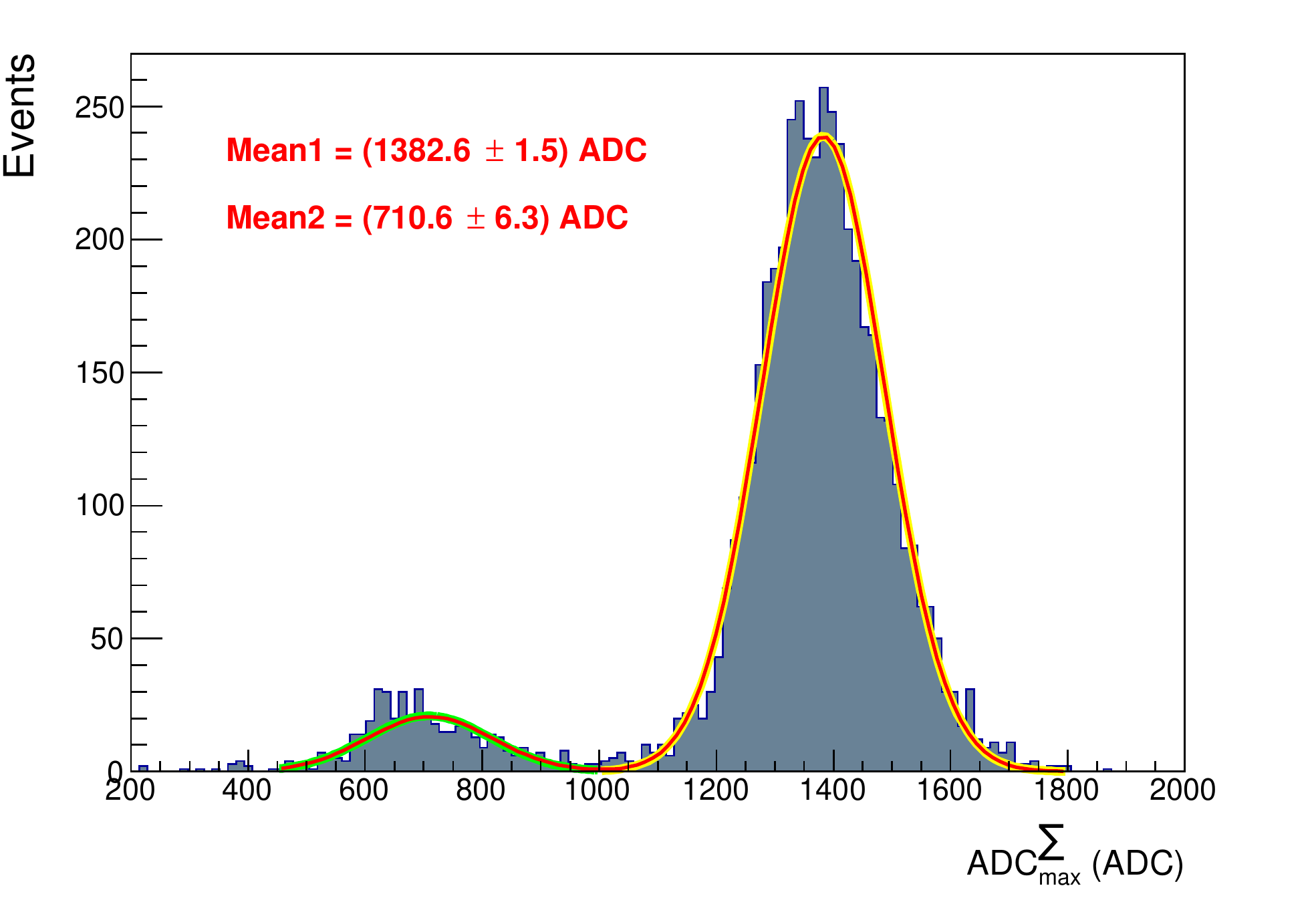}
  \caption{$^{55}$Fe spectrum reproduced from $ADC^{\Sigma}_{max}$ extracted from all the events in one pad.}
  \label{fig:FeSpectrum_sum}
\end{figure}

\begin{figure}[hbt!]
     \centering
     \begin{subfigure}[b]{0.45\textwidth}
         \centering
         \includegraphics[width=\textwidth]{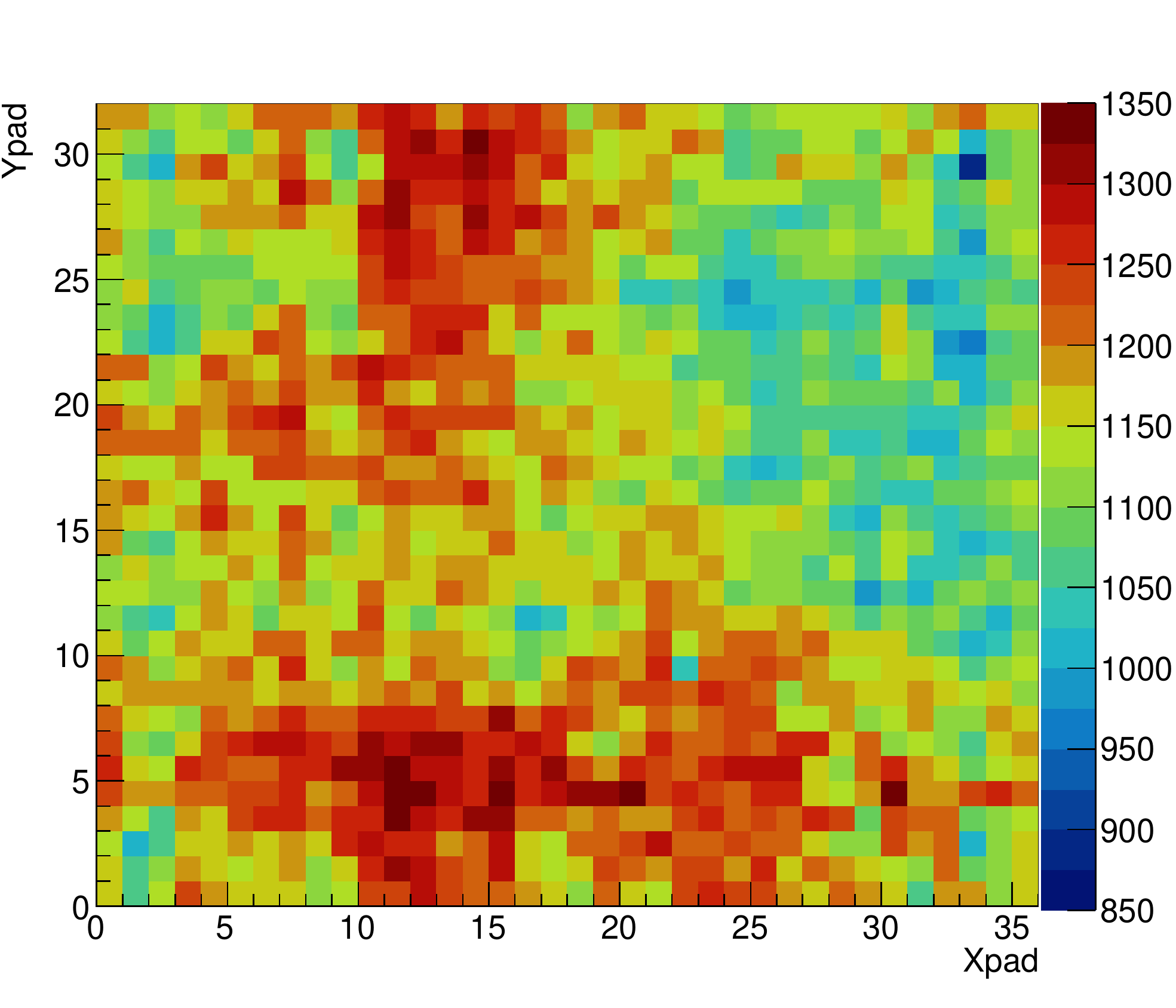}
         \caption{Gain map of ERAM-30}
         \label{fig:Gainmap_eram30_sum}
     \end{subfigure}
     \hfill
     \begin{subfigure}[b]{0.42\textwidth}
         \centering
         \includegraphics[width=1.25\textwidth]{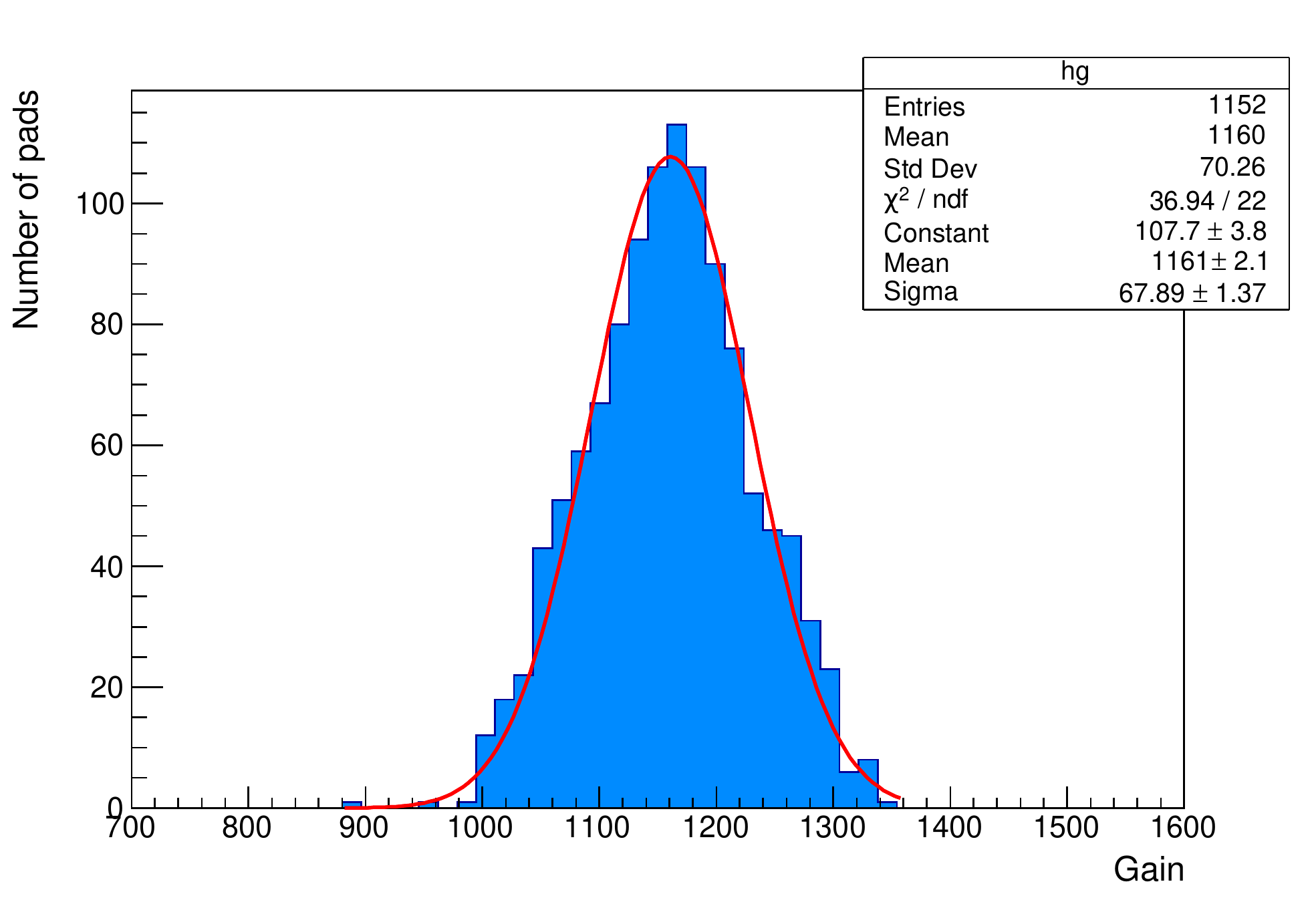}
         \caption{Gain distribution of ERAM-30}
         \label{fig:Gaindist_eram30_sum}
     \end{subfigure}
        \caption{Gain information of ERAM-30 obtained from the method described in sec. \ref{subsec:GainCalculationModel}.}
        \label{fig:Gain_eram30_sum}
\end{figure}

\begin{figure}[hbt!]
     \centering
     \begin{subfigure}[b]{0.45\textwidth}
         \centering
         \includegraphics[width=\textwidth]{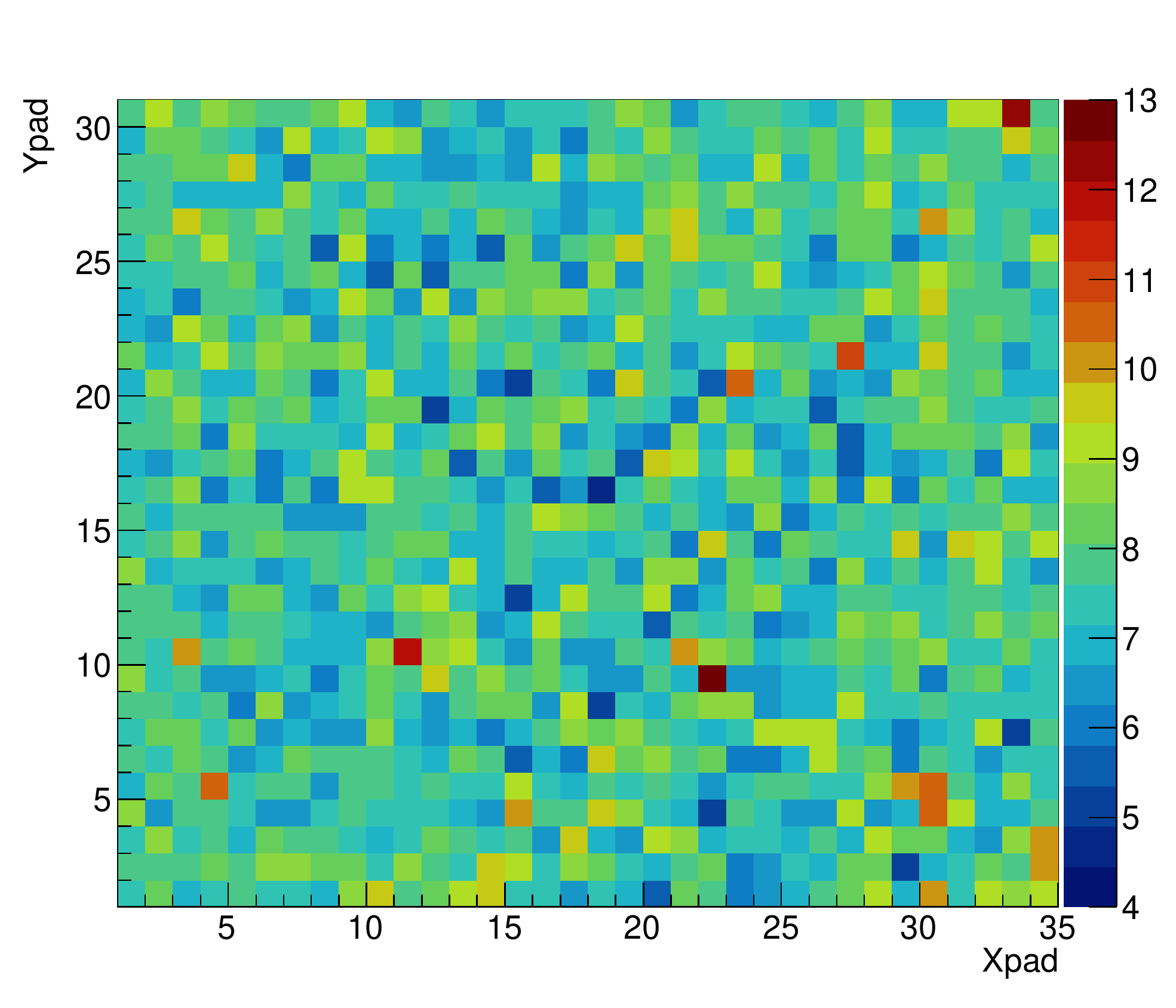}
         \caption{Resolution map of ERAM-30}
         \label{fig:Resmap_eram30_sum}
     \end{subfigure}
     \hfill
     \begin{subfigure}[b]{0.42\textwidth}
         \centering
         \includegraphics[width=1.25\textwidth]{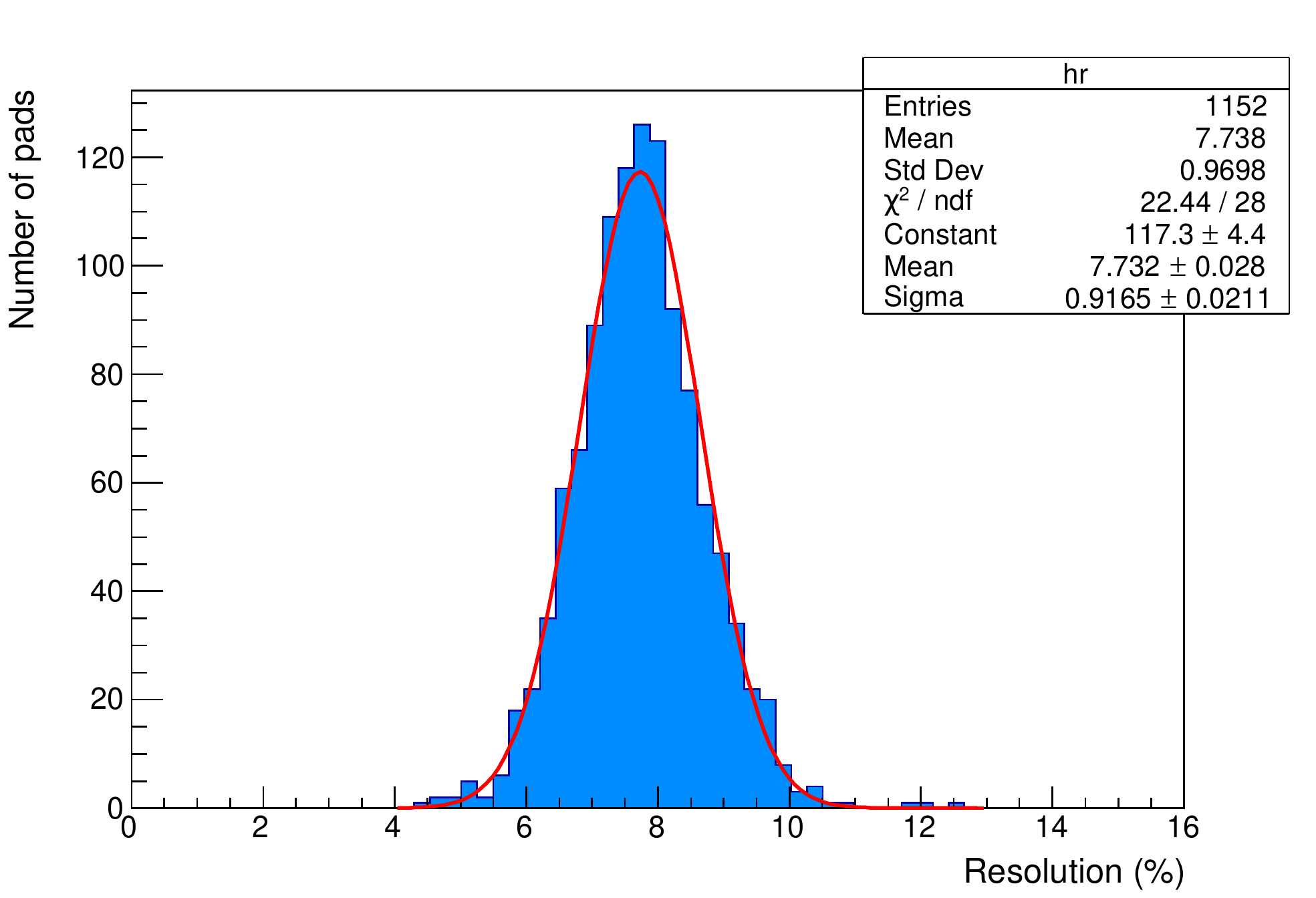}
         \caption{Resolution distribution of ERAM-30}
         \label{fig:Resdist_eram30_sum}
     \end{subfigure}
        \caption{Energy resolution corresponding to gain of ERAM-30 obtained from the method described in sec. \ref{subsec:GainCalculationModel}.}
        \label{fig:Resolution_eram30_sum}
\end{figure}

To quantify the gain and resolution of each pad, a fit is done over the peak of the reconstructed $^{55}$Fe spectrum. 
A typical $^{55}$Fe energy spectrum reconstructed using the gain method introduced in section~\ref{subsec:GainCalculationModel}, is presented in Figure~\ref{fig:FeSpectrum_sum}. Its associated Gaussian fit for the peak at 5.9~keV is superimposed. The corresponding Argon escape peak is also visible. The ratio between the two peak positions is 1.94.
From the peak position, the gain can be obtained using eq.~\ref{equ:GainMethod} where $Q_{Primary}=224~e$.
As for the energy resolution, it is defined as: $\frac{\Delta E}{E} = \frac{\sigma}{\mu}$, with $\sigma$ being the Gaussian standard deviation and $\mu$ being the fitted $K_{\alpha}$ line position. Figure~\ref{fig:Gainmap_eram30_sum} shows the 2D gain map of ERAM-30. This map exhibits local non-uniformity between pads up to 25\%. The 2D energy resolution map of ERAM-30 is shown in Figure~\ref{fig:Resmap_eram30_sum}. An energy resolution better than 10\% is obtained.

\begin{figure}[hbt!]
  \centering
  \includegraphics[width=0.47\textwidth]{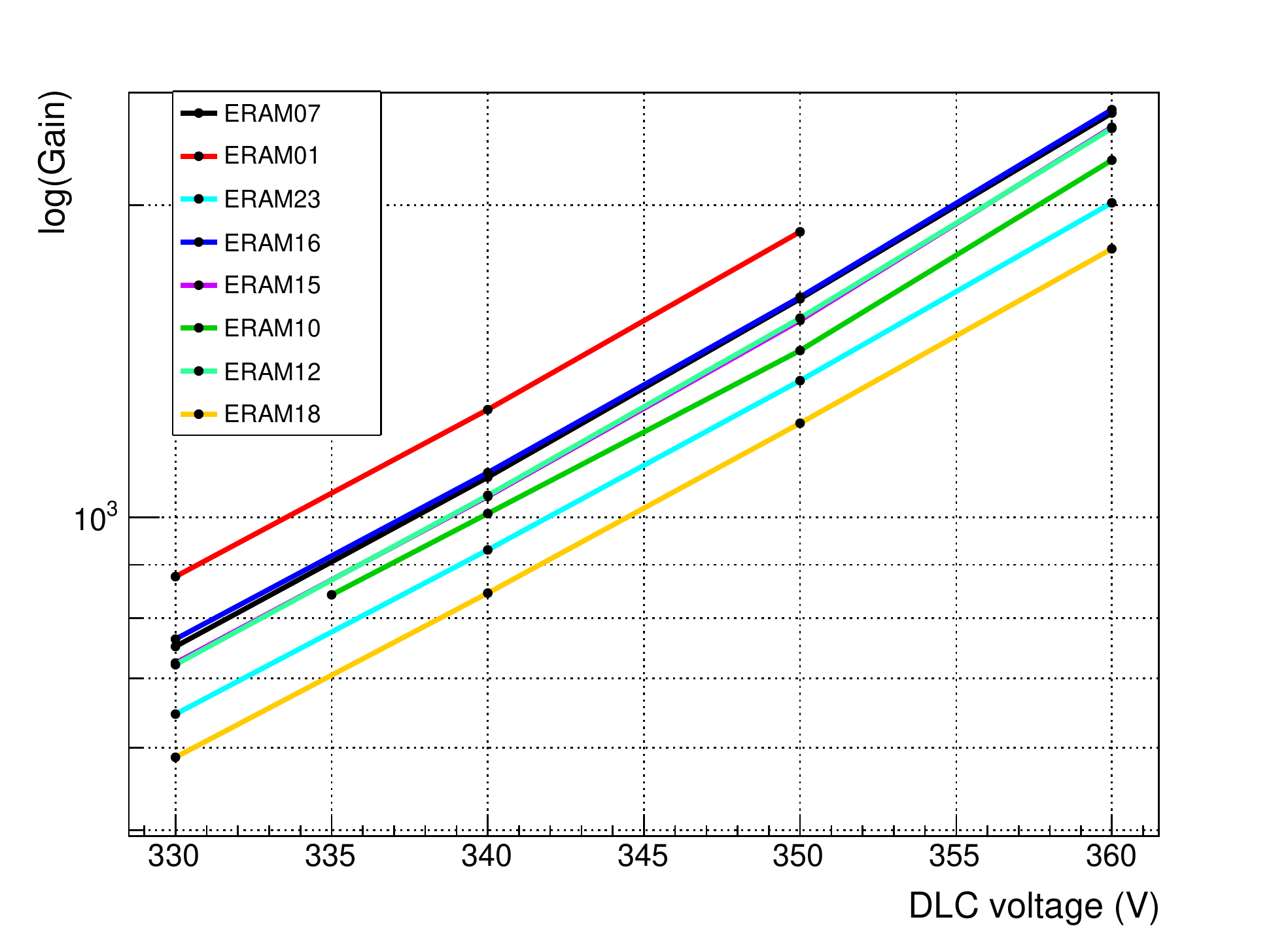}
  \caption{Variation of gain as a function of the applied DLC voltage for different ERAM detectors. The gain value is extracted for one pad in the center of the detector.}
  \label{fig:GainVsDLCVoltage}
\end{figure}

The gain is also studied as a function of the DLC voltage. Figure~\ref{fig:GainVsDLCVoltage} shows the gain as a function of the DLC voltage for different ERAM detectors for one pad at the center of the ERAM. The exponential increase of the gain as a function of the DLC voltage validates the proper functioning of the detector. 

\section{Signal Model}
\label{sec:simul_RCGain}
The signal model is based on the theory given in~\cite{Dixit:2003qg}, whose main points are summarized below.
The spatial spread of the charge on the resistive layer is governed by the diffusion equation:
\begin{equation}
 \frac{\partial \rho(r,t)}{\partial t} = \frac{1}{RC}\Delta\rho(r,t)
\end{equation} 
where $\Delta$ is the Laplace operator, $\rho(r,t)$ is the charge density function, $R$ is the surface resistivity of the layer and $C$ the surface capacitance determined by the spacing between the anode and readout planes. The $RC$ constant is denoted  $\tau=RC$.

For a punctual unit charge deposited at $r=0$ and $t=0$, the charge density as a function of radius $r$ and time $t$ is given by a Gaussian:
\begin{equation}
\label{eq:RMM}
\rho(r,t) = \frac{1}{4\pi (t/\tau)} e^{-r^{2}/4(t/\tau)}.
\end{equation}
The realistic initial charge profile is not a delta function, and may be better approximated
by a Gaussian distribution of finite width $\omega$ and total charge $Q_{e}$. In this case, the
anode surface charge density as a function of space and time is obtained by convoluting equation \ref{eq:RMM} with the initial Gaussian:
\begin{equation}
\label{eq:RMM1}
\rho(r,t) = \frac{Q_{e}}{2\pi \sigma^{2}(t)} e^{-r^{2}/2\sigma^{2}(t)}, 
\end{equation}
where $\sigma(t) = \sqrt{\dfrac{2t}{RC}+\omega^{2}}$.

The induced charge on a rectangular pad below the resistive layer can be calculated by integrating the charge density function over the pad area:
\begin{equation}
\begin{split}
Q_{pad}(t)=\frac{Q_{e}}{4}\times\left[erf( \frac{ x_\textup{high}-x_{0}}{\sqrt{2}\sigma(t)}) -erf( \frac{ x_\textup{low}-x_{0}}{\sqrt{2}\sigma(t)} )\right]\times \\ 
\left[erf( \frac{ y_\textup{high}-y_{0}}{\sqrt{2}\sigma(t)}) -erf( \frac{ y_\textup{low}-y_{0}}{\sqrt{2}\sigma(t)} )   \right]
\label{equ:Qdiffusion}
\end{split}
\end{equation}
where $Q_{e}$ is the initial charge after multiplication, and ($x_{0}$, $y_{0}$) is the position of initial charge deposition. Finally, $x_\textup{high}$, $x_\textup{low}$, $y_\textup{high}$,  $y_\textup{low}$ are the pad boundaries.  

The signal is also affected by electron arrival time spread and position spread due to longitudinal and transverse diffusion respectively. For the X-~ray measurements in the 3~cm drift distance, the charge cluster arriving at the anode had a transverse RMS width of about 540~$\mu$m and a longitudinal RMS spread in time of about 4.5~ns. In all the following studies, $\omega$ associated to the transverse diffusion term is fixed at 540~$\mu$m and the longitudinal diffusion is neglected. 

\begin{figure}[hbt!]
  \centering
  \includegraphics[width=0.75\textwidth]{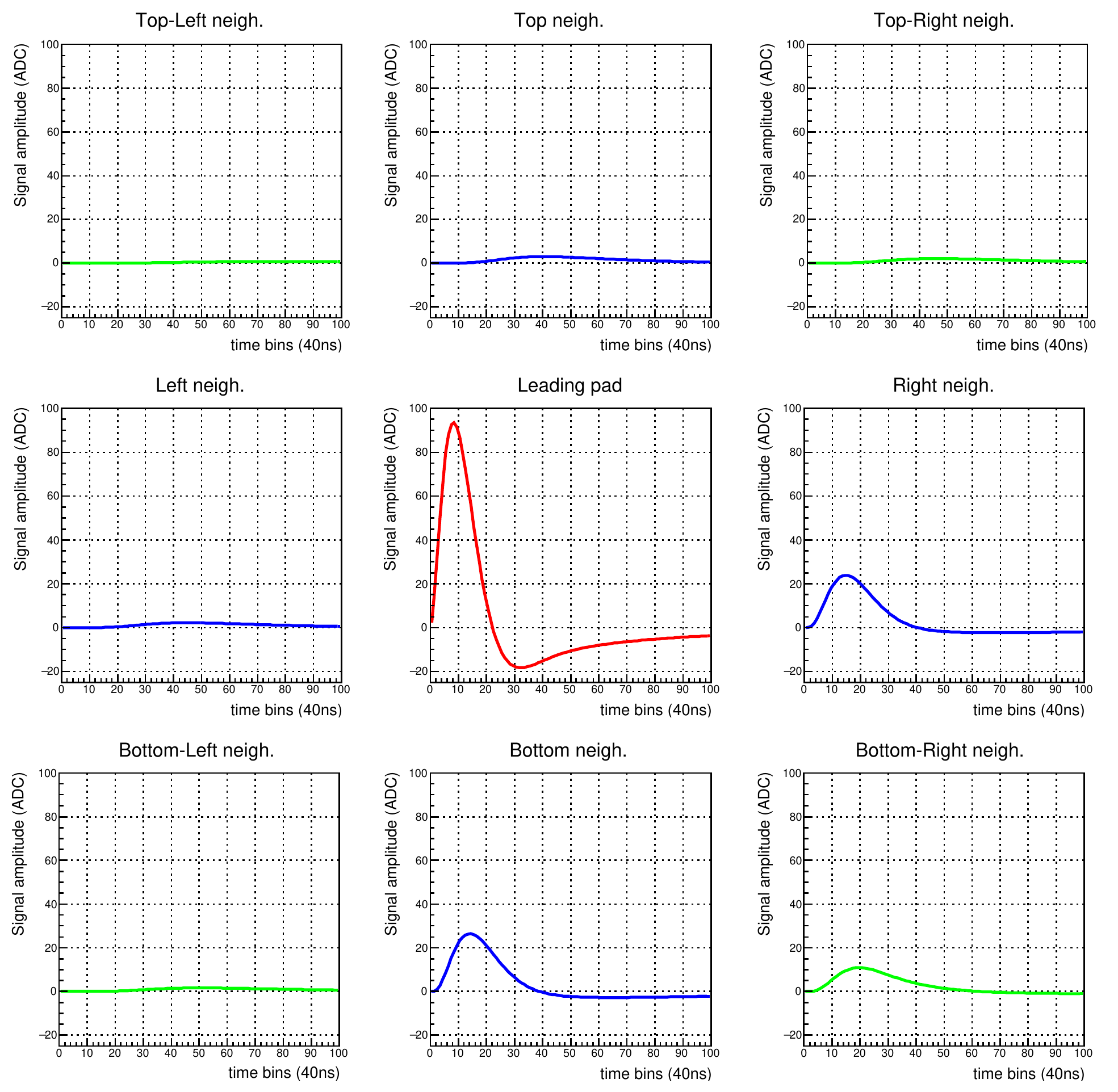}
  \caption{Simulation of charge diffusion convoluted with the electronics response using eq.~\ref{equ:signal_function} , when 24845 $e^{-}$ are deposited at a position (pad width/3, -pad length/3) w.r.t center of the leading pad with $RC$ = 100~ ns/mm$^{2}$.}
  \label{fig:exampleSignal9pads}
\end{figure}

In order to account for the response of electronics to induced charge, the charge diffusion function (eq.~\ref{equ:Qdiffusion}) is convoluted with the derivative of electronics response function (eq.~\ref{equ:ElResponse2Dirac}) to obtain the charge signal function $S(t)$ defined as:
\begin{equation}
S(t) \;=\; Q_{pad}(t) \; \circledast \; \dv{(ADC^{D}(t))}{t} 
  \label{equ:signal_function}
\end{equation}
Figure~\ref{fig:exampleSignal9pads} shows an example of convolution using eq.~\ref{equ:signal_function}, when 24845 electrons are deposited at a position corresponding to~(pad width/3,~-pad length/3)~w.r.t center of the leading pad with $RC$ = 100 ns/mm$^{2}$. A clear difference is seen between the magnitude and time of amplitudes in neighbouring pads closer to the charge deposition point and those that are farther away.
\subsection{Simultaneous fit of waveforms using X-ray data}
\label{subsec:simul_fit}
\begin{figure}[hbt!]
     \centering
     \begin{subfigure}[b]{0.4\textwidth}
         \centering
         \includegraphics[width=1.25\textwidth]{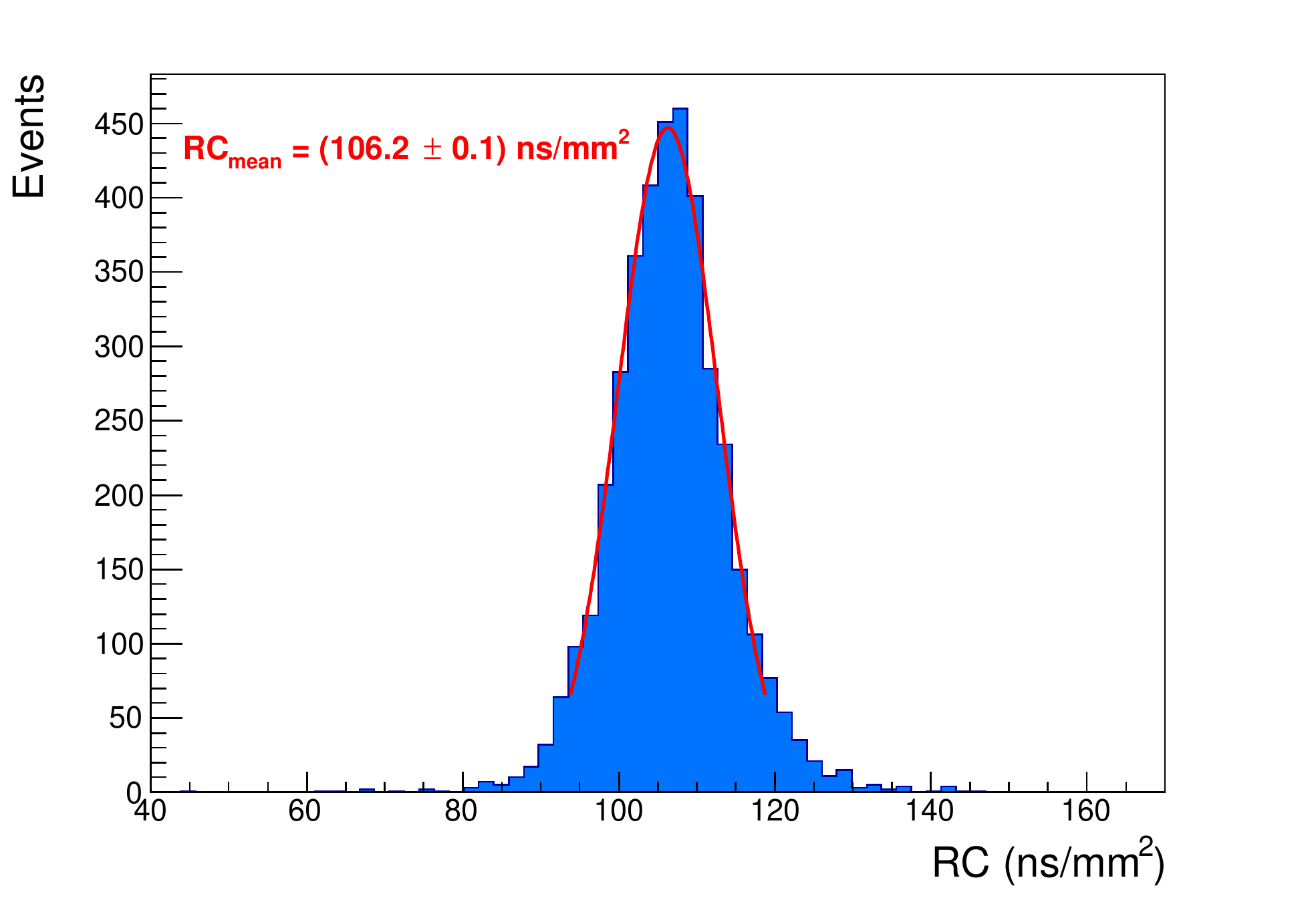}
         \caption{$RC$ distribution of one pad.}
         \label{fig:RCdist_1pad}
     \end{subfigure}
     \hfill
     \begin{subfigure}[b]{0.47\textwidth}
         \centering
         \includegraphics[width=\textwidth]{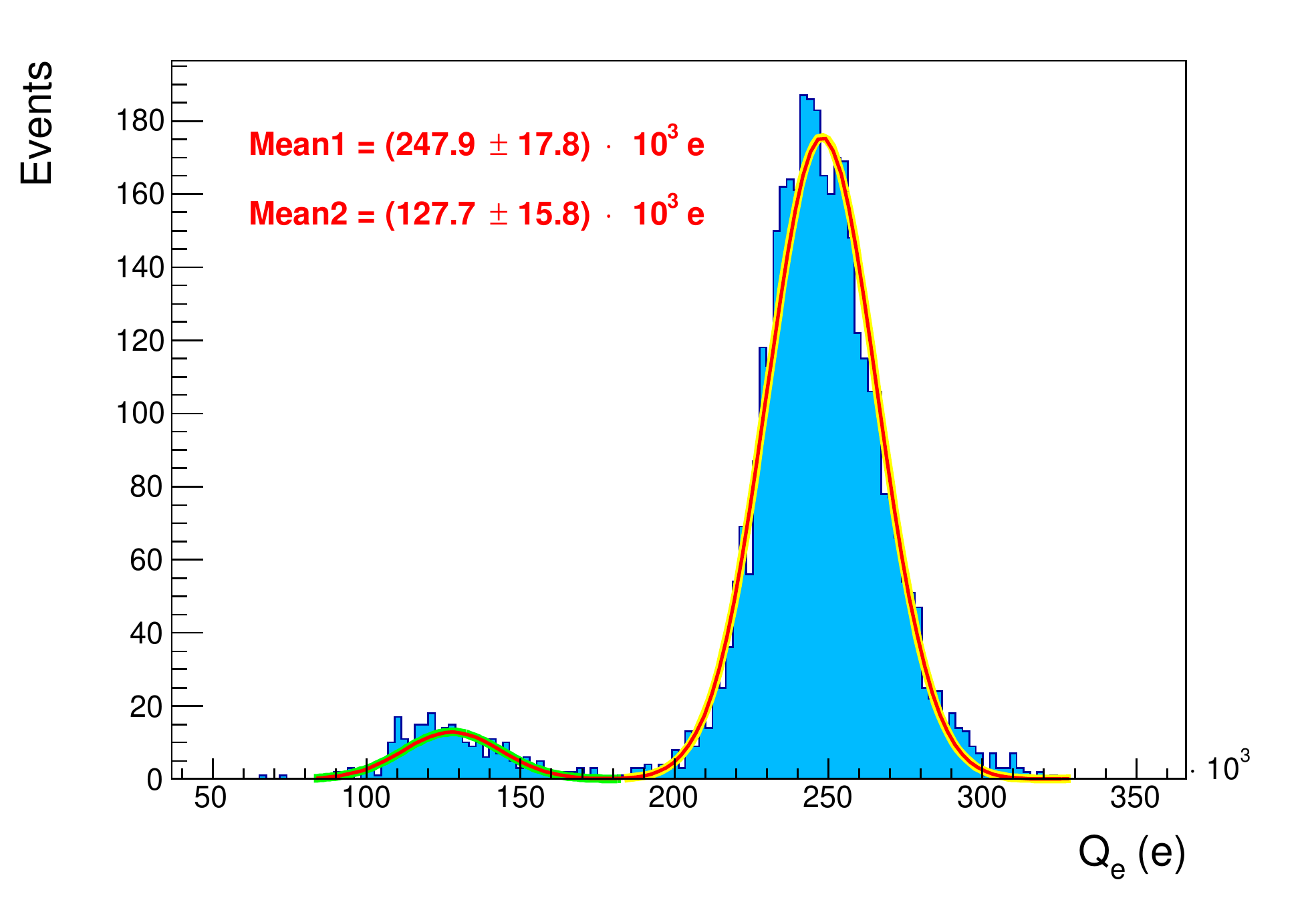}
        \caption{ $Q_{e}$ distribution of one pad.}
         \label{fig:Gaindist_1pad}
     \end{subfigure}
\caption{$RC$ and $Q_{e}$ distributions obtained from fitting all events in one pad.}
\label{fig:RCGain_1pad}
\end{figure}

\begin{figure}[hbt!]
     \centering
     \begin{subfigure}[b]{0.4\textwidth}
         \centering
         \includegraphics[width=1.25\textwidth]{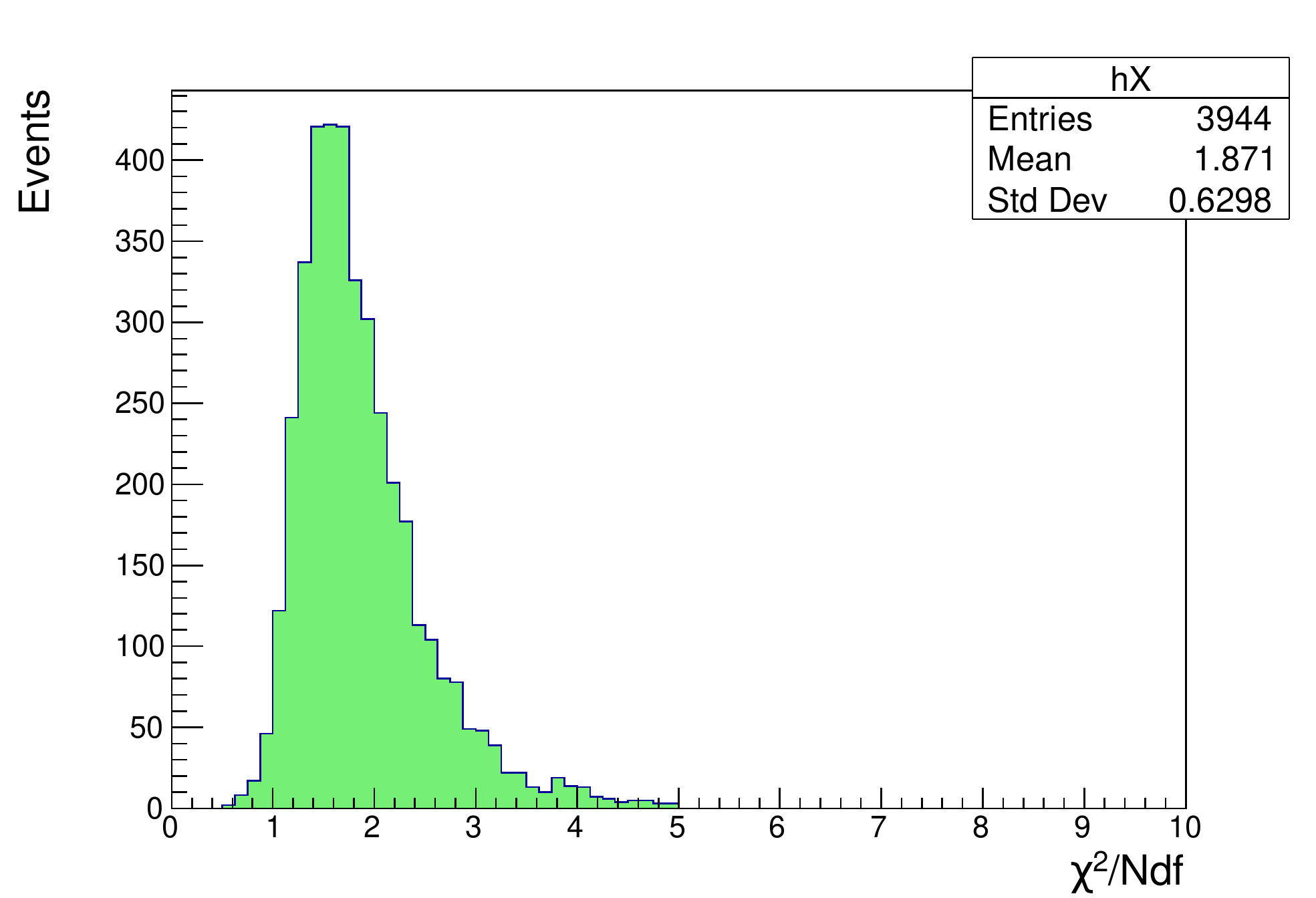}
     \end{subfigure}
     \hfill
     \begin{subfigure}[b]{0.4\textwidth}
         \centering
         \includegraphics[width=\textwidth]{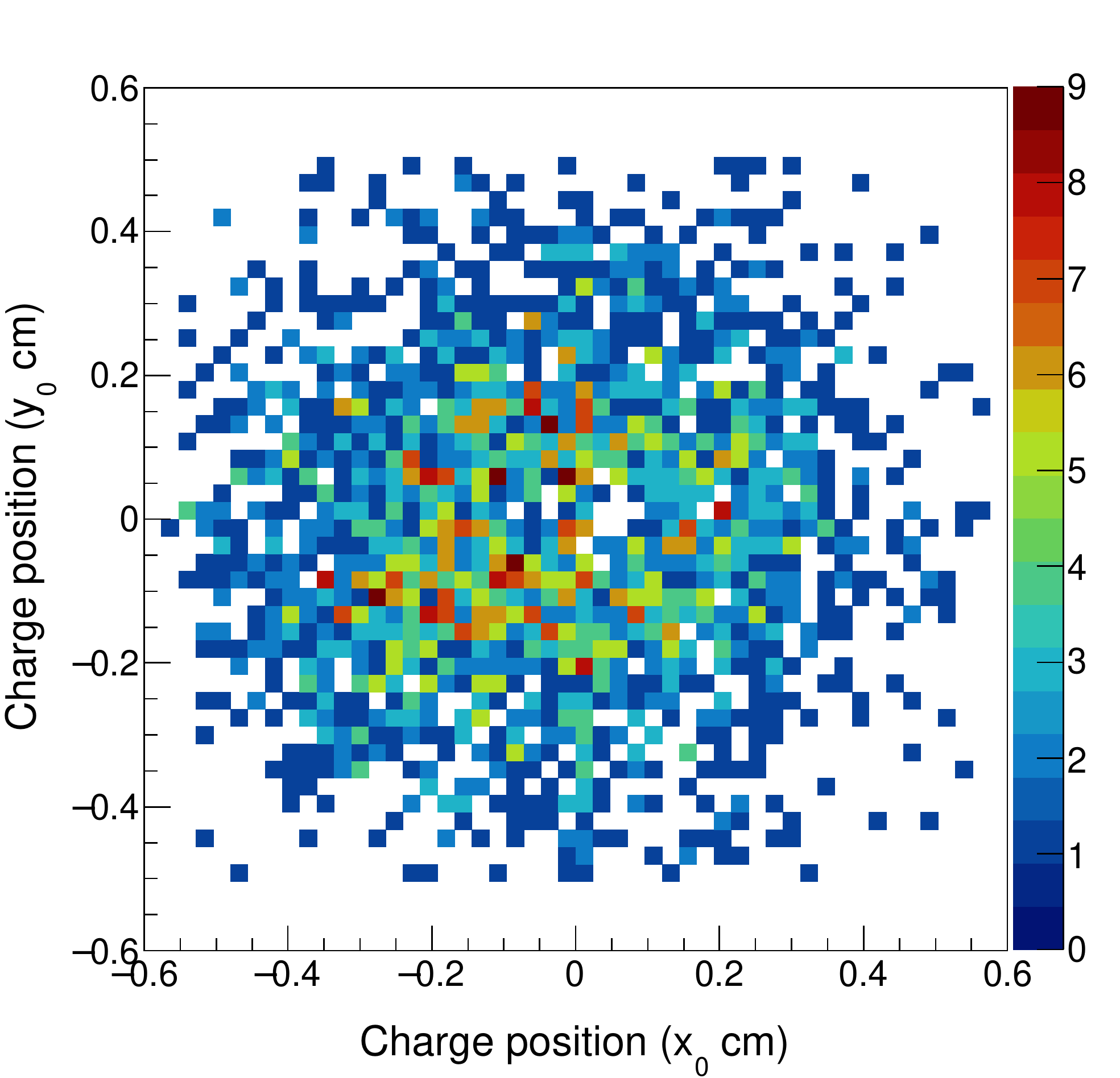}
     \end{subfigure}
        \caption{Distribution of $\chi^{2}$/Ndf   (left) and fitted position of charge depositions ($x_{0}$, $y_{0}$) (right) for all the fitted events in one pad.}
        \label{fig:Xndf_position}
\end{figure}

All the waveforms generated in one X-ray event are fitted simultaneously with the signal function (eq.~\ref{equ:signal_function}) to extract the $RC$ and gain of the leading pad. The fit is based on $\chi^{2}$ minimisation.
Here, an event is defined as an instance where an X-ray photo-electron causes the deposition of charge in the leading pad and its neighboring pads due to charge spreading phenomena. The fit uses at least three waveforms within $3\times3$ matrix of pads around the leading pad to have enough constraints on ($x_{0}$, $y_{0}$) position. To avoid inclusion of noise in the fitting process, waveforms with maximum of amplitude less than 70~ADC are discarded. The typical RMS of the pedestals is 6. This value is used as an error on the signal amplitude. \\
\begin{figure}[hbt!]
  \centering
  \includegraphics[width=0.75\textwidth]{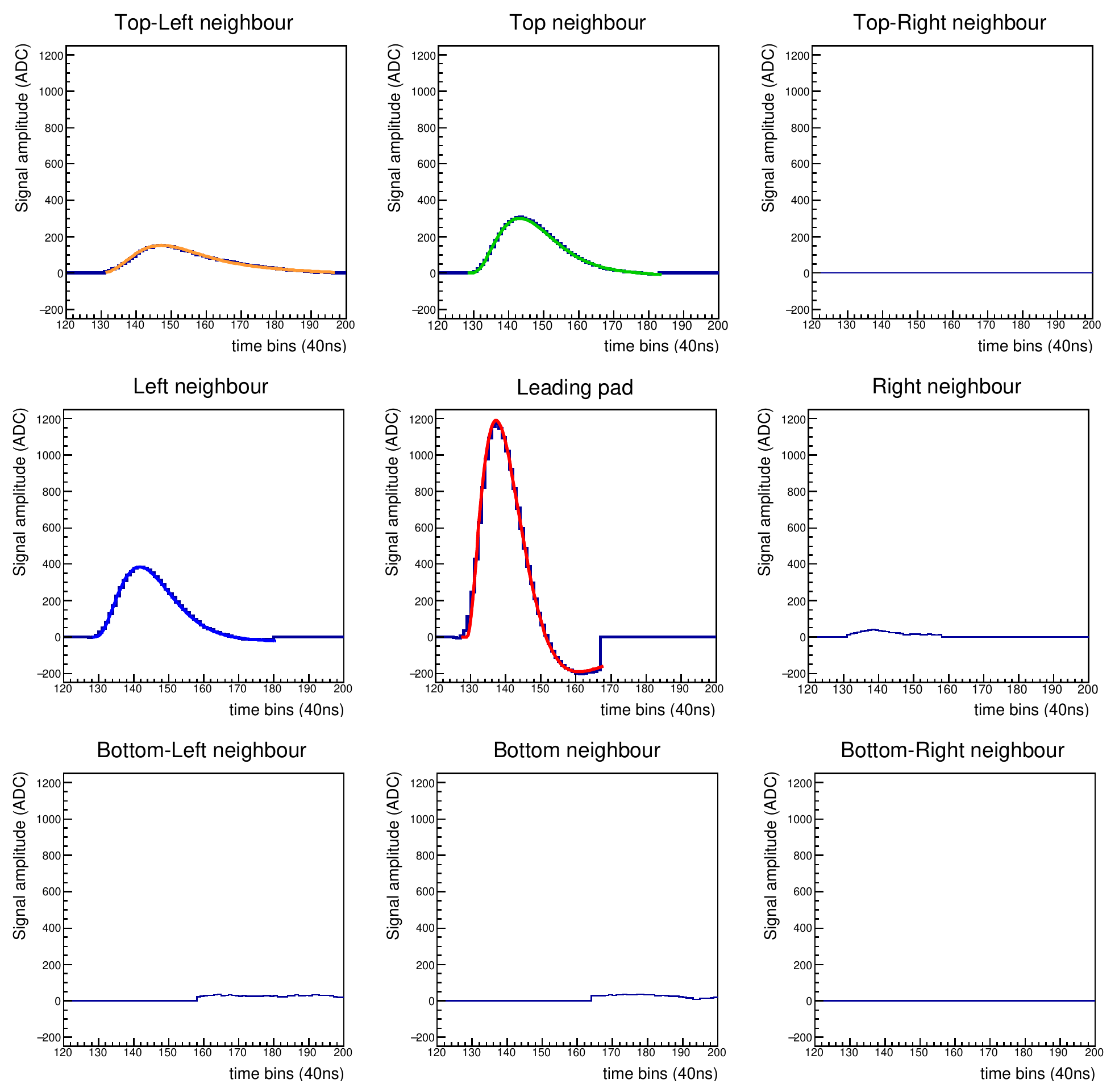}
  \caption{An example of a simultaneous fit of four waveforms of an X-ray generated event. \underline{Fit results:} 
  $RC = (146.6 \pm 1.6$~ns/mm$^{2}$), $Q_{e} = (327.6 \pm 1.8)\times 10^{3} e$, ($x_{0}, y_{0}$)=~(-0.442 cm, 0.352 cm) (w.r.t center of leading pad) , $\chi^{2}$/Ndf~=~1.08.} 
  \label{fig:simul_fit-example1}
\end{figure}
The fit is performed event by event with the electronics response parameters $w_s$, $Q$ fixed to the values obtained from the study described above. The model has five parameters:
\begin{itemize}
    \item $t_{0}$ is the time of charge deposition in the leading pad,
    \item ($x_{0}$, $y_{0}$) are the positions of initial charge deposition,
    \item $RC$ of the readout pad - glue - resistive foil network, and
    \item $Q_{e}$ is the initial charge after multiplication in amplification gap.
\end{itemize}
An example of a simultaneous fit of four waveforms generated in an X-ray event along with their fit results is shown in Figure~\ref{fig:simul_fit-example1}. 
Upon fitting all the events in one pad, a Gaussian distribution of $RC$ is obtained from the extracted $RC$ values, as shown in Figure~\ref{fig:RCdist_1pad}. The mean value of the $RC$ distribution over all the events is considered as the global $RC$ value of that pad. Figure~\ref{fig:Gaindist_1pad} shows the fitted parameter $Q_{e}$ which shape is found to reproduce the $^{55}$Fe spectrum and thus can be used to extract the gain. This topic is further elaborated in section \ref{sec:Gainmap-simulfit}. Figure~\ref{fig:Xndf_position}  (left) shows the distribution of $\chi^{2}/$Ndf of all the simultaneous event fits in one pad and Figure~\ref{fig:Xndf_position}(right) is a 2D mapping of estimated positions of all the charge depositions due to X-ray photon-electrons targeted at the leading pad. The shape of the charge position distribution reflects the circular aperture of the collimator that houses the X-ray source.


\section{$RC$ map from fit of waveforms using X-ray data}
\label{sec:RCmap}
\begin{figure}[hbt!]
     \centering
     \begin{subfigure}[b]{0.45\textwidth}
         \centering
         \includegraphics[width=\textwidth]{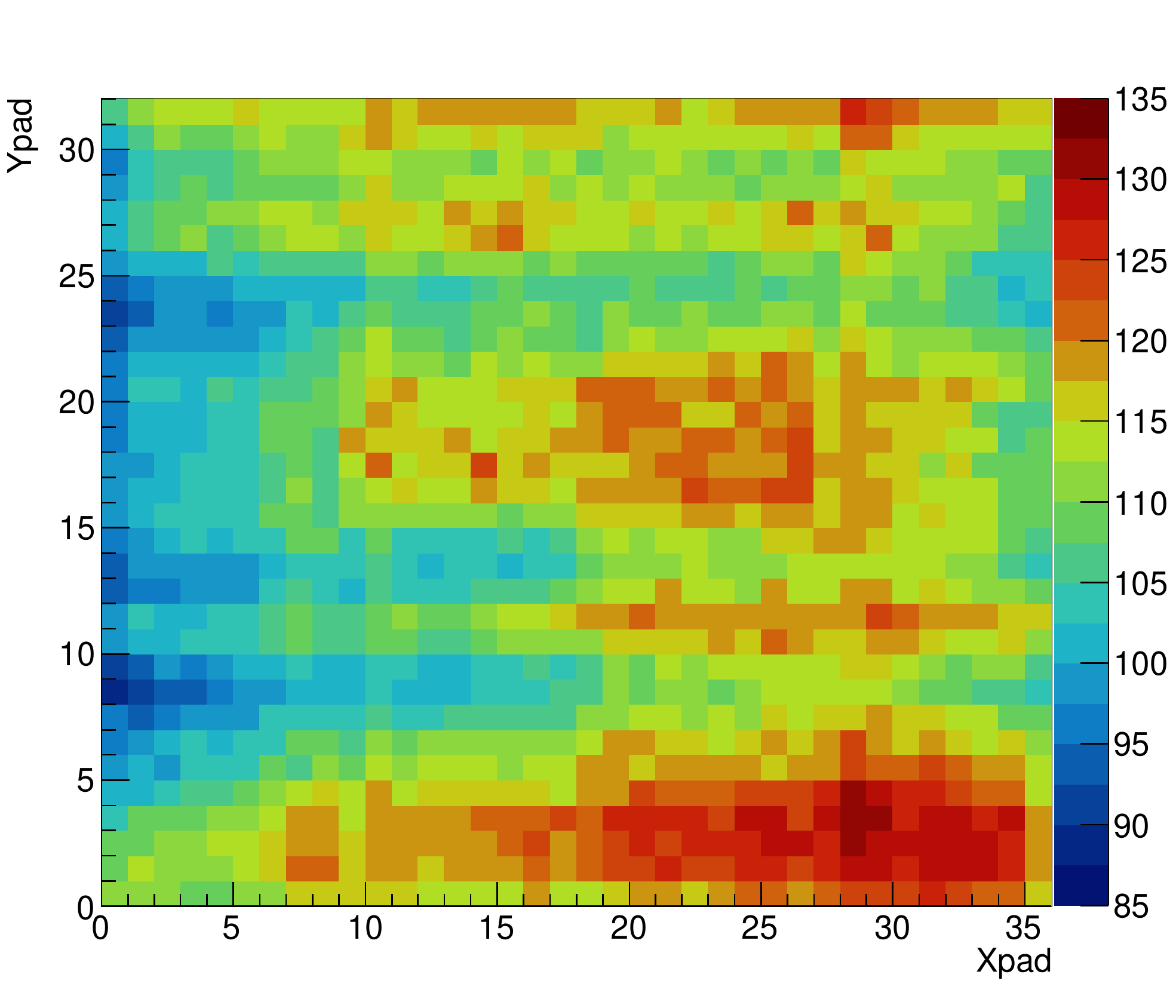}
         \caption{$RC$ map of ERAM-30.}
         \label{fig:RCmap_eram30}
     \end{subfigure}
     \hfill
     \begin{subfigure}[b]{0.4\textwidth}
         \centering
         \includegraphics[width=1.25\textwidth]{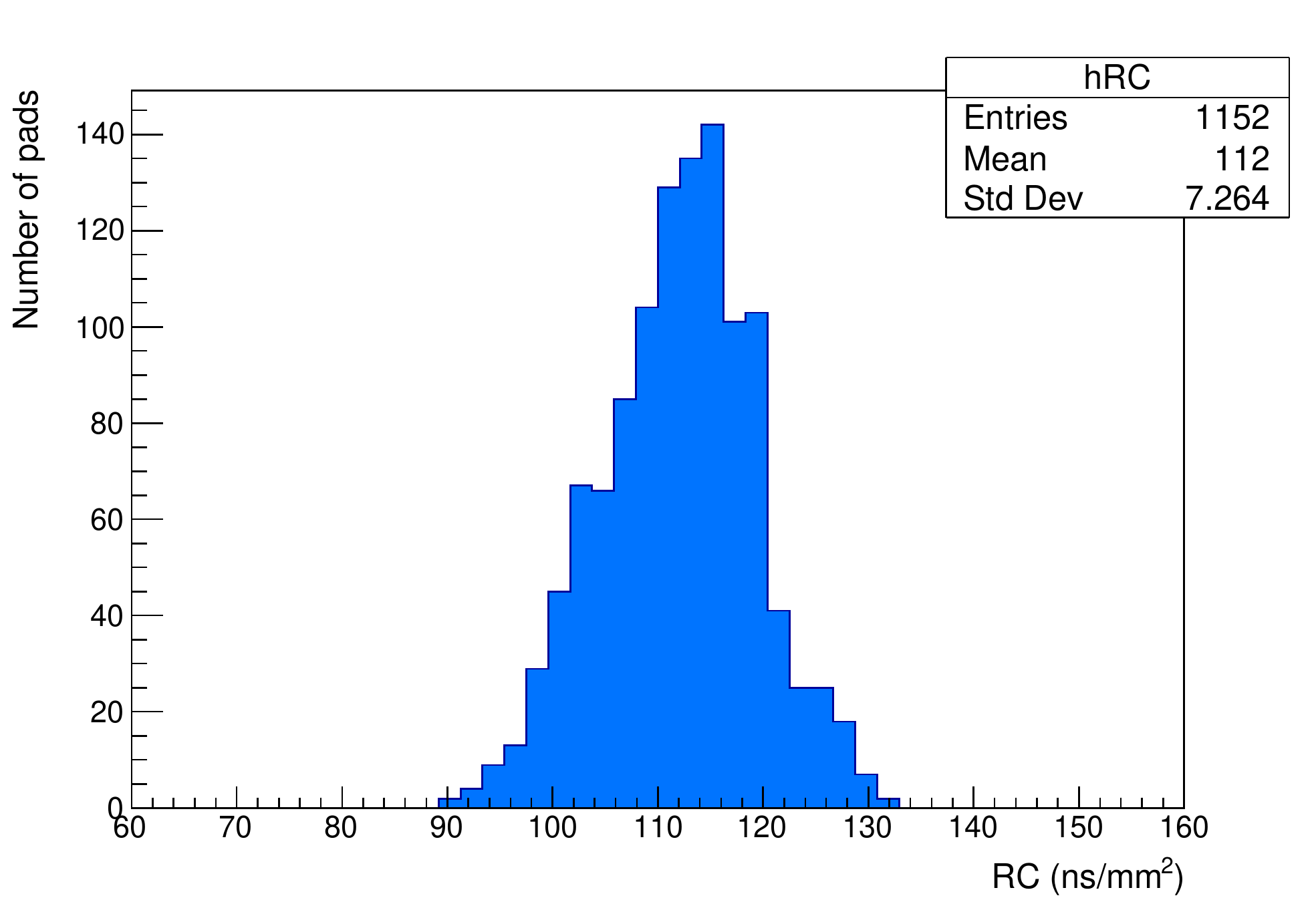}
         \caption{$RC$ distribution of all the pads of ERAM-30.}
         \label{fig:RCdist_eram30}
     \end{subfigure}
        \caption{$RC$ information extracted from ERAM-30 test bench data.}
        \label{fig:RC_eram30}
\end{figure}

The fitting process described in section \ref{subsec:simul_fit} is repeated for events generated in all the pads to obtain the 2D $RC$ map and a mean $RC$ value representative of the ERAM scanned under the X-ray test bench. Figures \ref{fig:RCmap_eram30} and \ref{fig:RCdist_eram30} show the $RC$ map and $RC$ distribution of all the pads of ERAM-30 respectively. The 2D $RC$ map exhibits homogeneity in horizontal direction that will be discussed in section ~\ref{sec:RCfeatures}. However, global variations in $RC$ map up to 35\% are observed.
\\

\subsection{$RC$ maps of different ERAMs}
\begin{figure}[hbt!]
  \centering
  \includegraphics[width=0.99\textwidth]{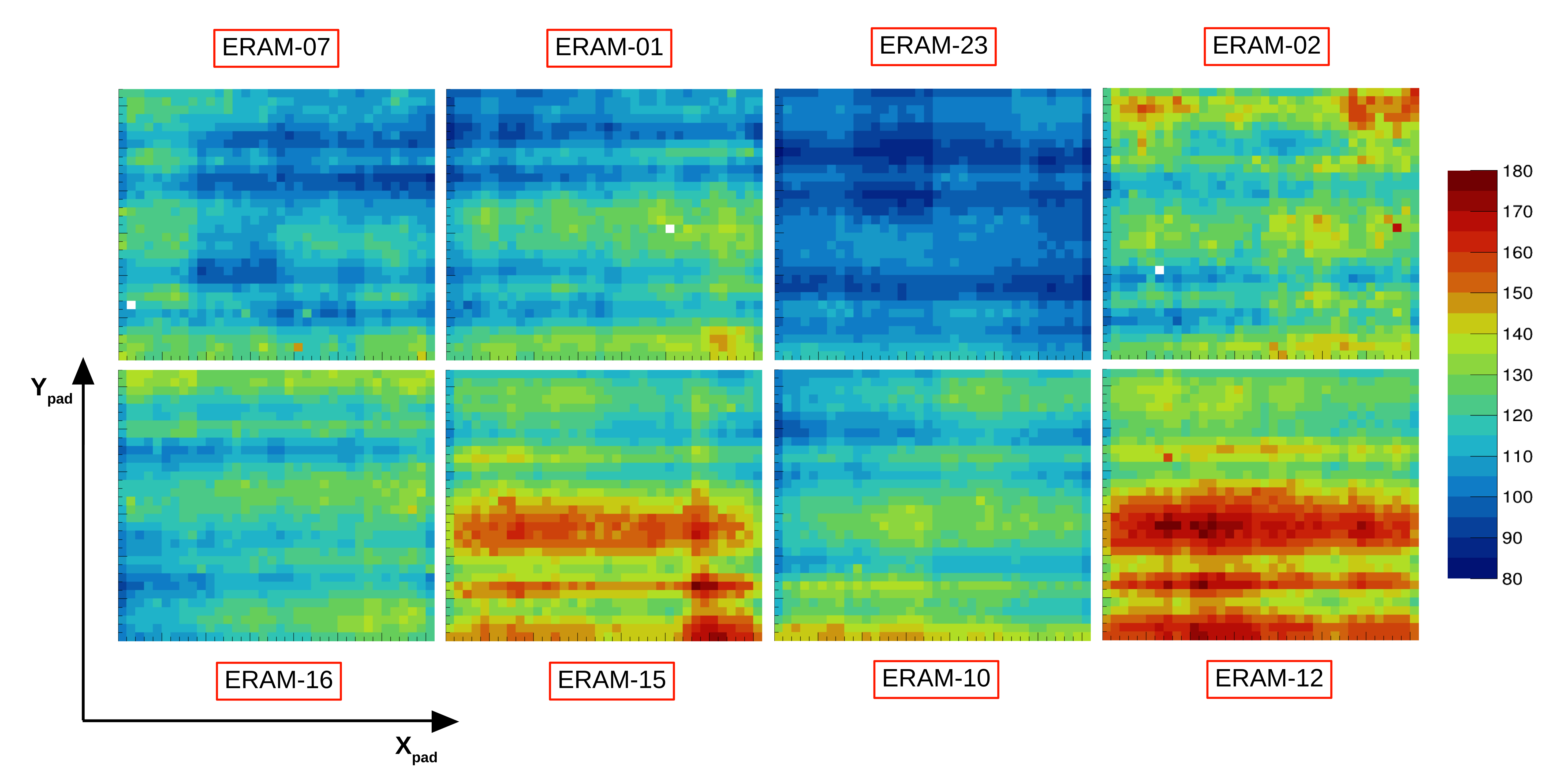}
  \caption{$RC$ maps of 8 ERAMs tested together in a field cage prototype at CERN test beam in September 2022.}
  \label{fig:Testbeam_RCmaps}
\end{figure}

A summary of $RC$ maps of 23 ERAMs that have been studied so far is given in table~\ref{tab:Table_RCGainmaps}.  Figure~\ref{fig:Testbeam_RCmaps} depicts the $RC$ maps of eight other ERAMs that have been characterized. DLC foils of surface resistivity of 400~k$\Omega/\square$ and glue thickness of 150~$\mu$m have been used for the production of these ERAMs. Mean $RC$ values of all these ERAMs lie in the range of 100~ns/mm$^{2}$ to 150~ns/mm$^{2}$. 

Two ERAMs with different values of surface resistivity and glue thickness than the usual specifications have also been studied. One of the ERAMs (ERAM-18) has the same glue thickness, but half the surface resistivity of a typical ERAM. Theoretically, its $RC$ value should be half of that of a typical ERAM, which was indeed found to be the case as it was measured to exhibit a $RC$ value of 70~ns/mm$^{2}$. Another ERAM (ERAM-29) has half the glue thickness and half the surface resistivity of a typical ERAM, thus its $RC$ should remain unchanged. ERAM-29 was evaluated to have a mean $RC$ value of 102~ns/mm$^{2}$, which is within the range of standard $RC$ values as expected. $RC$ maps of all the ERAMs that have been characterized so far possess well-defined uniformity in horizontal direction. 
\begin{table}[hbt!]
\begin{center}
\begin{tabular}{|c|c|c|} 
 \hline
 \small
ERAM & Mean $RC$  & RMS of $RC$  \\
& (ns/mm$^{2}$) & (ns/mm$^{2}$)\\
\hline 
01 & 114.9 & 12.1 \\ 
02 & 124.2 & 12.2 \\ 
03 & 116.4 & 8.0 \\ 
07 & 111.3 & 9.7 \\
09 & 115.7 & 9.9 \\
10 & 119.9 & 9.0 \\
11 & 122.4 & 6.7 \\
12 & 144.2 & 15.1 \\ 
13 & 114.9 & 8.1 \\ 
14 & 100.5 & 8.2 \\ 
15 & 134.9 & 14.6 \\ 
16 & 119.3 & 8.4 \\
17 & 122.1 & 6.7 \\
\hline
18 & 68.98 & 4.3 \\
\hline
19 & 109.5 & 5.6 \\
20 & 111.1 & 7.1 \\
21 & 97.7 & 5.9 \\
23 & 100.2 & 6.0 \\ 
24 & 105.9 & 6.9 \\ 
26 & 115.3 & 6.1 \\
28 & 109.6 & 5.5 \\
29 & 102.0 & 5.9 \\ 
30 & 112 & 7.3  \\ 
\hline
\end{tabular}
\caption{Summary of global $RC$ values  of 23 ERAMs. All ERAMs have been produced using DLC resistivity of 400~ k$\Omega/\square$ and glue thickness of 150~$\mu$m except ERAM-18 which has the same glue thickness, but half the surface resistivity and ERAM-29 which has half the glue thickness and half the surface resistivity. }
\label{tab:Table_RCGainmaps}
\end{center}
\end{table}
\subsection{Understanding $RC$ map features}
\label{sec:RCfeatures}
\begin{figure}[hbt!]
\begin{minipage}{.48\linewidth}
\centering
\subfloat[]{\label{main:a}\includegraphics[scale=.27]{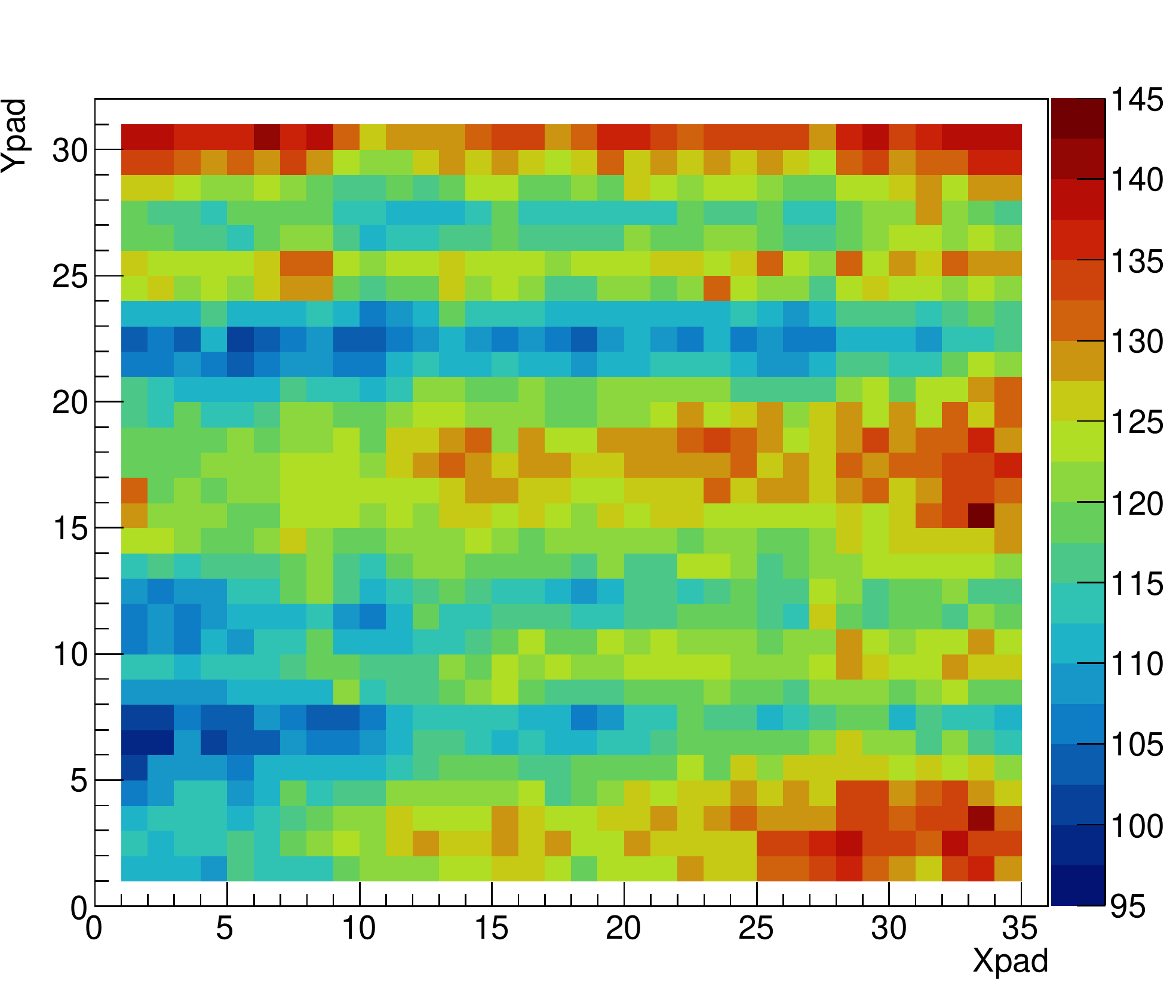}}
\end{minipage}%
\begin{minipage}{.48\linewidth}
\centering
\subfloat[]{\label{main:b}\includegraphics[scale=.27]{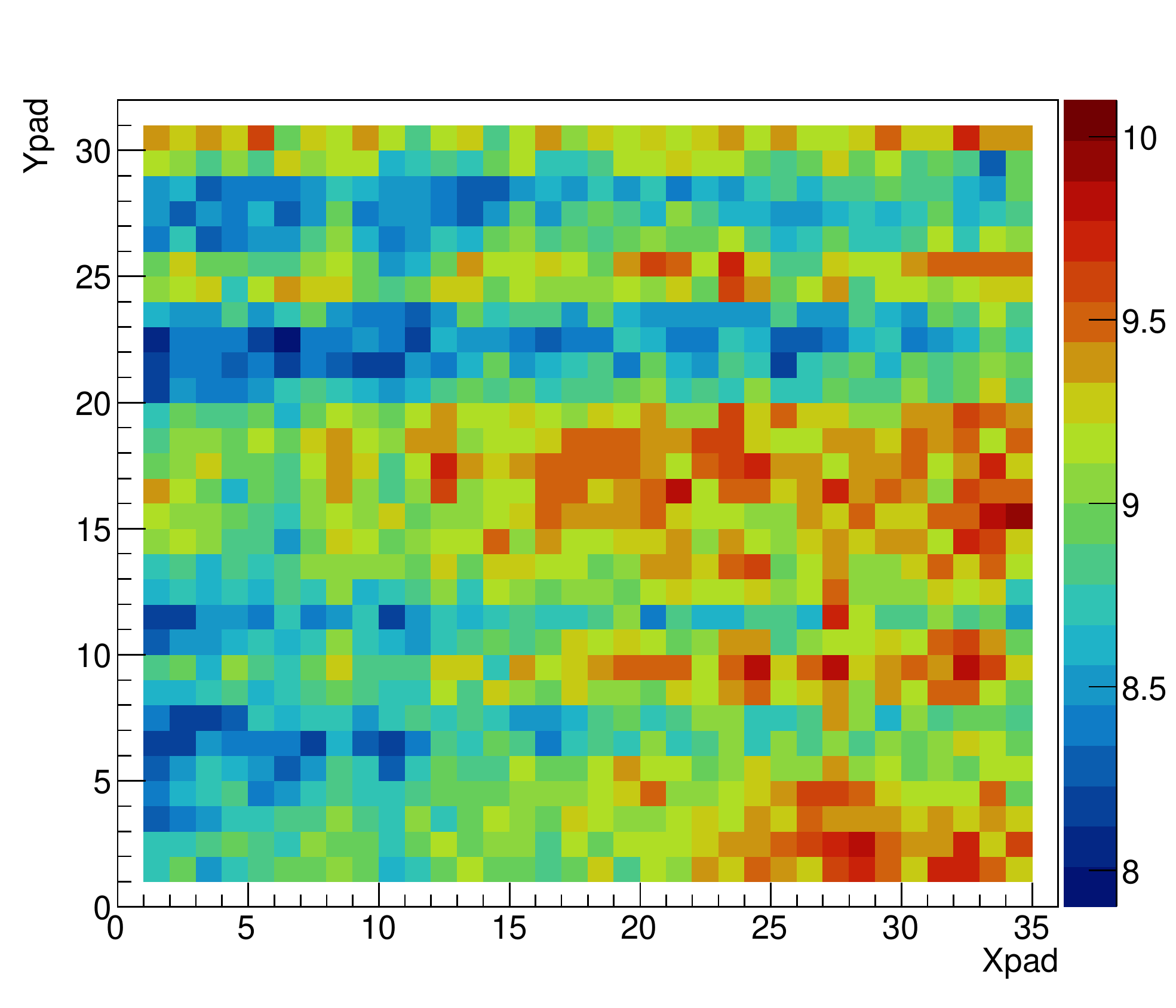}}
\end{minipage}\par\medskip     
\centering
\begin{minipage}{.48\linewidth}
\centering
\subfloat[]{\label{main:c}\includegraphics[scale=.27]{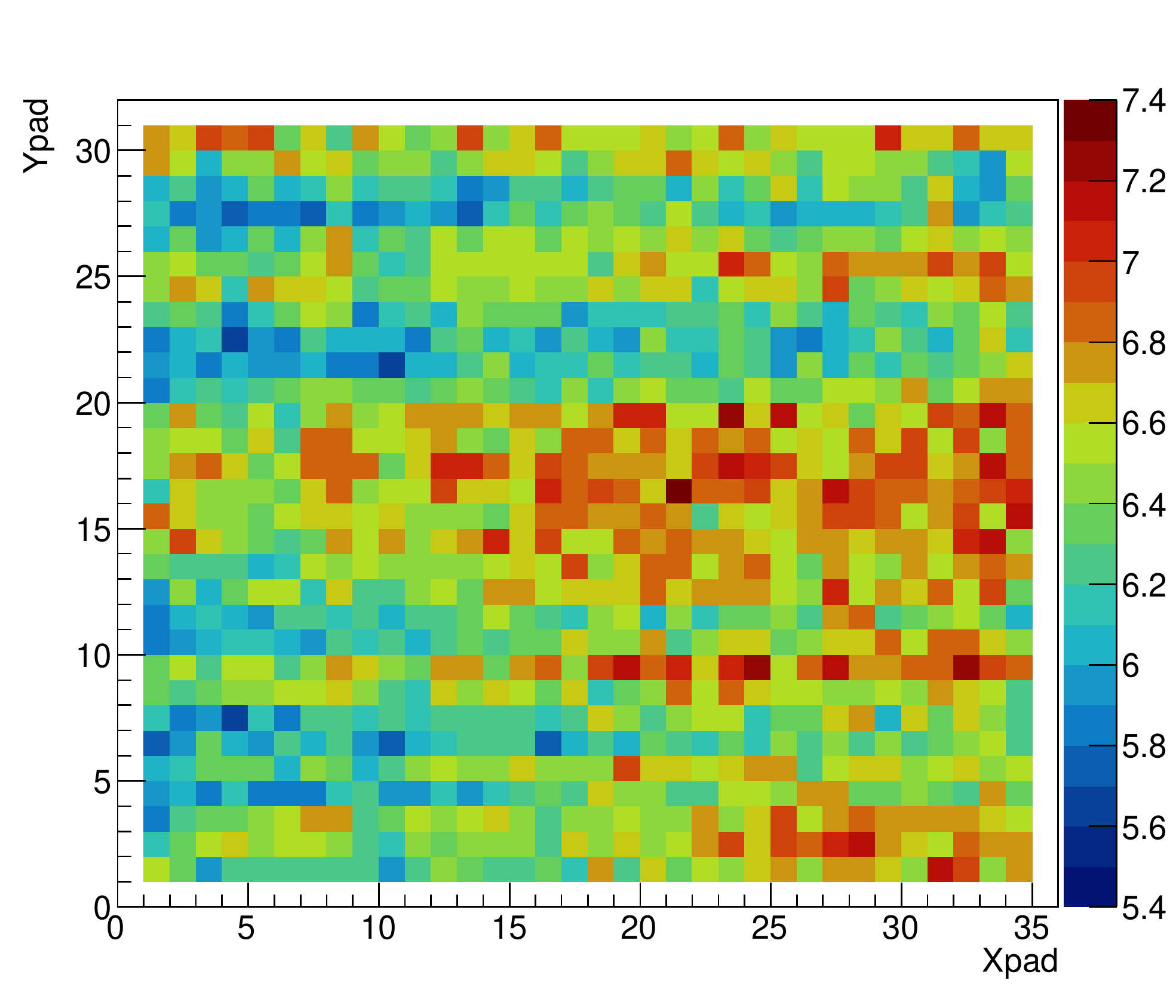}}
\end{minipage}%
\hfill
\begin{minipage}{.5\linewidth}
\centering
\subfloat[]{\label{main:d}\includegraphics[scale=.4]{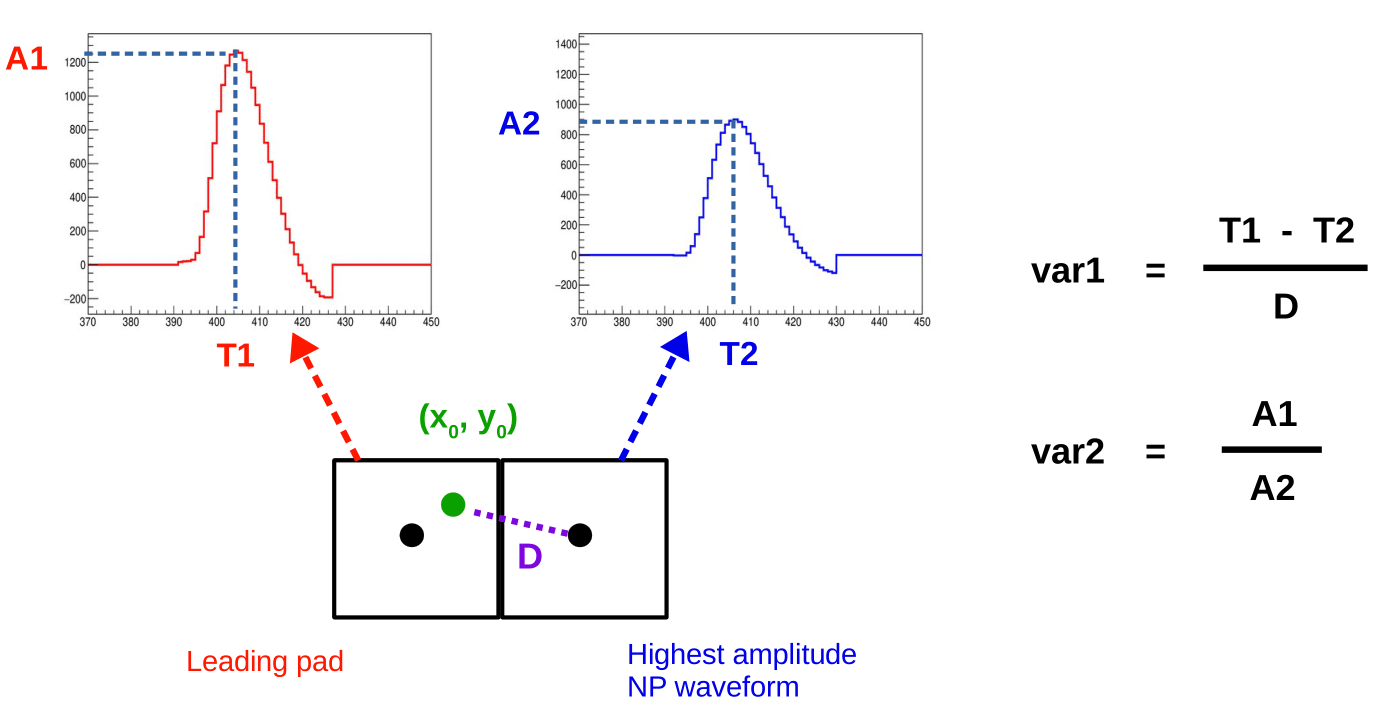}}
\end{minipage}
\caption{Comparing the features of an $RC$ map (a) with the maps of two different basic-level variables (b) and (c) for ERAM-16. Variables var1 and var2 described in plot (d) are used to construct the maps (b) and (c) respectively.}
\label{fig:simpVar_eram16}
\end{figure}

In order to understand if the horizontal structures observed in the $RC$ maps are intrinsic to the device or an artefact of the fit, basic-level observables sensitive to charge spreading are reconstructed. Two such variables that are related to charge spreading and can serve as an indicator of variations in $RC$ value from pad to pad are shown in Figures~\ref{main:b} and \ref{main:c}. 

The schema shown in Figure~\ref{main:d} explains the variables used to construct both maps. The variable var1, used to produce the map in Figure~\ref{main:a}, is defined as the ratio between the difference of $T_{max}$ (time at which the amplitude occurs in a waveform) of the leading pad waveform and that of highest amplitude neighbouring pad waveform, and the distance of charge deposition point $(x_{0}, y_{0})$ from the center of neighbouring pad with highest amplitude waveform. In order for the variable to remain independent of the simultaneous fit model, the charge deposition point is computed using the center of charge method. While the variable var2, used to produce the map in Figure~\ref{main:c}, is defined as the ratio of amplitude of the leading pad waveform to the amplitude of the highest amplitude neighbouring pad waveform. 

Maps of both basic-level variables exhibit the key features of the $RC$ map with varying degrees of precision. The position and magnitude of regions of high and low $RC$  are reproduced by the basic-level variable maps. Hence, the observed features in $RC$ maps are intrinsic to the device.

\subsection{Comparison with expected $RC$ value}
\begin{figure}[hbt!]
     \centering
     \begin{subfigure}[b]{0.44\textwidth}
         \centering
         \includegraphics[width=\textwidth]{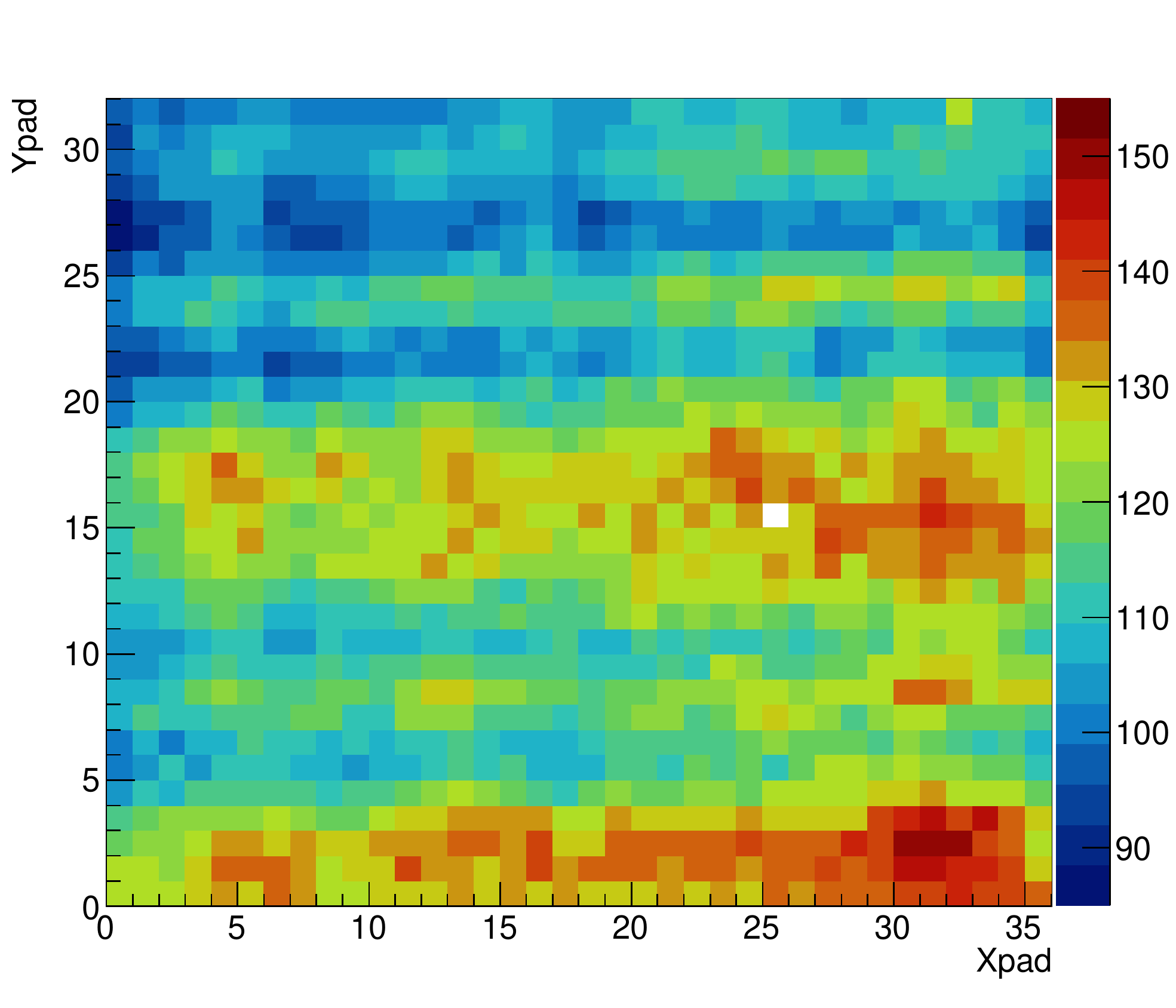}
     \end{subfigure}
     \hfill
     \begin{subfigure}[b]{0.42\textwidth}
         \centering
         \includegraphics[width=\textwidth]{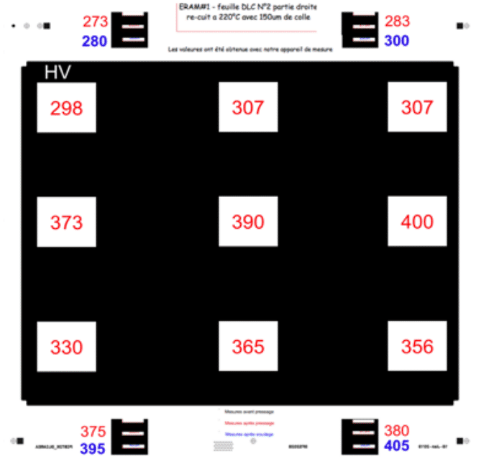}
     \end{subfigure}
        \caption{ERAM-01 $RC$ map   (left), resistivity measurements of ERAM-01 using a probe designed by CERN PCB workshop (right). ERAM-01 resistivity after DLC pressing is shown in red. After gluing the detector, the resistivity measurements are shown in blue. The measurements can be performed in only four positions outside the active area of the detector. }
        \label{fig:RCcorrelationResistivity}
\end{figure}

The correlation between the resistivity measurements and the 2D $RC$ map has been studied. 
The resistivity of DLC foils is measured during detector production using a probe designed by the CERN PCB workshop. Figure~\ref{fig:RCcorrelationResistivity} shows resistivity measurements of ERAM-01 in nine different positions after laminating the DLC foil on the detector PCB. The horizontal structures in $RC$ map are clearly correlated to resistivity measurements: the higher the resistivity value, the higher is the estimated $RC$ value. 
In order to relate the mean $RC$ value extracted from the fit with the expected one, $RC$ is calculated assuming a simple plane capacitance $C$ determined by the Apical (polyimide film) and glue thickness. In this case, the constant capacity $C=\epsilon_{0}\epsilon_{r}/{d}$ can be calculated, where $\epsilon_{0}$ is the vacuum permittivity and $\epsilon_{r}$ is the relative permittivity of the Apical and the glue material. The distance $d$ is defined as the thickness of the Apical and the glue.

After gluing the detector ERAM-01, the resistivity measurements can be performed in only four positions outside the active area of the detector. The $R$ measurements vary between 280 and 405~k$\Omega/\square$ as illustrated in Figure~\ref{fig:RCcorrelationResistivity} which leads to the expected $RC$ varying from 53 to 77 ns/mm$^{2}$. The calculated $RC$ is about a factor of 2 lower than the $RC$ extracted from the fit. 

A possible explanation of this discrepancy could be the ion field on the DLC~\cite{Luca:LUCAREF}. The ion field can greatly affect long distance charge spreading by limiting its spread. This phenomena is called \textit{Shepherd~Dog} effect. Using a toy model, the $RC$ value is indeed found to be a factor 2.7 greater than the one obtained without considering the ion contribution. This explanation is actually not completely proven and further studies are planned to understand this effect.
\subsection{Systematic uncertainties on $RC$}
\subsubsection{Effect of DLC Voltage on $RC$}
\label{sec:DLCRC}
\begin{figure}[hbt!]
  \centering
  \includegraphics[width=0.5\textwidth]{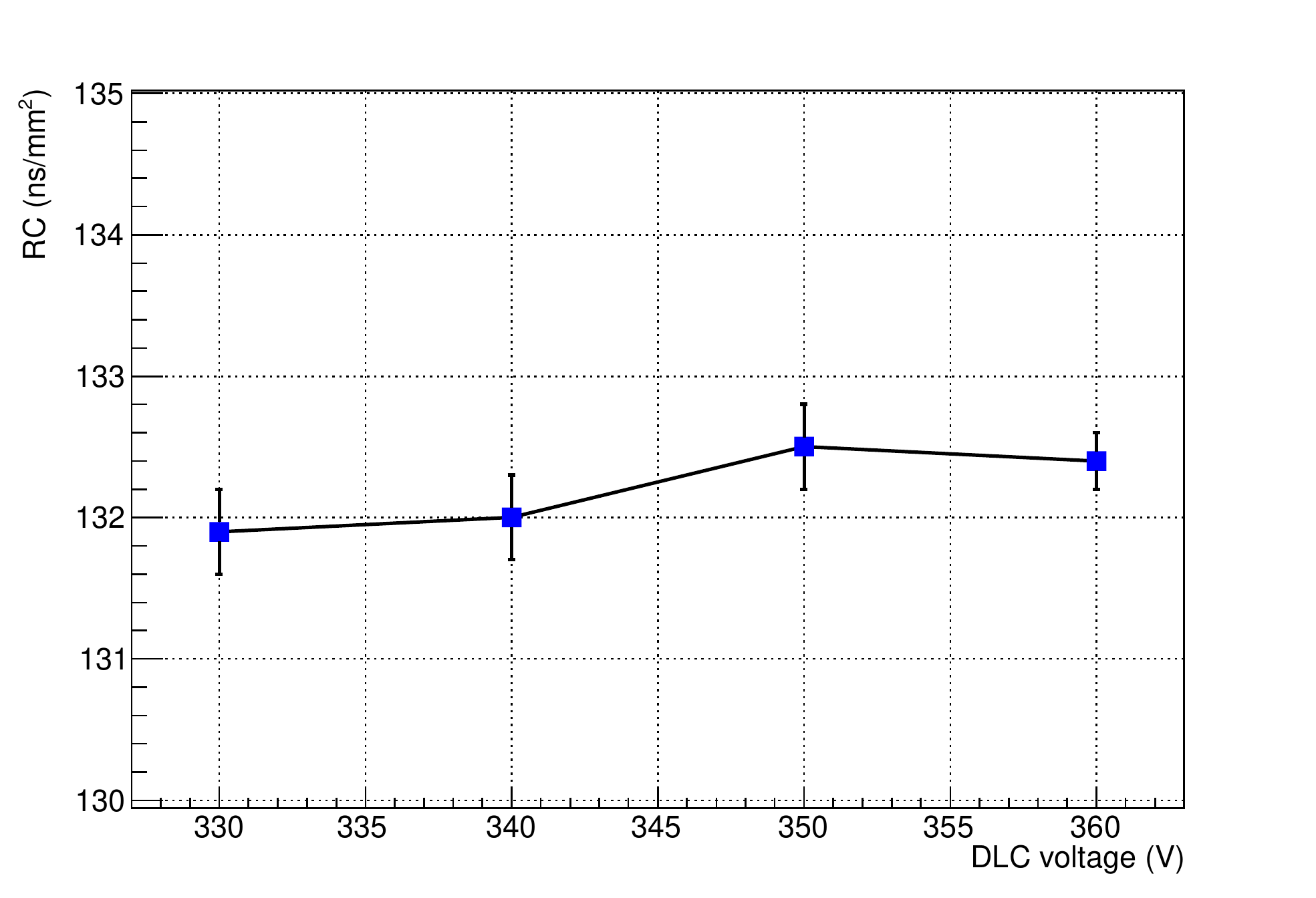}
  \caption{Effect of DLC voltage on $RC$ value.}
  \label{fig:RCvolt}
\end{figure}

The effect of the DLC voltage on the measurement of $RC$ was studied by fitting the events generated in one pad during scans at different DLC voltages. Figure~\ref{fig:RCvolt} depicts the result of this study. Changes in the mean $RC$ values w.r.t applied DLC voltage are negligible. This study demonstrates that there is no apparent correlation between $RC$ and gain values. 

\subsubsection{Effect of electronics  on $RC$}
\label{sec:fecRC}
The electronics response parameters $Q$ and $w_{s}$ were set as constants in the simultaneous fit model for all the results shown above. But as seen in Figures \ref{fig:ElectronicsResponse_Q} and \ref{fig:ElectronicsResponse_ws}, $Q$ and $w_{s}$ values can vary up to 4\% between individual pads and the mean value considered for the entire ERAM. Hence, it is important to study the effect of these parameters on charge signal, and consequently on $RC$ to understand the consequences of keeping $Q$ and $w_{s}$ fixed to a central value for an entire ERAM. 
This effect was studied for ERAM-03 with its set of front-end cards. For each of the pads involved in the simultaneous fit of an event, specific $Q$ and $w_{s}$ values corresponding to the FEC channels connected to those pads were used. A variation of 5\% in mean $RC$ of entire ERAM was observed when specific $Q$ and $w_{s}$ values were taken into account for each pad with respect to keeping them constant.
\section{Gain map from fit of waveforms using X-ray data}
\label{sec:Gainmap-simulfit}
\begin{figure}[hbt!]
  \centering
  \includegraphics[width=0.99\textwidth]{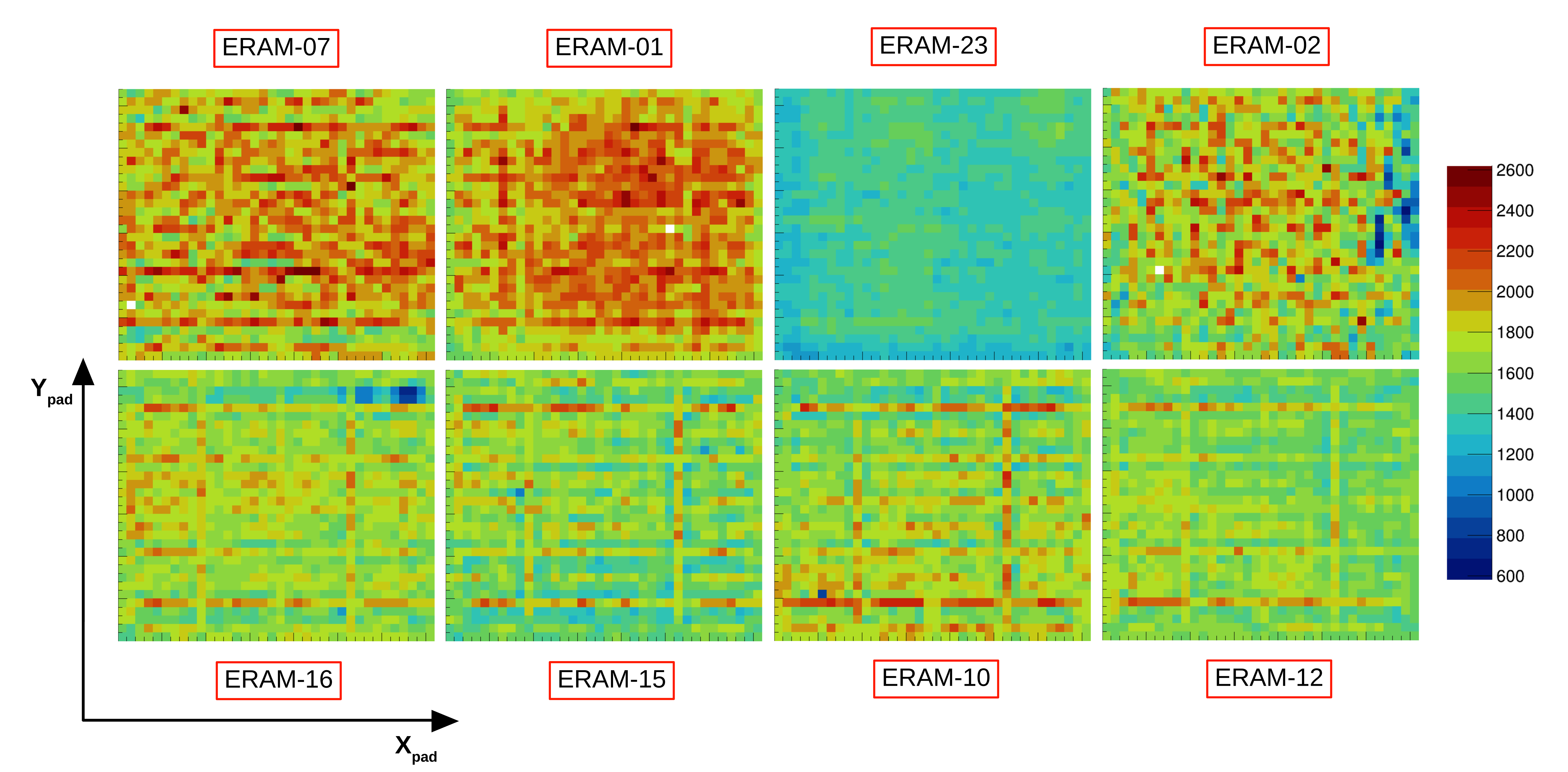}
  \caption{Gain maps of eight ERAMs tested together in a field cage prototype.}
  \label{fig:GainMaps8ERAMs}
\end{figure}

The gain can be directly extracted from the $Q_{e}$ distribution (Figure~\ref{fig:Gaindist_1pad}) obtained from the simultaneous fit by taking into account the number of primary electrons according to eq.~\ref{equ:Gain-ChargeRatio}.
Figure~\ref{fig:GainMaps8ERAMs} depicts the 2D gain maps of eight other ERAMs that have been characterized. It is found that the absolute gain of the ERAM detectors changes with the design. This effect is still under investigation. 

\begin{figure}[hbt!]
  \centering
  \includegraphics[width=0.8\textwidth]{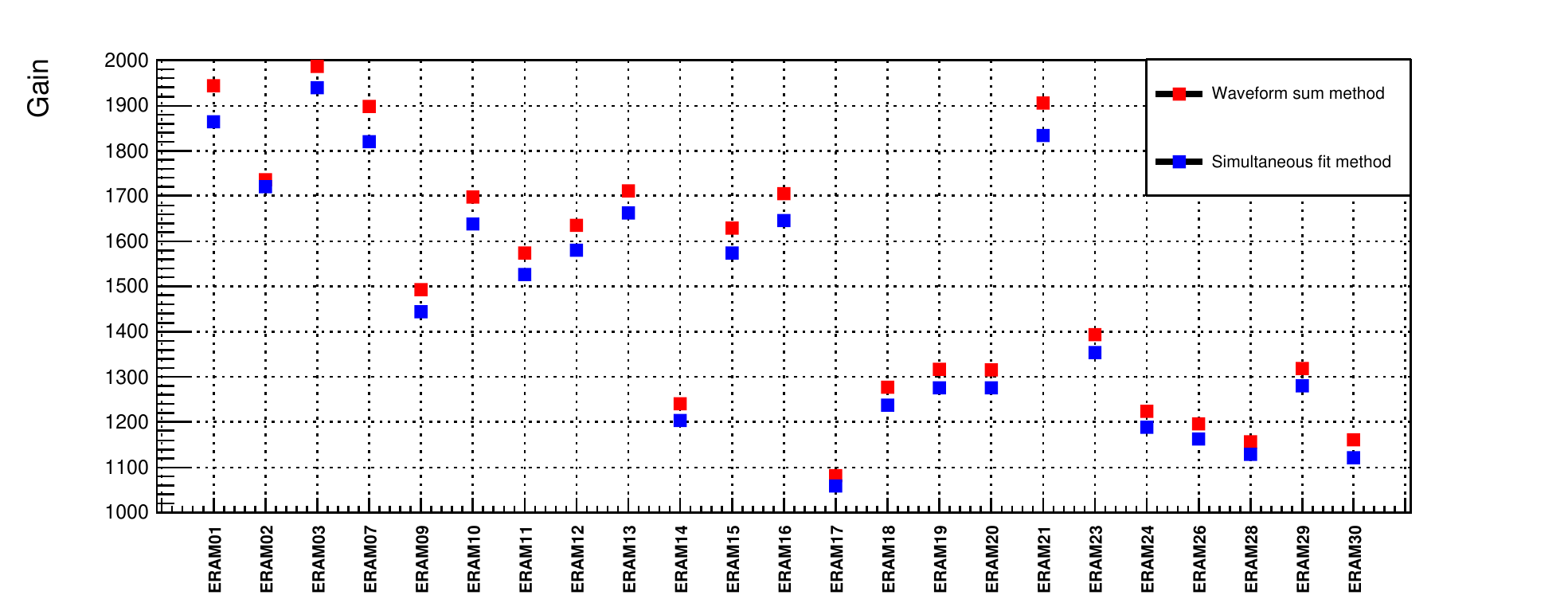}
  \caption{Comparison of gain extracted using the waveform sum and simultaneous fit methods for all the analyzed ERAMs.}
  \label{fig:G2_erams}
\end{figure}

Figure~\ref{fig:G2_erams} compares the gain obtained from the two different methods. The gain difference between the two methods for different ERAMs is less than 4\%. 
This systematic difference is due to the electronics effect in the simultaneous fit. In fact, the effect on the gain caused by using a fixed value of ($Q$, $w_{s}$) pair rather than accounting for pad by pad variations in $Q$ and $w_{s}$ values (as described in section~\ref{sec:fecRC}), was found to produce a variation of 5\% in the gain. The percentage variation in gain depends on the value of ($Q$, $w_{s}$) of each FEC channel relative to the constant value considered for the simultaneous fit. Hence, this variation would differ among different front-end cards.

\begin{figure}[hbt!]
  \centering
  \includegraphics[width=0.75\textwidth]{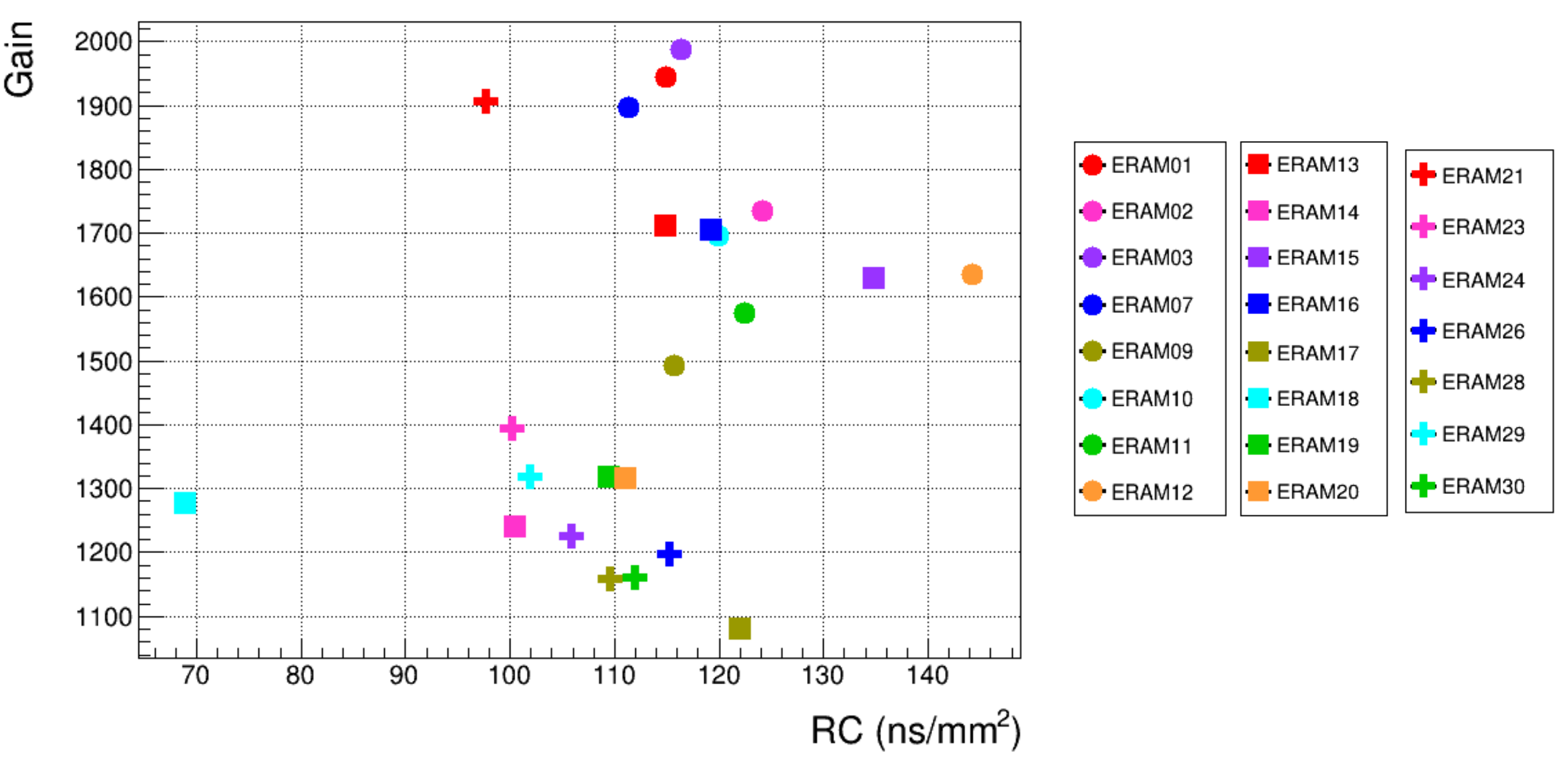}
  \caption{Dependence of mean $RC$ on mean gain of all the analyzed ERAMs.}
  \label{fig:RCG_erams}
\end{figure}

Figure~\ref{fig:RCG_erams} shows the dependence of the mean $RC$ on the mean gain values of these ERAMs. There is no obvious correlation between the $RC$  and absolute gain values. This observation confirms the result obtained in section~\ref{sec:DLCRC}.
\begin{figure}[hbt!]
     \centering
     \begin{subfigure}[b]{0.45\textwidth}
         \centering
         \includegraphics[width=\textwidth]{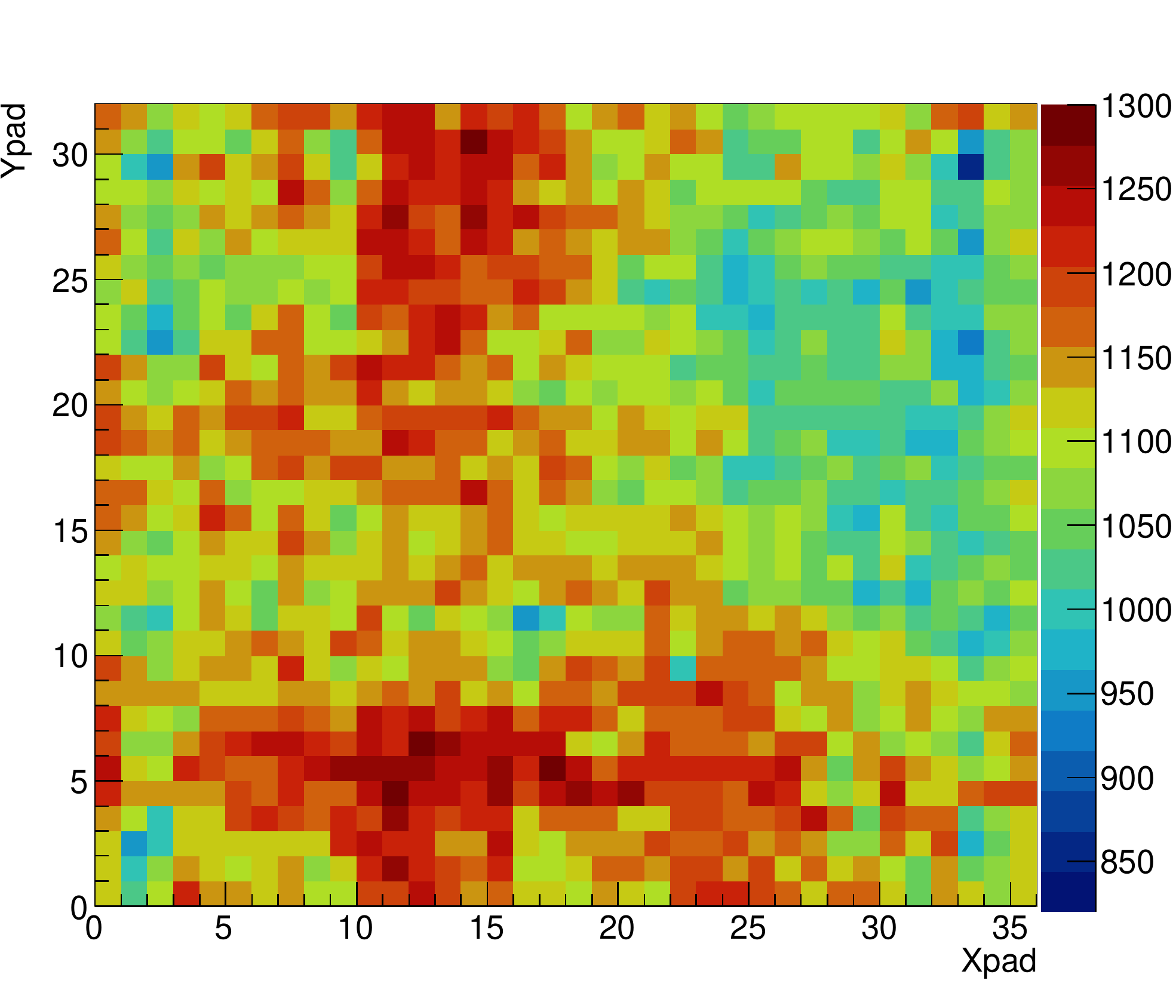}
         \caption{Gain map of ERAM-30 obtained from the simultaneous fit method.}
         \label{fig:Gainmap_eram30}
     \end{subfigure}
     \hfill
     \begin{subfigure}[b]{0.49\textwidth}
         \centering
         \includegraphics[width=1.1\textwidth]{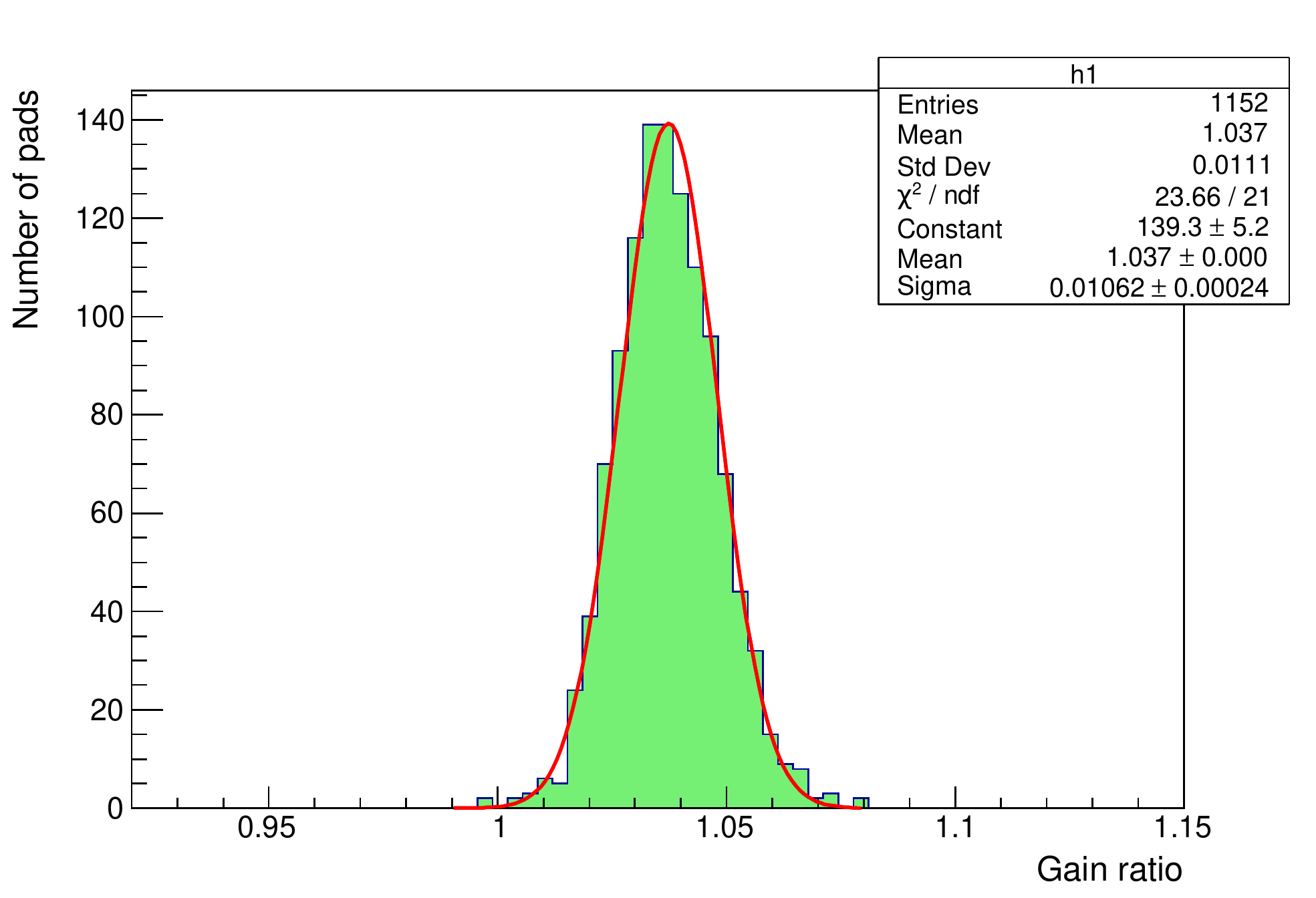}
         \caption{Ratio of the gain distributions of ERAM-30 obtained using two independent methods : the waveform sum method and simultaneous fit method.}
         \label{fig:Gain_ratio}
     \end{subfigure}
        \caption{Gain information extracted from ERAM-30 test bench data.}
        \label{fig:Gain_eram30}
\end{figure}
\subsection{Removing the PCB Soldermask}
\label{subsec:ProblemGain}
\begin{figure}[hbt!]
     \centering
     \begin{subfigure}[b]{0.33\textwidth}
         \centering
         \includegraphics[width=\textwidth]{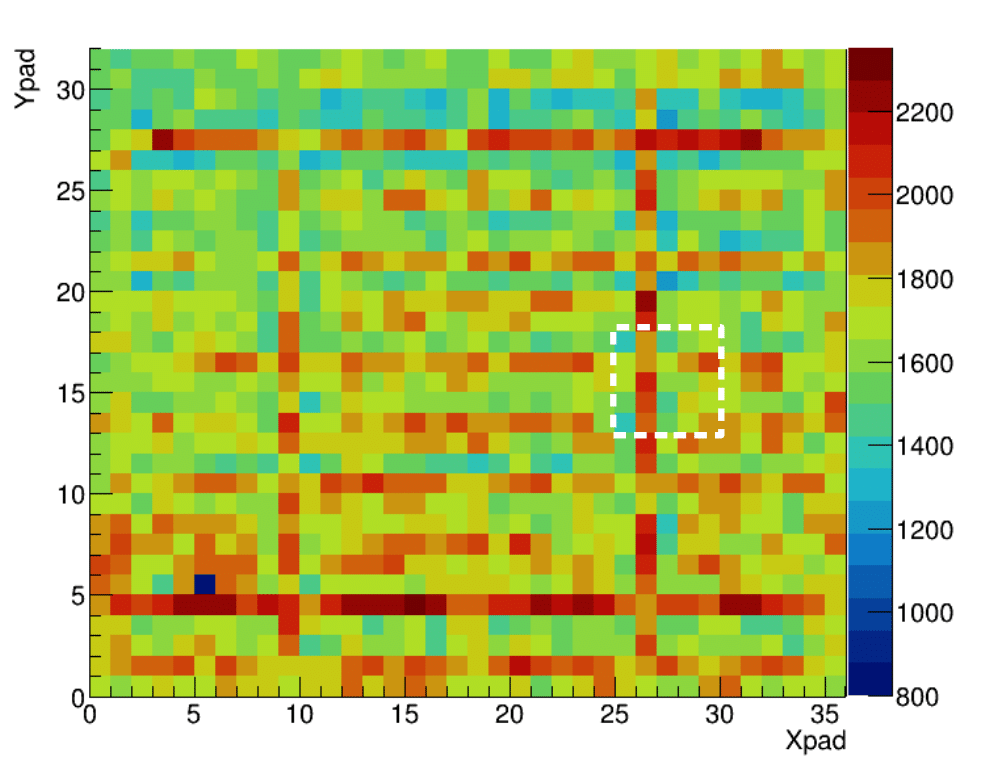}
     \end{subfigure}
     \hfill
     \begin{subfigure}[b]{0.33\textwidth}
         \centering
         \includegraphics[width=\textwidth]{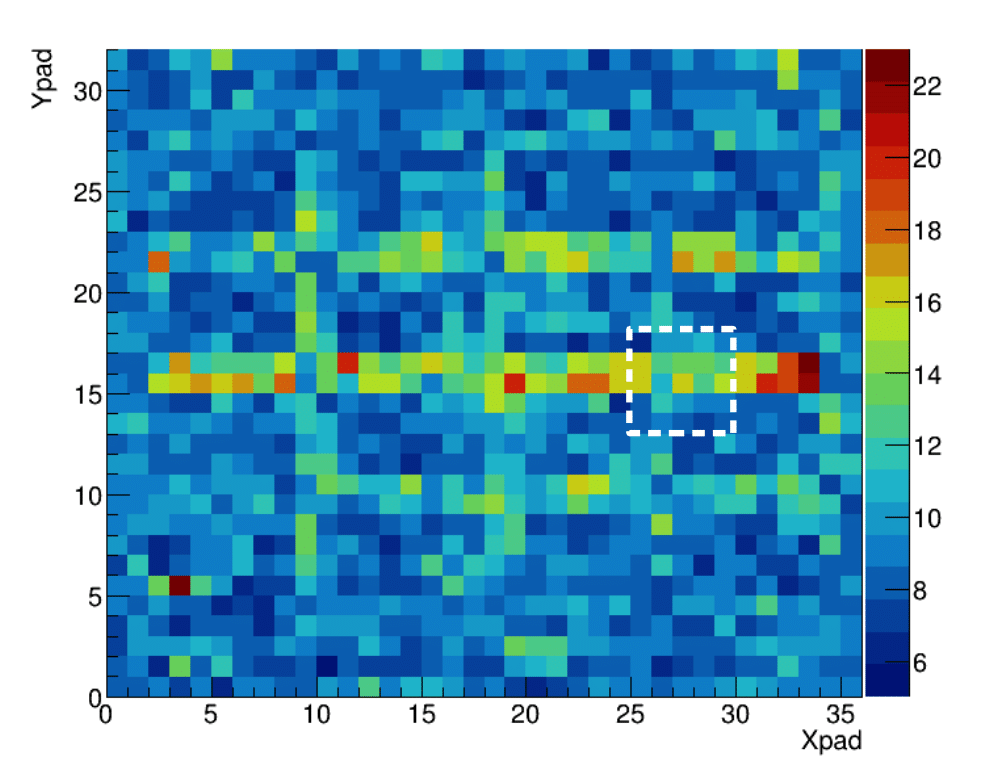}
     \end{subfigure}
          \hfill
     \begin{subfigure}[b]{0.31\textwidth}
         \centering
         \includegraphics[width=\textwidth]{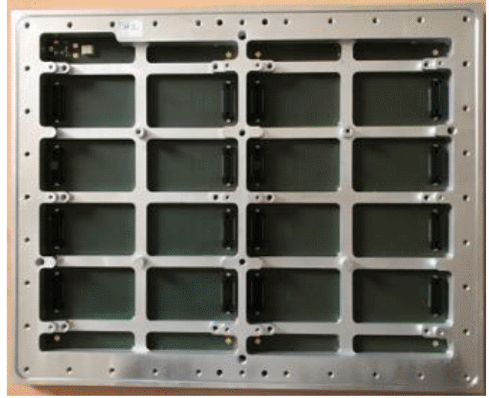}
     \end{subfigure}
        \caption{Left: 2D map of the relative gain in ADC of the ERAM-10 module; Middle: 2D map of the energy resolution in \% of the ERAM-10 module; Right: PCB top layer: the grey area are 20-35 $\mu$m thick copper + 50 $\mu$m soldermask while the cross hatched area is made of copper mesh only.}
        \label{fig:gainMapPCBsoldermask}
\end{figure}

\begin{figure}[hbt!]
  \centering
  \includegraphics[width=0.65\textwidth]{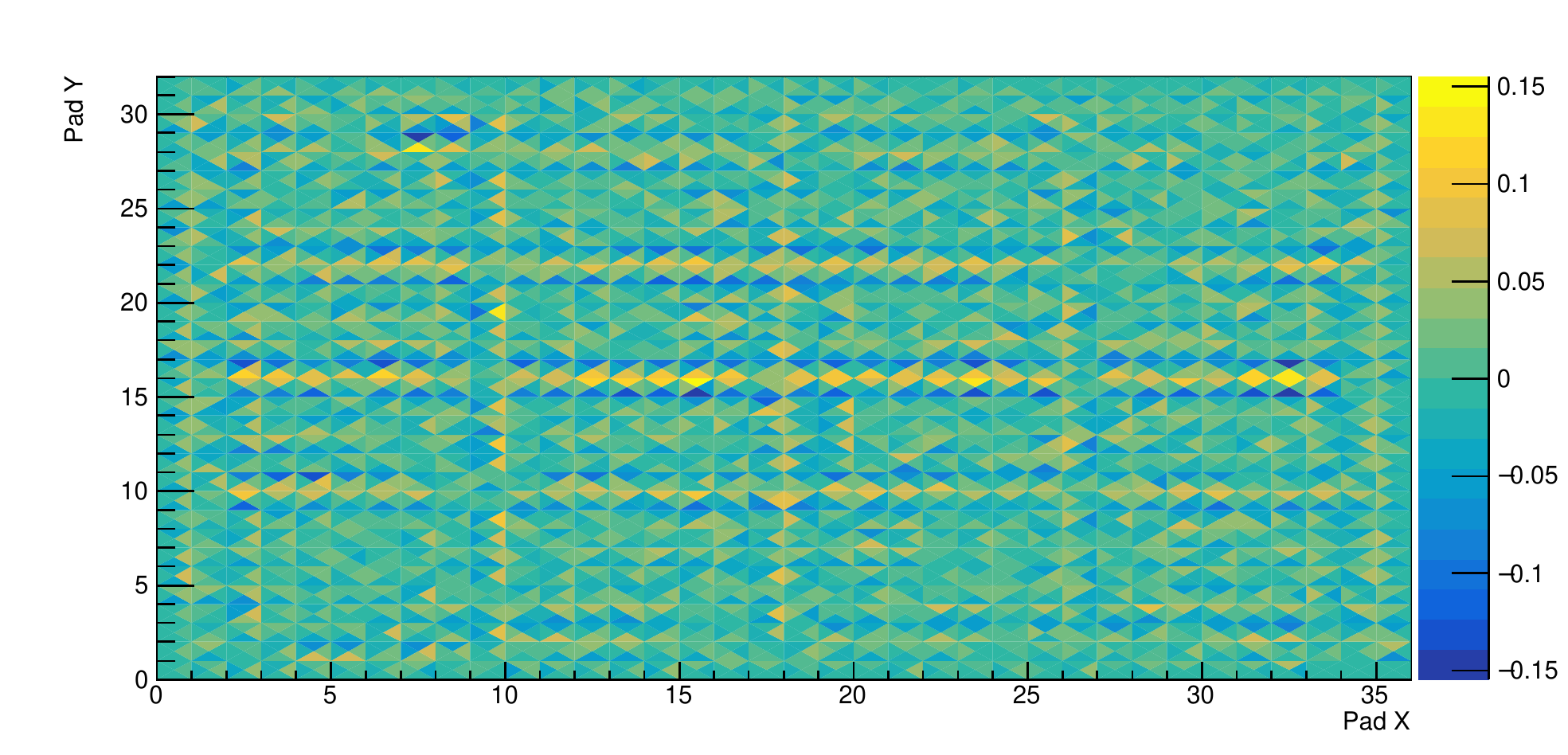}
  \caption{A map of gain non-uniformity within a pad. The $z$ axis represents the relative shift of the mean amplitude reconstructed in the top, bottom, left or right region of the pad under study with respect to the mean amplitude of the pad. }
  \label{fig:ERAM11plot}
\end{figure}

The 2D gain maps of some ERAM modules have shown grid pattern strangely similar to the shape of the soldermask of the PCB top layer as illustrated by Figure~\ref{fig:gainMapPCBsoldermask}.\\
A basic-level variable is used to study the uniformity of the gain within a given pad. It is defined as the relative shift of the mean amplitude reconstructed in the top, bottom, left or right region of the pad under scrutiny with respect to the mean amplitude of that pad. This variable highlights the non-uniformity of gain inside a pad taking the mean pad amplitude as a reference. A non-homogeneity of the gain up to 30\% within a pad is observed along the area showing the grid pattern as illustrated in Figure~\ref{fig:ERAM11plot}. 

\begin{figure}[hbt!]
  \centering
  \includegraphics[width=0.8\textwidth]{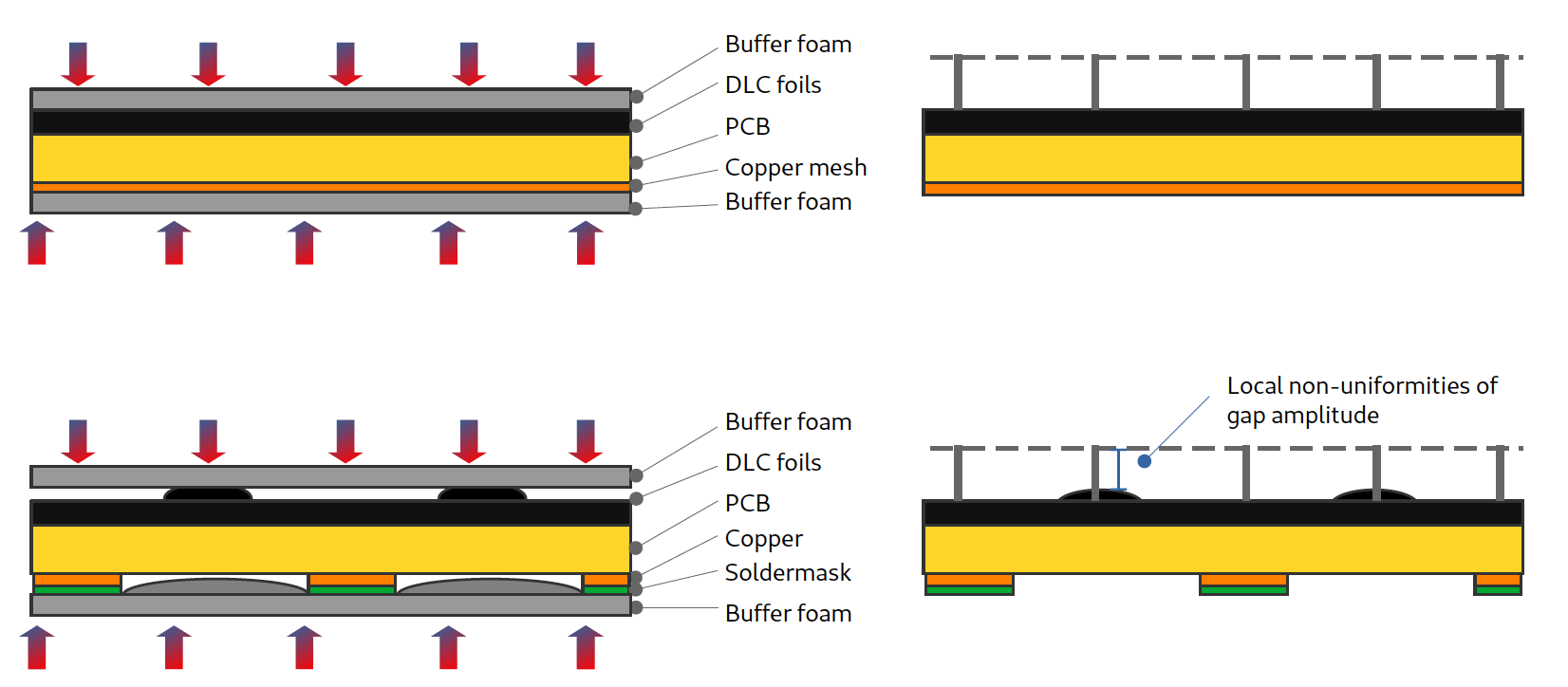}
  \caption{Schematic view of the DLC pressing on the PCB during detector assembly resulting in the non-uniformities observed on the 2D gain and energy resolution maps. The arrows represent the mechanical constraints applied which are evenly distributed when the soldermask is removed and replaced by the copper mesh.}
  \label{fig:PCBsoldermask}
\end{figure}

\begin{figure}
     \centering
     \begin{subfigure}[b]{0.44\textwidth}
         \centering
         \includegraphics[width=\textwidth]{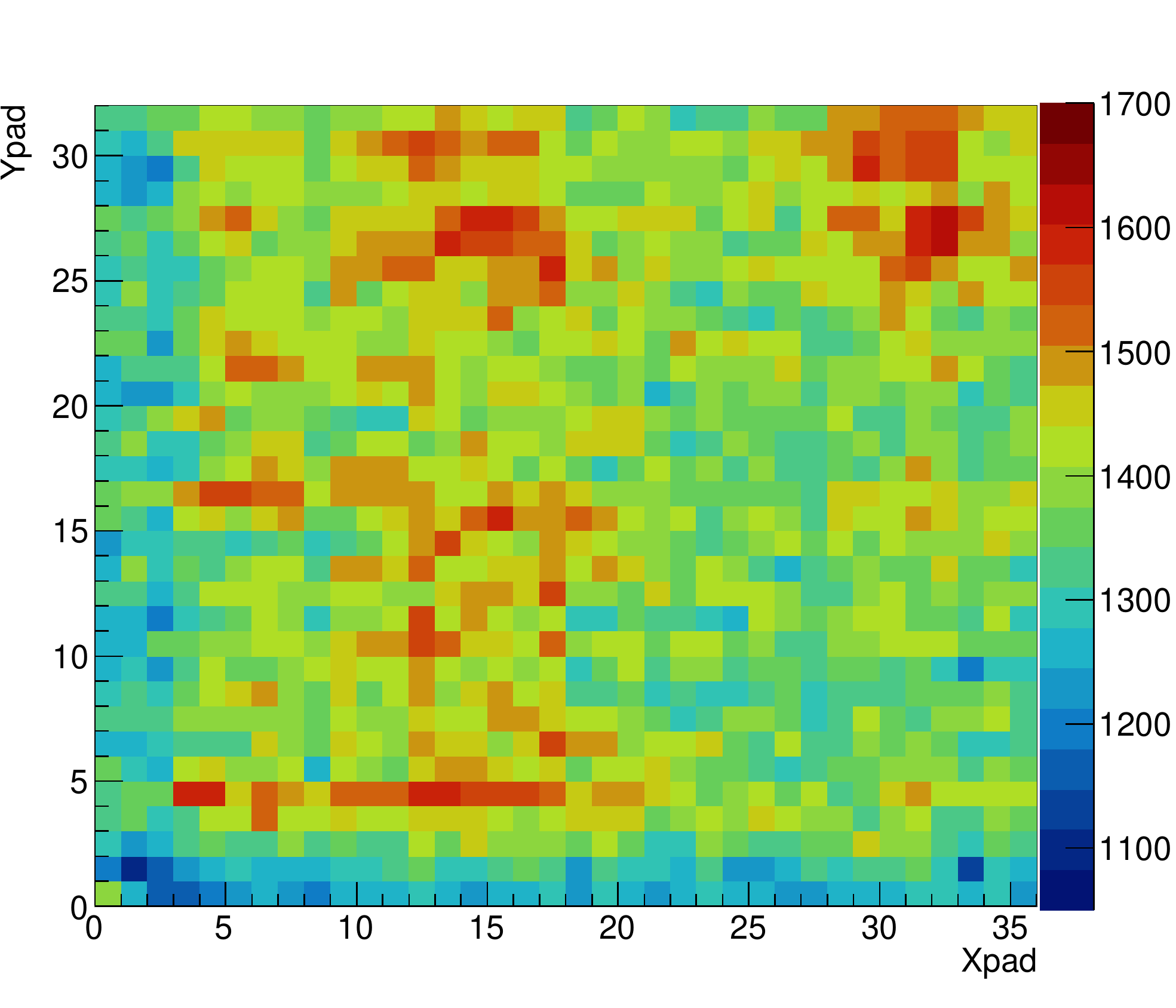}
     \end{subfigure}
     \hfill
     \begin{subfigure}[b]{0.44\textwidth}
         \centering
         \includegraphics[width=\textwidth]{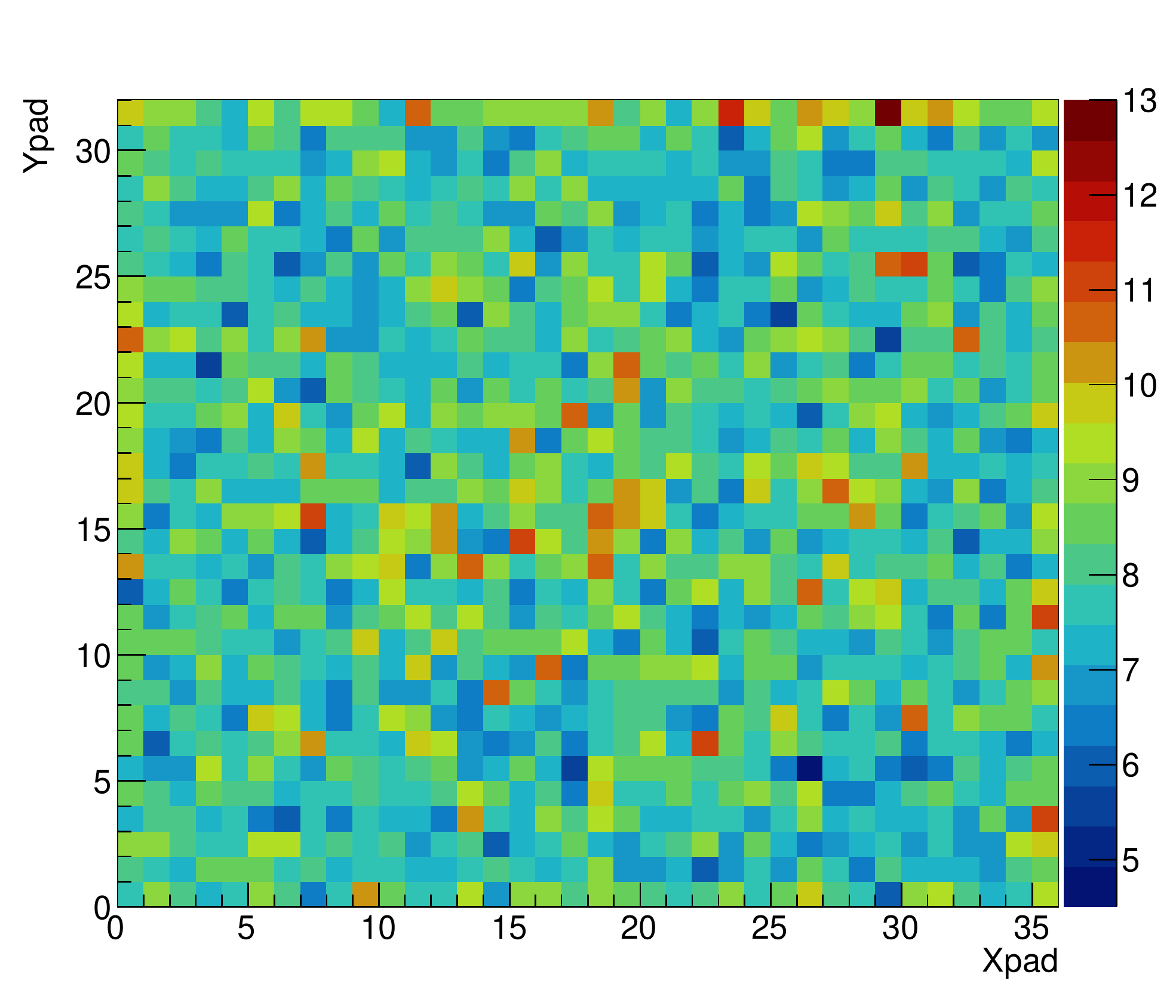}
     \end{subfigure}
        \caption{2D map of the relative gain in ADC   (left) and the energy resolution in \% (right) after modification of the PCB bottom layer. To be compared with Figure~\ref{fig:gainMapPCBsoldermask}.}
        \label{fig:gainMapAfterRemovingPCBsoldermask}
\end{figure}

Indeed, as illustrated by Figure~\ref{fig:PCBsoldermask}, when pressing the DLC on the PCB during the detector assembly, the non-uniformity of the PCB bottom layer results in an unequal distribution
of mechanical constraints leading to a reduction of the amplification
gap aligned with the stiffener grid. Considering the electric field in the amplification gap, a variation of only few microns is enough to explain the measured gain fluctuations. To solve this problem, the soldermask and copper plates are replaced by a uniform copper mesh. Figure~\ref{fig:gainMapAfterRemovingPCBsoldermask} shows the performances of the detector after correction of the PCB bottom layer.

\subsection{Gain variations within a pad}
\begin{figure}[hbt!]
  \centering
  \includegraphics[width=0.8\textwidth]{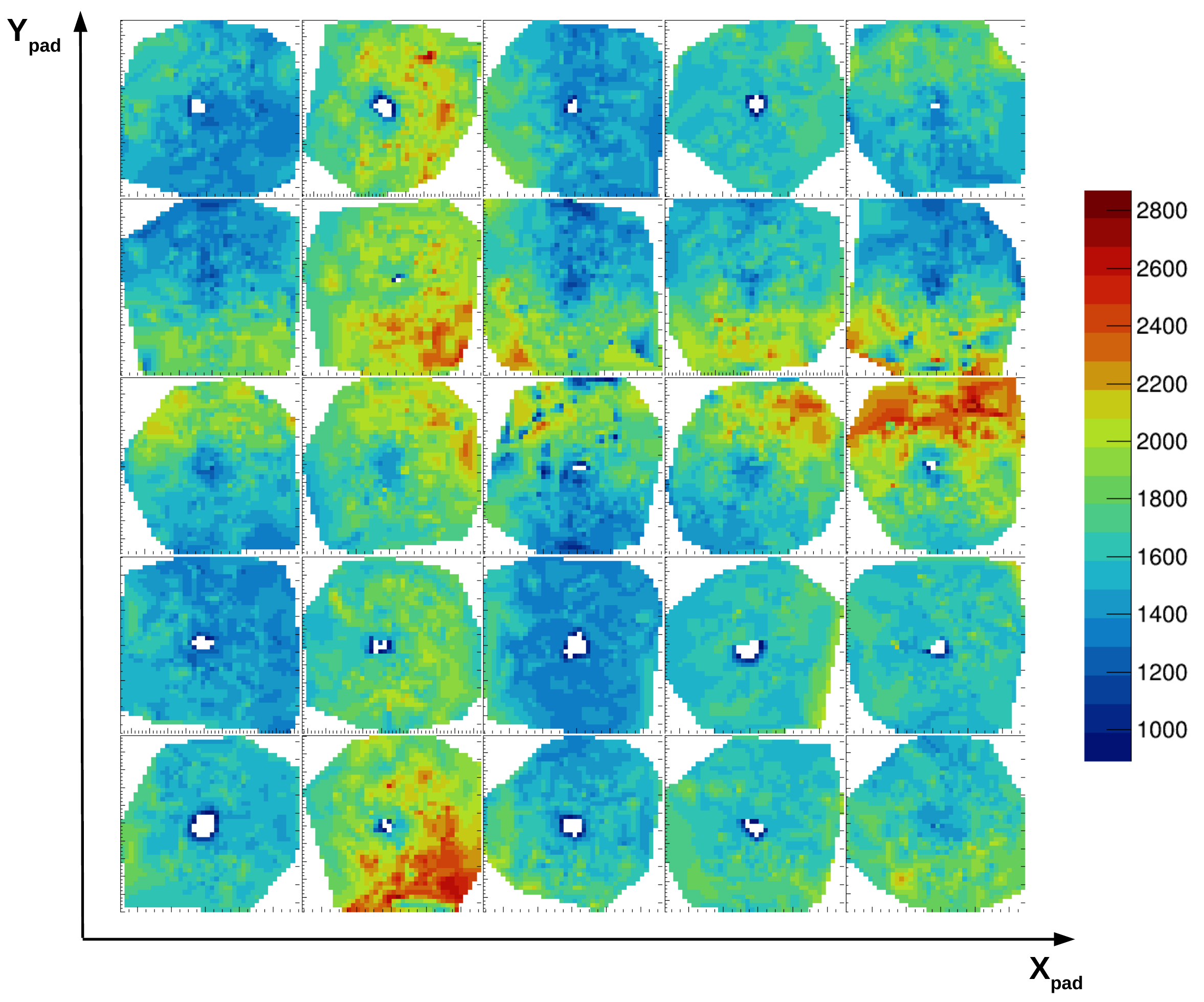}
  \caption{Visualization of gain non-uniformity within pads located over PCB stiffener, with high granularity, made possible due to simultaneous fit method.}
  \label{fig:GainNonUnifomityInOnePad}
\end{figure}

Using simultaneous fit results, the gain variations within a given pad can be studied with a high level of detail. This can be achieved by plotting the gain corresponding to the position of each charge deposition point ($x_{0}$, $y_{0}$) in a pad. 
 Using this capability, changes in gain within a pad caused by the underlying soldermask (section~\ref{subsec:ProblemGain}) can be probed. \\
Figure~\ref{fig:GainNonUnifomityInOnePad} shows the high granularity gain map of each pad in the $5\times5$ grid shown as white dashed squares in the gain and resolution maps in Figure~\ref{fig:gainMapPCBsoldermask}. The pads in the grid that lay on top of the soldermask bars have a different gain structure than that of other pads, as it can also be seen in Figure~\ref{fig:ERAM11plot}. This is especially evident for the pads lying atop the horizontal bar (second and third row from top in Figure~\ref{fig:GainNonUnifomityInOnePad}), for which a clear distinction in intensity of different gain regions is visible which demarcates the part of a pad with soldermask underneath, from the part without. This study was made possible thanks to the excellent resolution of the fit in reconstructing the charge position. 

\section{Effect of Environmental Conditions on Gain }
\label{sec:Gain_correction}
As mentioned in section~\ref{sec:setup}, gas environmental conditions within the X-ray chamber are closely monitored as variations in these conditions can have an effect on the gain. The following conditions have an appreciable effect on gain: gas temperature, chamber pressure and relative gas humidity. In order to observe their effects on the gain, an ERAM was scanned twice at  different times through the same gas flow rate.

\begin{figure}[hbt!]
  \centering
  \includegraphics[width=\textwidth]{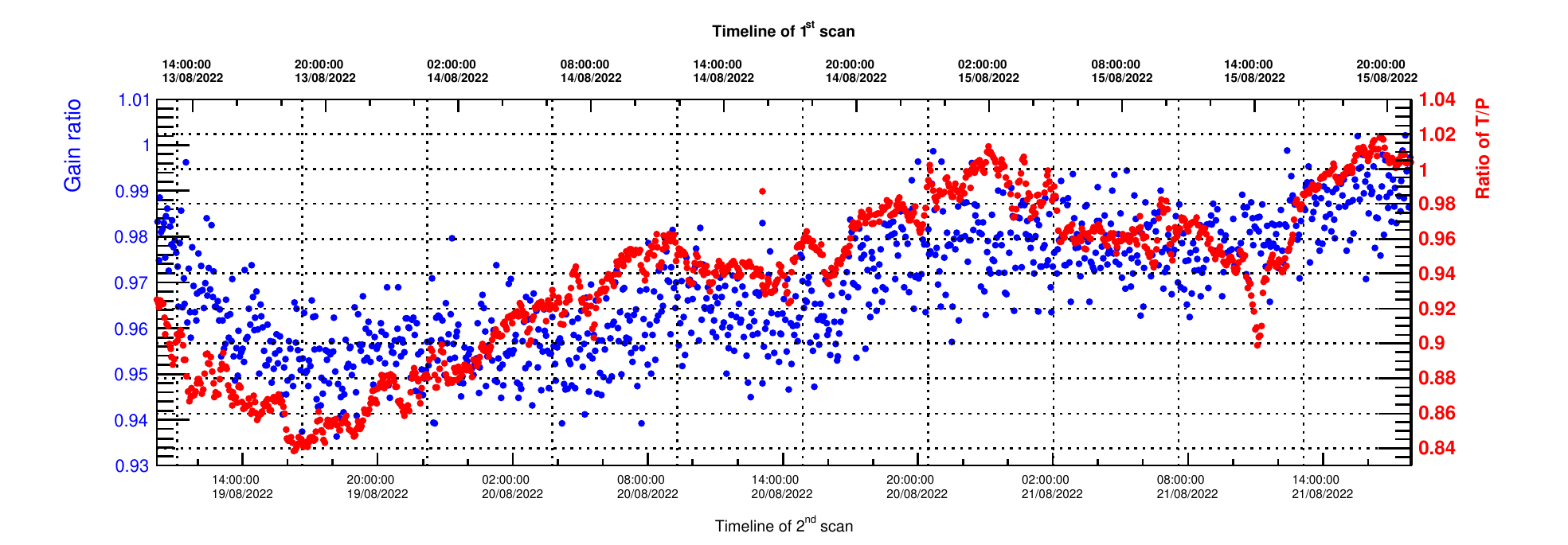}
  \caption{Effect of  $T/P$ on gain of an ERAM. The top and bottom $x$-axes represent the timelines of the two full detector scans.}
  \label{fig:Gain_TP-effect}
\end{figure}
The effect that the ratio of temperature over pressure ($T/P$) has on gain is shown in Figure~\ref{fig:Gain_TP-effect}. The left $y$-axis and points in blue represent the ratio of gain computed in each pad during two different scans, while the right $y$-axis and points in red represent the ratio of  $T/P$ recorded while each pad had been scanned at two different instances. The top and bottom $x$-axes represent the timelines of the two full detector scans. A direct correlation between gain and  $T/P$ is clearly visible in Figure~\ref{fig:Gain_TP-effect}.

\begin{figure}[hbt!]
  \centering
  \includegraphics[width=\textwidth]{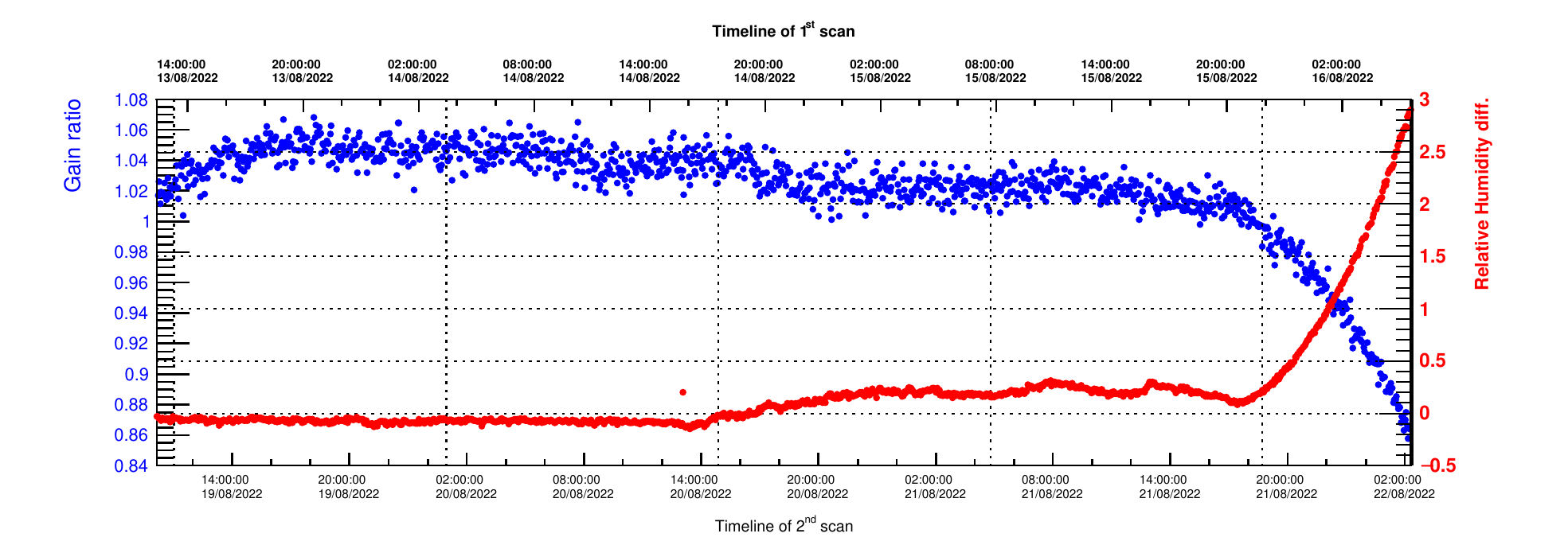}
  \caption{Effect of relative humidity on gain of an ERAM. The top and bottom $x$-axes represent the timelines of the two full detector scans.}
  \label{fig:Gain_RH-effect}
\end{figure}

The gas conditions are tightly controlled in the X-ray test bench to avoid any major fluctuations. Relative humidity is one such condition whose levels are closely gauged, and has a set upper limit above which an X-ray scan is not carried out. Despite these measures, in a rare instance, the humidity started rising rapidly close to the end of a full detector scan, which served as a good case study to observe its effect on gain. Figure~\ref{fig:Gain_RH-effect} depicts the aforementioned event w.r.t a normal repeated scan, and its consequent effect on gain. The plot follows the same convention as described in the previous paragraph, except that the right $y$-axis and red points represent the difference in relative humidity between the atypical full scan and a typical one. 
As seen in Figure~\ref{fig:Gain_RH-effect}, when the humidity starts rising near the end of the full scan, the gain starts to drop drastically. By the time the full scan ended, the relative humidity had reached 3\%, which caused an approximately 14\% drop in gain.

\begin{figure}[hbt!]
  \centering
  \includegraphics[width=0.55\textwidth]{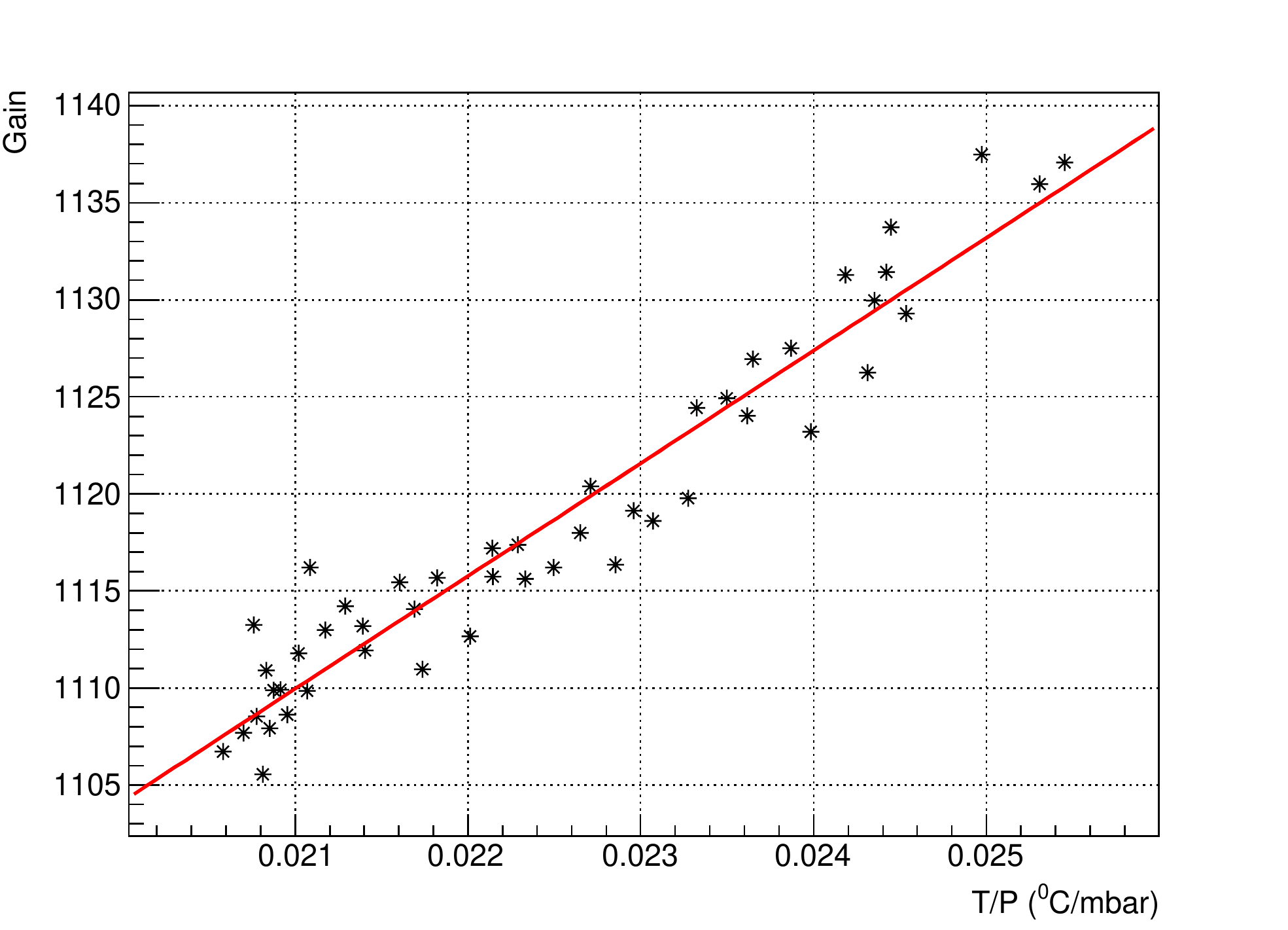}
  \caption{Effect of T/P on the gain of one pad. The lowest value of (T,P) recorded during the scan was (19.8 $^{\circ}$C, 959.5 mbar) and the highest value was (24.5 $^{\circ}$C, 963.7 mbar).} 
  \label{fig:Gain_T-effect}
\end{figure}

In order to quantify the variations of the gain caused by variations of temperature and pressure, 
 the same pad should be scanned for an extended period of time. For this study, a pad was scanned for 24 hours, and its data was divided into 48 datasets of events collected for a duration of 30 minutes each. The gain was extracted from each of the datasets and compared with the average of temperature and pressure values recorded during that time window. Figure~\ref{fig:Gain_T-effect} depicts a direct and linear relation between gain and the ratio of temperature and pressure. 

\section{Conclusions}
\label{sec:conclusion}
The production of encapsulated resistive anode bulk Micromegas modules is well underway. The produced modules are tested and validated using a X-ray test bench at CERN before their installation in the HA-TPC chambers of the T2K experiment. 
The X-ray test bench is used to characterize the detectors by scanning
each pad individually and therefore precisely measure the uniformity of the gain and energy resolution over the pad plane. An energy resolution of about 10\% was measured. \\
A detailed physical model has been developed to describe the charge dispersion phenomena on a resistive Micromegas anode. This model allows for the simultaneous extraction of 2D gain and $RC$ maps of the modules using X-ray data. Excellent agreement is found between the data and the model. The $RC$ and gain maps uniformity are studied in detail.
The measured $RC$ and gain information will be used to feed the HA-TPC simulation and reconstruction.
\section*{Acknowledgements}
We acknowledge the support of CEA and CNRS/IN2P3, France; DFG, Germany; INFN, Italy; National Science Centre (NCN) and Ministry of Science and Higher Education (Grant No. DIR/WK/2017/05), Poland; the Spanish Ministerio de Economıa y Competitividad (SEIDI - MINECO) under Grant No. PID2019-107564GB-I00 (IFAE, Spain). IFAE is partially funded by the CERCA program of the Generalitat de Catalunya.

This work was supported by P2IO LabEx (ANR-10-LABX-0038 – Project “BSMNu”) in the framework "Investissements d’Avenir" (ANR-11-IDEX-0003-01), managed by
the Agence Nationale de la Recherche (ANR), France. 
 In addition, the participation of individual researchers and institutions has
 been further supported by H2020 Grant No. RISE-GA822070-JENNIFER2 2020, MSCA-COFUND-2016 No.754496, ANR-19-CE31-0001, RFBR grants \#19-32-90100, the Secretariat for Universities and Research of the Ministry
 of Business and Knowledge of the Government of Catalonia and the European Social Fund (2022FI\textunderscore B 00336) and from the program Plan de Doctorados Industriales of the Research and Universities Department of the Catalan Government (2022 DI 011).





\bibliographystyle{elsarticle-num}
\bibliography{bibliography}

\end{document}